\documentclass[iop,numberedappendix,twocolappendix]{emulateapj}
\usepackage{amsmath}
\usepackage{natbib}
\newcommand{\rhosink}{\ensuremath{\rho_{\rm s}}}
\newcommand{\rhoacc}{\ensuremath{\rho_{\rm acc}}}
\newcommand{\rexcl}{\ensuremath{r_{\rm ex}}}
\newcommand{\racc}{\ensuremath{r_{\rm acc}}}
\newcommand{\eout}{\ensuremath{\epsilon_{\rm acc}}}

\graphicspath{{figures/}}

\submitted{}
\shorttitle{}
\shortauthors{Haugb{\o}lle, Padoan, Nordlund}

\begin{document}
\title{The stellar IMF from Isothermal MHD Turbulence}

\author{Troels Haugb{\o}lle}
\affil{Centre for Star and Planet Formation, Niels Bohr Institute \& Natural History Museum of Denmark, University of Copenhagen,
{\O}ster Voldgade 5-7, DK-1350 Copenhagen, Denmark; haugboel@nbi.ku.dk}
\author{Paolo Padoan}
\affil{Institut de Ci\`{e}ncies del Cosmos, Universitat de Barcelona, IEEC-UB, Mart\'{i} i Franqu\`{e}s 1, E08028 Barcelona, Spain}
\affil{ICREA, Pg. Llu\'{i}s Companys 23, 08010 Barcelona, Spain; ppadoan@icc.ub.edu}
\author{{\AA}ke Nordlund}
\affil{Centre for Star and Planet Formation, Niels Bohr Institute \& Natural History Museum of Denmark, University of Copenhagen,
{\O}ster Voldgade 5-7, DK-1350 Copenhagen, Denmark; aake@nbi.ku.dk}

\begin{abstract}
We address the turbulent fragmentation scenario for the origin of the stellar initial mass function (IMF), using a large set of numerical
simulations of randomly driven supersonic MHD turbulence. The
turbulent fragmentation model successfully predicts the main features of the observed stellar IMF assuming an isothermal equation
of state without any stellar feedback. As a test of the model, we focus on the case of a magnetized isothermal gas, neglecting stellar
feedback, while pursuing a large dynamic range in both space and time scales covering
the full spectrum of stellar masses from brown dwarfs to massive stars. Our simulations represent a generic $4\,\mathrm{pc}$ region within a typical
Galactic molecular cloud, with a mass of $3000\,\mathrm{M_{\odot}}$ and an rms velocity 10 times the isothermal sound speed and 5 times
the average Alfv\'{e}n velocity, in agreement with observations. We achieve a maximum resolution of $50\,\mathrm{au}$ and a maximum duration
of star formation of $4.0\,\mathrm{Myr}$, forming up to a thousand sink particles whose mass distribution closely matches the observed stellar IMF. A large set of medium-size
simulations is used to test the sink particle algorithm, while larger simulations are used to test the numerical convergence of the IMF and the
dependence of the IMF turnover on physical parameters predicted by the turbulent fragmentation model. We find a clear trend toward numerical
convergence and strong support for the model predictions, including the initial time evolution of the IMF. We conclude that the physics of
isothermal MHD turbulence is sufficient to explain the origin of the IMF.
\end{abstract}


\section{Introduction}
The origin of the stellar initial mass function (IMF) is still not fully understood. Many processes, such as magnetic support, radiative and
mechanical feedbacks from young stars, density enhancements or pressure support from supersonic turbulence, dynamical interactions
between accreting stars, or competitive accretion, may affect the mass distribution of stars. Their relative importance varies with
environment and is still disputed. Numerical simulations of star formation have started to reveal the mechanisms controlling the
star formation rate (SFR), thanks to systematic parameter studies based on large sets of simulations with a modest dynamic range of
scales \citep[e.g.][]{Padoan+Nordlund11sfr,Padoan+12sfr,Federrath+Klessen12}, or, more recently, with large-scale simulations where
many star-forming regions are formed ab initio and the time evolution and scatter of the SFR can also be investigated \citep{Padoan+17sfr}.
While the SFR is already converged in these simulations, a much larger dynamic range is needed to achieve a numerically
converged IMF covering the whole spectrum of stellar masses, from brown dwarfs to massive stars. The computational cost of such
experiments is beyond the reach of most studies of star formation, so stellar IMFs from numerical simulations are scarce and generally
of low statistical significance. Typically limited to $\sim$100 stars, these simulations barely constrain the IMF turnover and do not
yield enough massive stars to probe the Salpeter range. Unable to pursue the necessary range in space and time scales, numerical
studies often opt for increasing the physical complexity, for example, including radiative and mechanical feedbacks, in the hope of
discerning the effect of new processes on the IMF, despite the small sample size and dubious numerical convergence.

The first numerical IMF with a large enough number of sink particles to probe both the IMF turnover, including brown dwarfs, and the
Salpeter range was presented in \citet{Padoan+14acc}, where we simulated randomly driven, supersonic MHD turbulence in a 4 pc
region with a total mass of 3000 M$_{\odot}$ and a maximum resolution of 50 au. After a long initialization phase without self-gravity,
the simulation was evolved with self-gravity, generating 1288 sink particles over a period of 3.2 Myr (2.7 free-fall times). The mass
distribution of the sink particles was consistent with a Chabrier IMF \citep{Chabrier05} at low masses and a power law with Salpeter's slope
\citep{Salpeter55} above 1-2 M$_{\odot}$. In previous numerical studies, the size of the simulated region, $L$, the duration of star formation
(between the first and the last sink particles), $t_{\rm SF}$, and the total number of sink particles, $N_{\rm *}$, were respectively $\sim$10,
$\sim$100 and $\sim$10 times smaller than in \citet{Padoan+14acc}. The radiation hydrodynamic (HD) smoothed particle hydrodynamics (SPH) simulations by \citet{Bate12}
and \citet{Bate14} described a 500 M$_{\odot}$ region with $L=0.4$ pc and yielded at most $N_{\rm *}=183$ over a time $t_{\rm SF}\sim 0.09$ Myr.
The HD grid-based simulations by \citet{Krumholz+12imf}, representing a 1000 M$_{\odot}$ region, had $L=0.46$ pc, $N_{\rm *}=158$, and an even
shorter star formation time, $t_{\rm SF}\sim 0.02$ Myr. The more recent MHD simulations by \citet{Myers+14imf}, modeling a region of the
same size and mass, yielded at most $N_{\rm *}=92$ over a time $t_{\rm SF}\sim 0.05$ Myr.

The earlier barotropic SPH simulations by \citet{Bate09}, with 500 M$_{\odot}$ in a region with $L=0.4$ pc, achieved a large number of sink
particles, $N_{\rm *}=1254$, and a slightly longer star formation time, $t_{\rm SF}\sim 0.15$ Myr. However, that IMF peaked at 0.02 M$_{\odot}$,
an order of magnitude below the peak of the Chabrier IMF, so most of the sink particles had very low masses, which are created only in the
absence of a magnetic field or radiation feedback \citep{Bate12}, and would not exist in nature. An even earlier SPH simulation by \citet{Bonnell+03},
modeling 1000 M$_{\odot}$ in a region with $L=1$ pc, reached $N_{\rm *}\approx 400$ and $t_{\rm SF}\sim 0.35$ Myr. This IMF peaked around
0.3 M$_{\odot}$ and produced a realistic Salpeter range. However, lacking both magnetic field and radiation feedback, the realistic peak was
almost certainly the result of the very limited number of SPH particles, $5\times 10^5$ compared with $3.5\times 10^7$ in \citet{Bate09}. With
such a low number of SPH particles, the IMF was incomplete just below its peak and, more importantly, the turbulent velocity field could not be
resolved well enough to describe the smallest scales of the turbulent fragmentation process responsible for the IMF turnover. Similar
considerations apply to the more recent SPH simulations by \citet{Ballesteros_Paredes+15imf}, modeling a region of the same size, same mass, and
similar duration of star formation as in \citet{Bonnell+03}.

SPH and grid-based simulations without a magnetic field have been used to claim that the radiative feedback from accreting protostars
(perhaps even the mechanical feedback from stellar outflows) is necessary to reproduce the observed IMF \citep{Bate09,Bate12,Krumholz+12imf,Bate14}.
MHD simulations focusing on the effect of the magnetic field strength on star formation have included the radiative feedback as well, without exploring the isothermal
case \citep{Myers+14imf}, and thus could not establish the relative importance of the magnetic field and radiation feedback in controlling the IMF.
According to the turbulent fragmentation models
\citep{Padoan+97imf,Padoan+Nordlund02imf,Hennebelle+Chabrier08imf,Hennebelle+Chabrier09imf,Padoan+Nordlund11imf,Hopkins12imf},
the IMF originates primarily as the consequence of supersonic turbulence and can be reproduced under an isothermal gas approximation. Thus, it is important to
carefully test the case of supersonic MHD turbulence of an isothermal gas, using as realistic as possible initial conditions and pursuing a large dynamic range
in space and time scales in order to probe the whole IMF with a large statistical sample.

As mentioned above, in \citet{Padoan+14acc} we already succeeded in reproducing the observed IMF with a simulation of
isothermal MHD turbulence that represented a significant step forward in terms of the size of the simulated region, the length of the star formation time,
and the number of sink particles. This model has since been refined with slightly better hydrodynamics and more optimal parameters for the sink particle model
\citep{Frimann+16b,Jensen+Haugboelle17}. In this work, we carry out a more systematic study of the isothermal MHD case, using a large number of simulations
to test the dependence of the results on the numerical parameters of the sink particle model (Sections 2 and 3 and Appendix~\ref{A:sinks}), to verify the numerical convergence
of the IMF (see \S~\ref{sec_conv}), and to test the predicted variability of the IMF due to variations of physical parameters (virial parameter or mean density) or
resulting from the early time evolution of the IMF (see \S~\ref{sec_var}). We also use observed properties of molecular clouds (MCs) to constrain the
environmental dependence of the IMF peak (\S~\ref{sec_env}) and stress that the IMF time evolution predicted by our turbulent fragmentation model
\citep[][hereafter PN02]{Padoan+Nordlund02imf}, a consequence of the relatively long timescale of massive star formation \citep{Padoan+Nordlund11imf,Padoan+14acc}, 
is not necessarily predicted by other turbulent fragmentation models \citep{Hennebelle+Chabrier08imf,Hennebelle+Chabrier09imf,Hopkins12imf}, where massive stars result 
from the collapse of massive cores, in line with the scenario of massive star formation by \citet{McKee+Tan02,McKee+Tan03}. However, we do not see an accelerated 
accretion rate as the stars gain mass, so the results of our simulations are also at odds with the predictions of the competitive accretion scenario 
\citep{Bonnell+01comp,Bonnell+Bate06competitive}.

This work focuses on star formation under physical conditions typical of Galactic MCs, i.e.~\emph{turbulent} regions of cold molecular gas following the observed
velocity--size relation (with a large scatter). Different conditions in more extreme environments may require a careful consideration of detailed processes
that are neglected here, such as radiation and mechanical feedbacks.

\section{Numerical Methods}\label{sec:methods}
The simulations are carried out with a locally developed version of the public adaptive mesh refinement (AMR) code {\sc ramses} \citep{Teyssier02} that includes random turbulence
driving and a robust algorithm for sink particles. Compared to the public version, it has been heavily modified to scale better on
supercomputers with large number of cores per node, using an OpenMP/MPI hybrid parallelization, improved MPI load balancing,
and with special attention to minimizing and bundling MPI communication. We have also improved the stability of the HLLD in high Mach
number flows (see Appendix~\ref{A:MHD}), optimized the conjugate gradient method used for solving self-gravity, and improved the consistency
of gravitational forces and the stability of sink particle orbits when using subcycling in time (see App.~\ref{A:gravity}). For the models considered in this
paper we use periodic boundary conditions and an isothermal equation of state.

To initialize the turbulent state, we first run without self-gravity for $\sim$20 dynamical times, starting with uniform
density and magnetic fields and a random acceleration with power only at wavenumbers $1 \le k\le 2$ ($k=1$ corresponds
to the computational box size). The driving force keeps the rms sonic Mach number,
${\mathcal M}_{\rm s}\equiv \sigma_{\rm v}/c_{\rm s}$, at an approximate constant value, where $\sigma_{\rm v}$ is the
three-dimensional rms velocity and $c_{\rm s}$ is the isothermal speed of sound.

The driving is implemented as in
\citet{Padoan+Nordlund04bd,Padoan+Nordlund11sfr} and \citet{Padoan+12sfr}.
We use purely solenoidal driving, because of the large separation between the typical driving scale of supernovae
($\sim$70 pc) and the size of the simulated region (4 pc). In \citet{Pan+16} we showed that, in supernova driven turbulence,
the compressive ratio (compressive over solenoid power) at MC scales is consistent with that obtained in the inertial
range of randomly driven turbulence with purely solenoidal driving.

\begin{figure}[t]
\includegraphics[width=0.45\textwidth]{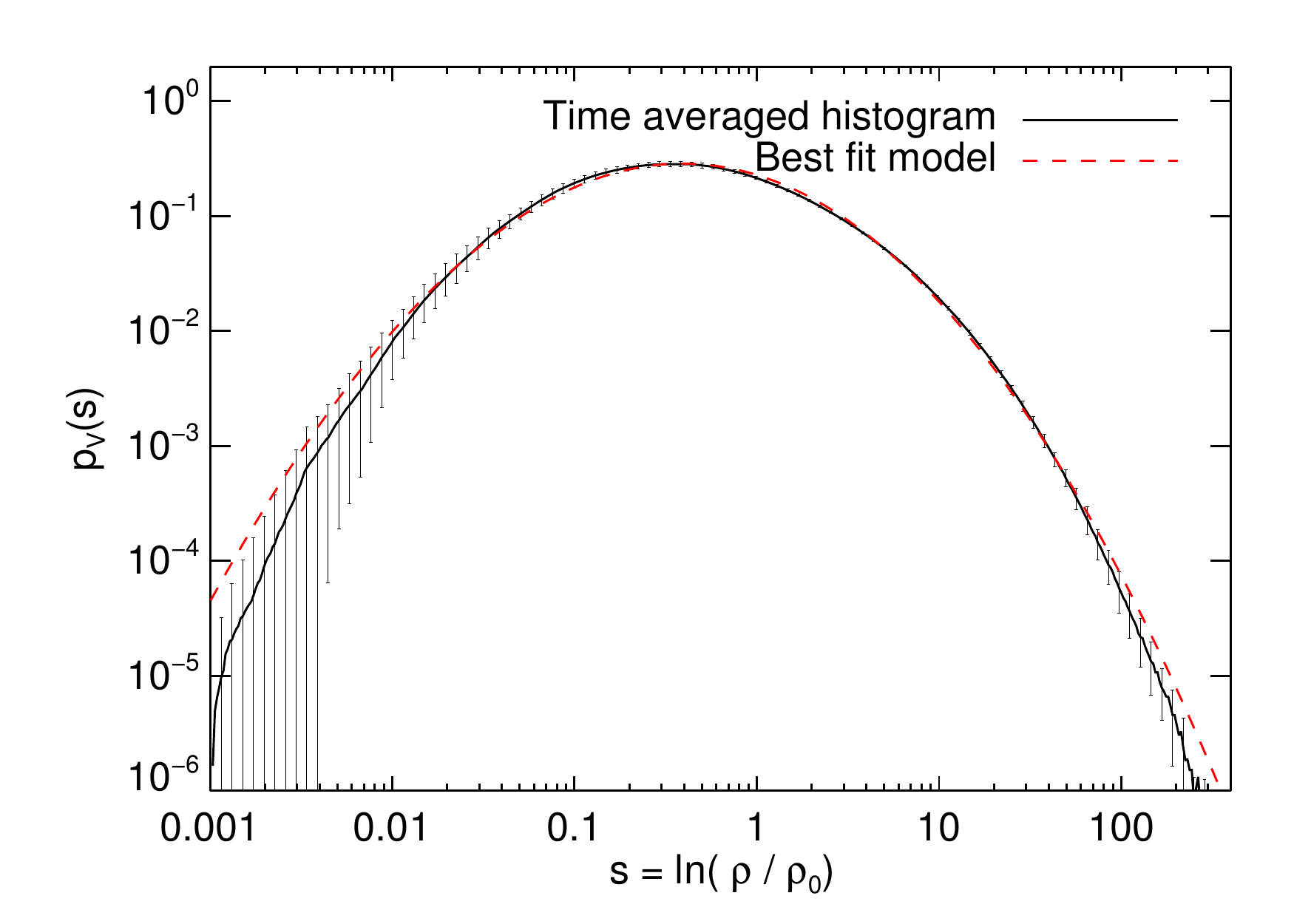}
\caption{Volume-weighted log-density probability distribution. The dashed red line is a best fit using the model proposed by
\citet{hopkins+13}. The histogram is time averaged over 10 snapshots taken from 6 to 24 turnover times, sampling the fully
developed turbulence, and is calculated from the initial run with $\mathcal M_{\rm s}=10$, $\mathcal M_{\rm a}=5$, solenoidal driving,
no self gravity, and a uniform resolution. Error bars indicate the standard deviation of variations between different
snapshots.}
\label{fig:pdf}
\end{figure}

The refinement strategy is based on the overdensity. We refine the root grid when the density reaches a certain level, typically
10 times the average density, and then refine using a Truelove refinement criterion \citep{Truelove+97},
where the grid is refined every time the density increases by a factor of four. This refinement strategy is retained when self-gravity
is turned on, though the number of AMR levels is increased to better resolve the gravitational collapse. Additionally, we expand
the most refined grid so that it is at least 8 cells across, to counteract the creation of many small island grids around density peaks.
In contrast to \citet{Kritsuk+06} and \citet{Schmidt+09}, we do not refine based on velocity or pressure gradients. However, we
verified that the tails of the density PDFs are sufficiently converged in the turbulence simulations used as initial conditions
(see figure \ref{fig:pdf}).

Gravitational collapse is central to simulations of star formation. As cloud cores in the turbulent flow become unstable and
start to collapse, the density rises by many orders of magnitude. With current computational capabilities it is very challenging to capture
the collapse to stellar densities even with very deep AMR hierarchies, and the correspondingly small time step would lead to prohibitively
expensive simulations \citep[though see][for examples where stellar densities are almost reached]{Nordlund+14,Kuffmeier+17b}.
In this work, we are not interested in the detailed stellar physics, and to correctly capture the physics at
larger scales of tens of astronomical units, we use sink particles to represent collapsed-gas regions. These should be formed robustly
where the gas has unequivocally started to collapse gravitationally, but the conversion from gas in a grid cell to a sink particle
representation cannot happen at an arbitrarily high density. An isothermal gas will fragment while collapsing,
and if the Jeans length is not sufficiently resolved, the gas can undergo numerical fragmentation \citep{Truelove+97,Truelove+98}.
The normalized Jeans length at a given density and numerical resolution, the Jeans number, is
\begin{equation}
L_J=\frac{\lambda_J}{\Delta x} = \sqrt{\frac{c_s^2 \pi}{G\rho}}\frac{1}{\Delta x}
\end{equation}
and it has been argued that one should have $L_{\rm J}\ge 4$  \citep{Truelove+97} everywhere.

To detect gravitational collapse in the gas and to convert the gas into a sink particle, we use a number of criteria:
\begin{itemize}
\item{Sink particles can only be created in cells where the gas density is above a threshold value $\rhosink$. We generally
make sure that this only happens at the highest AMR level and that the Jeans length is resolved. In most of the presented
runs the Jeans length is resolved with at least two cells at $\rhosink$ ($L_{\rm J,s}=2$), though tests have shown that even if the Jeans
length is slightly under resolved we do not generate artificially a higher amount of sinks \citep[in accordance with][]{Ostriker+13}.
This is likely because even if the gas has started to fragment a bit, the different fragments will afterward be accreted by the sink
particle. A typical value for $\rhosink$ in the runs presented here is $10^5-10^6$ times the average density.}
\item{The gravitational potential is required to have a local minimum at the cell where the sink particle will be created. This is
evaluated by smoothing with a 2x2x2 average and comparing with the potential in the 26 neighboring smoothed cells.}
\item{The velocity field has to be converging in the cell, $\nabla \cdot \boldsymbol{v} < 0$.}
\item{No other previously created sink particle can be present within an exclusion radius, \rexcl, of the cell where the new particle
is created.}
\item{In runs with an active energy equation, we also disallow creation of sinks from very warm gas. This can otherwise spur the
creation of sink particles in, e.g.,~the high-density expanding shell of a supernova.}
\end{itemize}
This sink particle recipe has  already been used in a number of papers
\citep{Padoan+12sfr,Nordlund+13,Nordlund+14,Padoan+14acc,Padoan+16SN_I,Padoan+16SN_III,Pan+16,Frimann+16a,Frimann+16b,Kuffmeier+16,Kuffmeier+17,Jensen+Haugboelle17,Padoan+17sfr}.
In the past, we had used a simpler recipe based only on a density
criteria in lower-dynamic-range simulations aimed at deriving the SFR \citep{Padoan+Nordlund11sfr}. Due to the much larger resolution in following works and the
added goal of predicting robustly the stellar IMF, we have switched to using all the above criteria. The individual conditions are similar to what has
been used by \citet{Federrath+10sinks, Bate+95sink, Krumholz+04sink, Ostriker+13,Bleuler+Teyssier14}. A major difference between our models and
many of the cloud-scale models in the literature is the very deep AMR grids we can afford. This makes it possible to use a high-density threshold, $\rhosink$,
while still resolving the Jeans length and ensures that the gas is already undergoing a physical collapse, at the point where the sink particle
is created.

A sink particle is born without any mass but will immediately acquire mass by accretion. Sink particles can accrete
gas from cells that are within the accretion radius, \racc{}, if the gas density is higher than a threshold value, $\rhoacc$. In the runs presented
below we are using two different accretion recipes. A simple recipe is used for the test runs, while a more complicated recipe,
capable of tracing the accretion down to very low densities, is used for the convergence runs. The stellar masses obtained with
the two recipes are practically identical (see Fig.~\ref{fig:wind-comp} in App.~\ref{A:sinks}), and the only difference is that the second recipe 
allows us to get a more precise picture of the accretion history at accretion rates down to $10^{-9} M_\odot\,{\rm yr}^{-1}$.

In the simple recipe, the amount of
gas accreted is such that the density in the cell is left to be just below $\rhoacc$. Typically, we set $\rhoacc$ to be $\rhosink / 2$.
The sink particles only accrete mass, momentum, and, if present, passive scalars. Thermal energy is removed from the gas in
proportion to the amount of mass removed, while magnetic field cannot be deposited on the sink particles. To avoid a pileup of
high-density gas just outside the accretion radius, \racc{}, it is necessary to have $\rhoacc \leq \rhosink$. The accretion distance
is typically set to one to two Jeans' lengths at $\racc$ around the sink particle.
This simple accretion recipe of removing gas to bring it below a critical density is similar to what we
have used in the past \citep{Padoan+Nordlund11sfr,Padoan+12sfr,Nordlund+13,Nordlund+14,Padoan+14acc}, and what
is used by \citet{Federrath+10sinks}.  At the very high densities and small $\racc$ we are considering, due to the deep AMR
hierarchy, the Bondi--Hoyle accretion radius is larger than $\racc$ for most of the lifetime of a sink particle, and correct accretion
rates are therefore naturally enforced, and we do not need the more complicated accretion recipes employed by other groups
\citep[e.g.][]{Krumholz+04sink,Ostriker+13}.

In the more complicated recipe, we set $\rhoacc$ to a much lower value, allowing for
accretion from gas at lower densities, but we also add the condition that the relative speed between
the sink particle and the cell is lower than $\sqrt{2}$ the Kepler velocity, to avoid artificial accretion of unbound gas.
We also taper off the efficiency of the accretion depending on how bound the gas in a cell is to the sink particle. In a simple picture
where a sink particle has an effective geometric cross section of $\pi R^2$, the accretion rate should be $\dot{m}_{\rm gas} = \pi R^2 \rho v$,
where $v$ is the relative speed between the gas and the sink particle. This formula is our starting point, but we modify it to take into
account that unless $R$ is the grid spacing in our code this would correspond to accretion at a distance.
Let $d$ be the distance between the center of the cell and the nearest sink particle and  $v_K = (G m_{\rm sink} / d)^{1/2}$ be the Kepler velocity.
The mass accretion rate in a single time step $\Delta t$ from a cell with density $\rho$ is then
\begin{equation}
\dot{m}_{\rm gas} = \frac{\Delta \rho \Delta V}{\Delta t}= \left\{ \begin{array}{ll}
\alpha_{\rm rate} \,\rho v_K \Delta x^2  f_v & {\rm if }\rho \le \rho_{\rm th} \\
0.5 \rho \Delta V / \Delta t & {\rm if }\rho > \rho_{\rm th}
\end{array}\right.
\end{equation}
where $\Delta \rho$ is the change in the gas density in the cell volume $\Delta V$, and $\rho_{\rm th}$ is a threshold density that we normally
set to $2\rhosink$ as a safety valve to avoid a large amount of gas piling up
faster than it is accreted inside the accretion radius. $\alpha_{\rm rate}=0.2$ controls how efficient accretion proceeds compared
to rotation. In the unrealistic case of spherical accretion with zero rotation $\alpha_{\rm rate}=1$. $f_v$ is a tapering function that limits the
rate depending on how far the cell is from the sink and how large the relative speed is,
\begin{equation}\label{eq:tapering}
f_v =  \left[1 - \left(\frac{d}{\racc}\right)^2 \right]\times \left\{ \begin{array}{ll}
0 & {\rm if }v \ge \sqrt{2} v_K \\
2 - \left(\frac{v}{v_K}\right)^2 & {\rm if }v_K < v < \sqrt{2} v_K \\
1 & {\rm if }v \le v_K
\end{array}\right.
\end{equation}

An optimal accretion algorithm should work as a transparent boundary condition such that no discontinuity is seen across the accretion radius.
We have tested the formula both in the low-resolution limit, where a sink particle is streaming through a low-density gas, and in the opposite limit
with very high resolution, where the sink is accreting from an accretion disk. In both cases, the above numerical parameters,
$\alpha_{\rm rate}=0.2$ and $\rho_{\rm th}=2 \rhosink$, result in a smooth transition between gas inside and outside the accretion radius, $\racc$.

A large fraction of the mass in protostellar systems is lost through winds and jets \citep{Matzner+McKee00cluster,Alves+07,Konyves+10}, launched from
small scales not resolved in the current simulations.
To account for this mass loss, we apply an efficiency factor, {\eout=0.5}, to the accreted mass and momentum of the sink particles in
the simulation, so that only $\eout \dot{m}_{\rm gas} dt$ of the mass is accreted to a given sink particle in a time interval $dt$ (the nonaccreted gas
fraction is simply removed from the simulation).
Without a more detailed knowledge of the typical mass loss in outflows, we prefer to stay with a single value of {\eout},
rather than expand the parameter space with yet another dimension.  In an older version of the
code, used for some of the smaller runs in App.~\ref{A:sinks}, we accreted the full amount of mass ($\eout=1$), and the accretion
rates and sink particle masses were readjusted after the simulation had finished running.
Since this overestimates the gravitational pull of the individual stars, one cannot simply rescale after the fact with the same factor. Two
otherwise- identical test runs have shown that to rescale from a run with $\eout=1$ to one with $\eout=0.5$ the appropriate factor is 2.4
(see Fig.~\ref{fig:wind-comp} in App.~\ref{A:sinks}). As discussed by \citet{Kuffmeier+17} in the context of deep zoom-in simulations, and
by \citet{Federrath+14jet}, in the context of jet and outflow models as a function of resolution, {\eout} depends on the maximum resolution
in the model. In our case with a maximum resolution of 50 au for the \emph{high} model, we do not resolve disks, except for the most massive
stars, and we are therefore not accounting for outflows. Thus, it is appropriate to use $\eout=0.5$ in all runs.

\section{Numerical Model} \label{sec_model}

\begin{table*}[t]
\setlength{\tabcolsep}{3.5pt}
{\centering
\caption{Numerical parameters of the large-scale runs exploring convergence and the dependence on $\alpha_{\rm vir}$}
\begin{tabular}{l|ccccccccrc|cccc|ccccc}
\hline\hline \\[-2ex]
 & \multicolumn{10}{c|}{Run parameters}  & \multicolumn{4}{c|}{Creation of Sinks} & \multicolumn{5}{c}{Accretion to Sinks}  \\
Run & Root & $N_{\rm AMR}$ &$\Delta x$  & $L_{\rm J}$ & $\rho_{\rm ref}$  & $M_{\rm box}$ & $\alpha_{\rm vir}$ & $t_{\rm end}$ & SFE & $N_{\rm sink}$ &$L_{\rm J,s}$  &  $\rho_{\rm s}$ & $\rho_{\rm s}$ & $r_{\rm ex}$ & $\rho_{\rm acc}$ & $\rho_{\rm acc}$ & $\rho_{\rm th}$ & $r_{\rm acc}$& {\eout} \\
 & Grid &  & au & & $\langle \rho \rangle$ & M$_\odot$ & & Myr & & & & cm$^{-3}$ & $\langle \rho \rangle$  & $\Delta x$ &cm$^{-3}$ & $\langle \rho \rangle$ & $\rho_{\rm s}$ & $\Delta x$ &\\ [0.8ex]
\hline \hline \\[-2ex]
\emph{16}   &  16$^3$  & 6 & 800 &  2.0 & 2 & 3000 & 0.83 & 1.6 & 13\% & 108 & 2 & $6.6\times10^6$ & $8.3\times10^3$ & 8  & 4227 & 5.3 & 2 & 4 & 0.5 \\  
\emph{32}   &  32$^3$  & 6 & 400 &  2.5 & 5 & 3000 & 0.83 & 1.8 & 13\% & 169 & 2 & $2.6\times10^7$ & $3.3\times10^4$ & 8 & 4227 & 5.3 &  2 & 4 & 0.5 \\  
\emph{low}  &  64$^3$  & 6 & 200 &  3.6 & 10 & 3000 & 0.83 & 2.4 & 13\% & 279 & 2 & $1.1\times10^8$ & $1.3\times10^5$ & 8 & 4227 & 5.3 & 2 & 4 & 0.5 \\  
\emph{med }& 128$^3$ & 6 &100 &  7.2 & 10 & 3000 & 0.83 & 2.5 & 13\% & 363 & 2 & $4.2\times10^8$ & $5.3\times10^5$ & 8 & 4227 & 5.3 & 2 & 4 & 0.5 \\  
\emph{high} & 256$^3$ & 6 & 50 & 14.4 & 10 & 3000 & 0.83 & 2.5 & 13\% & 410 & 2 & $1.7\times10^9$ & $2.1\times10^6$ & 8 & 4227 & 5.3 & 2 & 4 & 0.5 \\[0.8ex]  
\hline \\[-2ex]
\emph{light}& 256$^3$ & 6 & 50 & 14.4 & 20 &1500 & 1.67 & 4.0 & 5\% & 86 & 2 & $1.7\times10^9$ & $4.3\times10^6$ & 8 & 4227 & 5.3 & 2 & 4 & 0.5 \\  
\emph{heavy}& 256$^3$ & 6 & 50 & 14.4 & 5 & 6000 & 0.42 & 0.7 & 5\% & 614 & 2 & $1.7\times10^9$ & $1.1\times10^6$ & 8 & 4227 & 5.3 & 2 & 4 & 0.5 \\  
\emph{massive}& 256$^3$ & 6 & 50 & 14.4 & 2.5 &12000 & 0.21 & 0.3 & 3\% & 1223 & 2 & $1.7\times10^9$ & $5.3\times10^6$ & 8 & 4227 & 5.3 & 2 & 4 & 0.5 \\  
\end{tabular}\\
\label{tc}
}
\end{table*}

This paper is based on a large suite of simulations that describe the evolution of a generic MC piece of size $L_{\rm box}=4$ pc,
using isothermal supersonic turbulence, self-gravity, magnetic field, and a subgrid sink particle model for the gravitational collapse of the gas.
The simulations are characterized by three nondimensional numbers: the sonic Mach number, $\mathcal{M}_{\rm s}=\sigma_{\rm v}/c_{\rm s}$,
the Alfv\'{e}nic Mach number, ${\mathcal M}_{\rm a}=\sigma_{\rm v}/v_{\rm a}$, and the virial number, $\alpha_{vir}=5 \sigma_{\rm v}^2R / (3 G M)$,
which measure the relative strength of kinetic energy to thermal, magnetic, and gravitational energies, respectively. In the expression for the Alfv\'{e}nic Mach number,
$v_{\rm a}$ is the Alfv\'en speed corresponding to the mean magnetic field and the mean density in the computational domain. $R$ is a characteristic
size we use to define the virial number of a simulation, which we take to be $R=L_{\rm box}/2$, and $M$ is the total mass in the simulation, $M=M_{\rm box}$.

The basic parameters of our simulations cannot be chosen arbitrarily. They are constrained by observations and the physical conditions in the ISM,
and by what is feasible with current computational capabilities. At large scales of hundreds of parsecs, the interstellar medium (ISM) has a complex thermal
structure. On parsec scales, high-density cold molecular gas is shielded from UV radiation and cooled by dust and atomic lines balanced
by cosmic-ray and shock heating, forming an effectively isothermal medium from where stars are born. Thus, to be realistic, our model should
have a size smaller than $\sim$10 pc. On the other hand, because we aim at studying the emergence of the IMF including the high-mass Salpeter range,
the model should contain at least several thousand solar masses of gas. Numerically, the resolution needed to properly characterize the density fluctuations
induced by supersonic turbulence limits the sonic Mach number corresponding to the rms velocity and therefore the size of the box we can consider
(because of the velocity--size relation followed by MCs \citep{Larson81}, albeit with a significant scatter). Another
constraint is the desire to resolve small enough scales that we marginally resolve large disks around the sink particles.
In summary, the approximation of an isothermal gas sets an ultimate upper limit for a realistic box size,
while numerical constraints set a more stringent limit. The requirement of sufficient mass sets a lower limit.

Taking the above constraints into consideration, we have chosen $\mathcal{M}_{\rm s}=10$, ${\mathcal M}_{\rm a}=5$,
and $\alpha_{\rm vir}=0.83$ for the reference simulations. To convert to physical units, we assume an isothermal sound speed of 0.18 km s$^{-1}$,
corresponding to a temperature of $T\approx$10 K, and a mean molecular weight $\mu=2.37$, appropriate for cold MCs.
Given the nondimensional parameters, the rest depends on the physical box size. Choosing $L_{\rm box}=4$ pc (in line with a characteristic
velocity--size relation in MCs) corresponds to setting a total mass $M_{\rm box}=3000\, {\rm M}_\odot$, a mean density of 795 cm$^{-3}$,
and a mean magnetic field strength of 7.2 $\mu G$. The resulting mean column density is consistent with MC observations
\citep[e.g.][]{Heyer+01,Roman-Duval+10}, and the magnetic field strength with OH Zeeman splitting measurements  \citep[e.g.][]{Crutcher2012},
once density and magnetic fluctuations arising in the super-Alfv\'{e}nic turbulence are properly modeled \citep{Lunttila+08,Lunttila+09}.
This gives a free-fall time,  $t_{\rm ff}=\sqrt{3\pi/32G\rho_0}$, of 1.18 Myr and a dynamical or crossing time of $t_{\rm dyn} = 1.08$ Myr.
The virial parameter is known to control the SFR \citep{Krumholz+McKee05sfr,Padoan+12sfr,Padoan+14ppvi,Padoan+17sfr},
and is expected to influence the peak of the IMF (see \S \ref{sec_env}). Thus, we complement our base model, \emph{high}, with three
additional models, \emph{light}, \emph{heavy}, and \emph{massive} with $M_{\rm box}=1500 \,{\rm M}_\odot$, $M_{\rm box}=6000 \,{\rm M}_\odot$,
and $M_{\rm box}=12000 \,{\rm M}_\odot$ respectively.

The main properties of the five \emph{convergence models} with a total mass of $3000\, {\rm M}_\odot$ and the three models \emph{light},
\emph{heavy}, and \emph{massive} are listed in Table \ref{tc}. All models use six levels of refinement on top of the root grid.
These are the reference models discussed in the main sections of the paper. An additional 36 models
are discussed in App.~\ref{A:sinks}, where we explore the model dependence on the numerical parameters of our sink particle implementation.
Computationally, the project has required more than 50 million CPUh at four different HPC centers and has resulted in $\sim$100 TB of data.

The results from running the numerical tests discussed in Appendix~\ref{A:sinks} have guided us toward a definitive set of parameters for the sink particle
model. We require a density threshold for sink particle creation corresponding to resolving the Jeans length with at least two grid cells, $L_{\rm J,s}=2$.
Very close to a sink particle the flow will be artificially disturbed and we require an exclusion radius of $\rexcl=8$ cells to avoid the creation of spurious
sink particles, the occurrence of which can be tested with a very powerful method, the nearest--neighbor histogram (see Appendix~\ref{A:neighbor}).
In addition, it is important that the density at sink formation $\rhosink$ is significantly above the highest densities reached by the turbulence alone,
which sets a minimum
requirement of $\approx$$10^5$ for the gas overdensity at the highest AMR level, which in this case requires six levels of refinement and a gas density of
$\approx$$10^8$ cm$^{-3}$ (in our reference model \emph{high}, we actually have $\rhosink=1.7\times10^9$ cm$^{-3}$ and a minimum cell size of
$\Delta x=50$ au).
Finally, the model needs to have a high enough resolution that small-scale high-density cores, which are the progenitors of low-mass stars, are resolved
and can collapse; otherwise, a numerical IMF turnover would be created, dictated by the resolution. This requires both that the turbulence is sufficiently
well resolved to sample well the high-density tail of the gas density pdf, and that the Jeans length is sufficiently resolved to allow the gravitational collapse
and suppress numerical fragmentation \citep{Truelove+97}.

\section{IMF convergence} \label{sec_conv}
\begin{figure*}[t]
 \includegraphics[width=2.1\columnwidth]{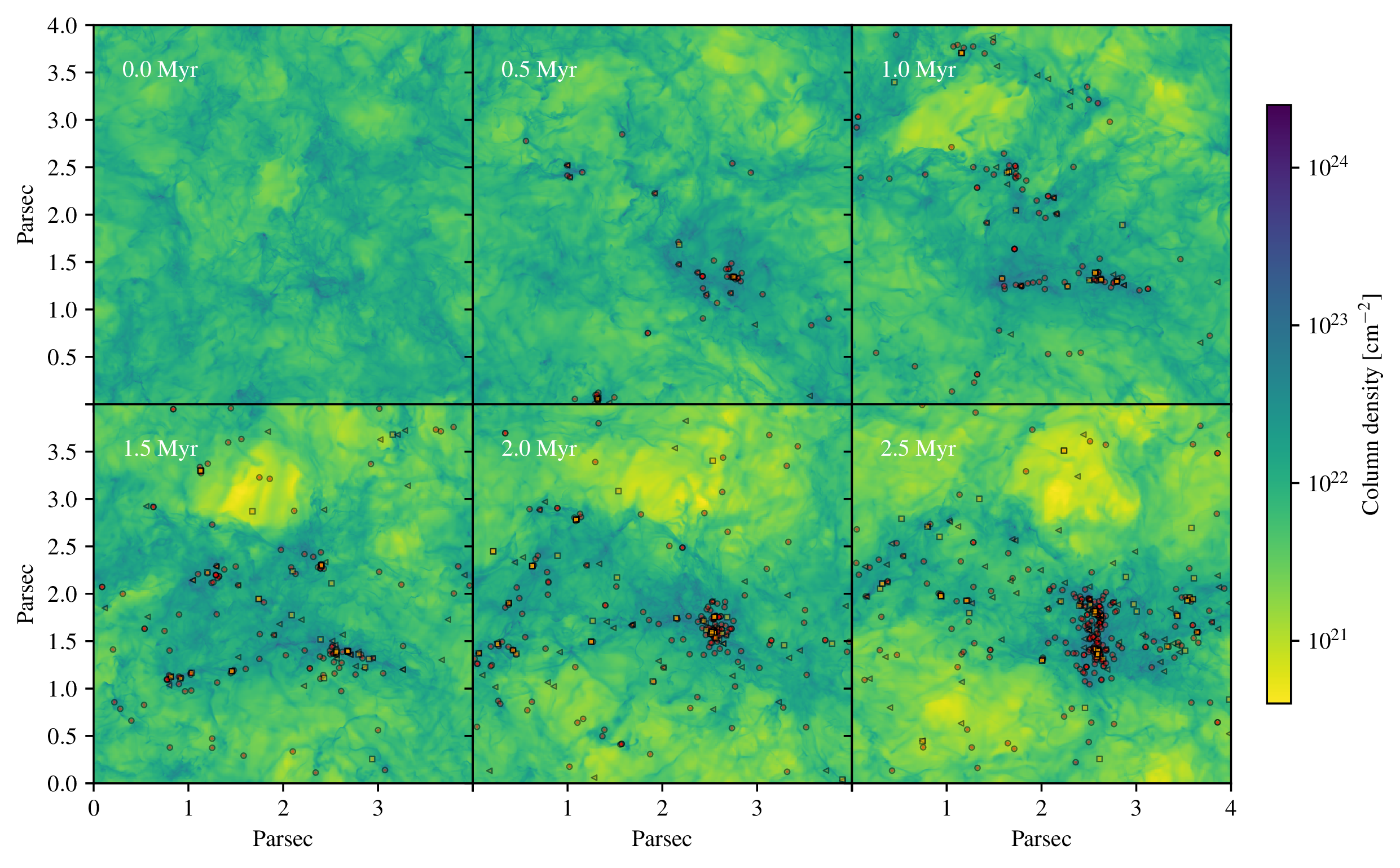}
  \caption{Column density from left to right and top to bottom at 0.5 Myr time intervals after the formation of the first star. The red circles mark the positions of stars with
  masses $m < 0.5 \mathrm{M}_{\sun}$, brown triangles are stars with $0.5 \mathrm{M}_{\sun} < m < 1.5 \mathrm{M}_{\sun}$, and orange
  squares indicate stars with $m > 1.5 \mathrm{M}_{\sun}$.
}
  \label{fig:coldens}
\end{figure*}

To test the numerical convergence of the IMF, we have carried out five runs, \emph{16}, \emph{32}, \emph{low}, \emph{med}, and \emph{high}, with a varying
root grid size of $16^3$, $32^3$, $64^3$, $128^3$, and $256^3$, respectively. and six levels of refinement (reaching a minimum cell size $\Delta x=50$ au in
the reference simulation \emph{high}). The corresponding minimum numerical values of the Jeans length (in units of $\Delta x$ of each simulation) are
$L_{\rm J}=2$, $2.5$, $3.6$, $7.2$, and $14.4$, while the parameters in the sink particle model are identical in all five runs (see Table~\ref{tc}). Notice
that instead of using the same overdensity threshold $\rho_{\rm ref}=10 \langle\rho\rangle$ for refinement for all the runs, giving a factor of two increase in
$L_{\rm J}$ between runs when the
spatial resolution is doubled, we have adopted a slightly more generous refinement strategy (lower density thresholds) for the runs \emph{16} and \emph{32}
to avoid resolving the Jeans length with less than two cells (see column with $\rho_{\rm ref}$ in Table~\ref{tc}). However, the threshold density for sink
creation, $\rho_{\rm s}$, does increase by a factor of
four between runs with increasing resolution by a factor of two, in order to keep the numerical value of the Jeans length at the density of sink creation,
$L_{\rm J,s}$, independent of resolution (as all other sink particle model parameters), $L_{\rm J,s}=2$. In these simulations, we have also imposed a small
maximum time-step size of 80 days at the highest level of refinement. The small time step size facilitates an accurate integration of sink particle orbits over
the long integration time of most of these runs.

\begin{figure}
  \includegraphics[width=\columnwidth]{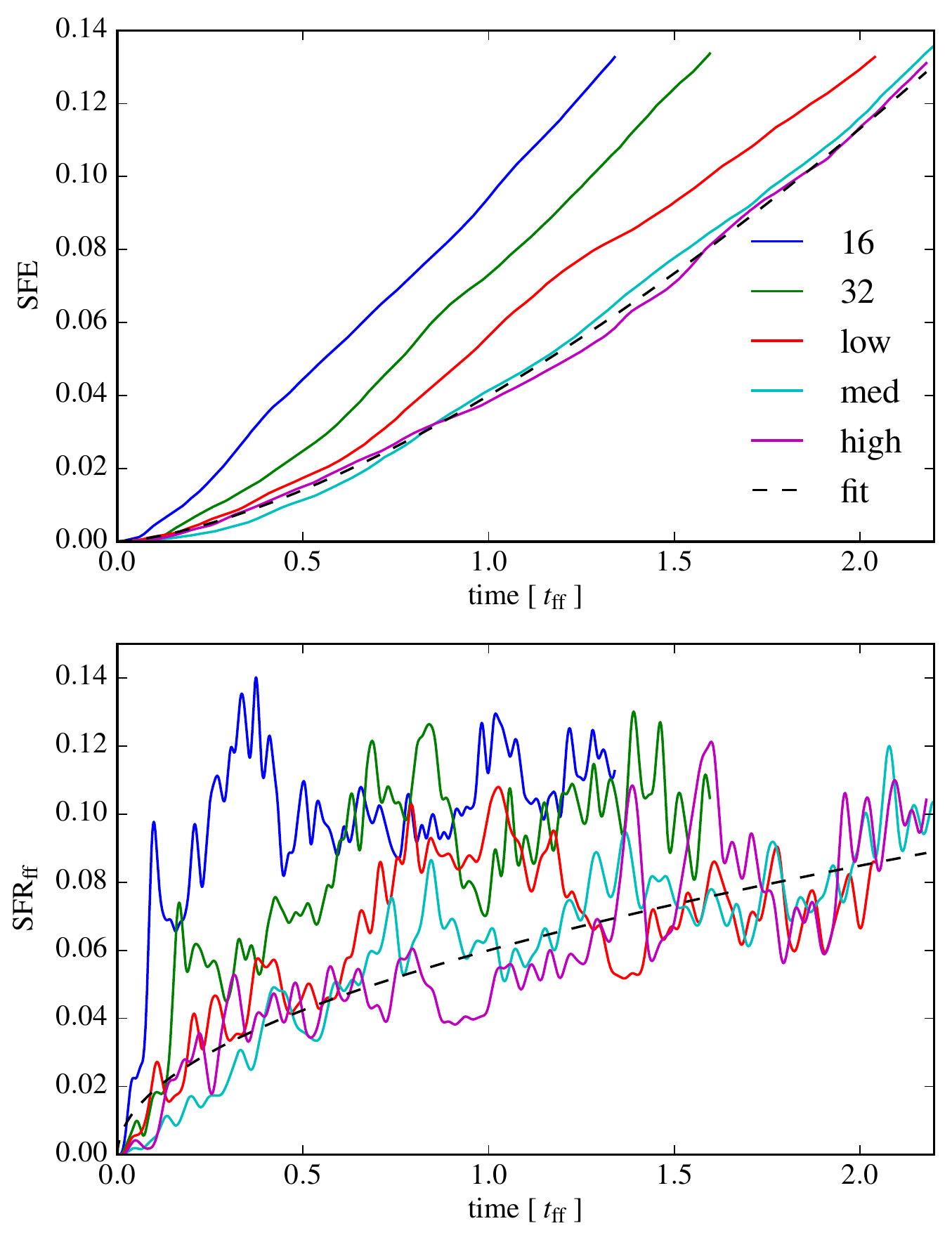}
  \caption{The evolution of the SFE (top) and SFR per free-fall time (bottom) as a function of time, measured in free-fall times
  for the convergence runs. The dashed line is a power-law fit SFE $=0.04 (t / t_{\rm ff})^{1.5}$ corresponding to
  SFR$_{\rm ff} = 0.06 ( t / t_{\rm ff})^{0.5}$. The free-fall time of the runs is $t_{\rm ff}=1.18$ Myr. The SFR$_{\rm ff}$ is highly intermittent
  on the very fine cadence of 80 days, which we use to record the sink particle properties, and has been low-pass filtered to aid readability.}
  \label{fig:SFE_conv}
\end{figure}

To illustrate the general evolution of the simulations, in Fig.~\ref{fig:coldens} we show the column density in the \emph{high} run at different
times. At the time when gravity is turned on, the density field is well mixed and well described by a lognormal pdf (see Fig.~\ref{fig:pdf}). At later times, the densest
gas decouples from the turbulent flow and starts to collapse. Some effect of self-gravity can also be seen at larger scales, such as the formation of a large ($\sim$1 pc),
elongated dense region that has assembled most of the high-density gas and most of the sink particles by the end of the simulation. This region is itself embedded in
a looser, but still discernible, overdense structure that stretches the full width of the computational volume (see the bottom right panel of Fig.~\ref{fig:coldens}).

\begin{figure}[t]
\includegraphics[width=\columnwidth]{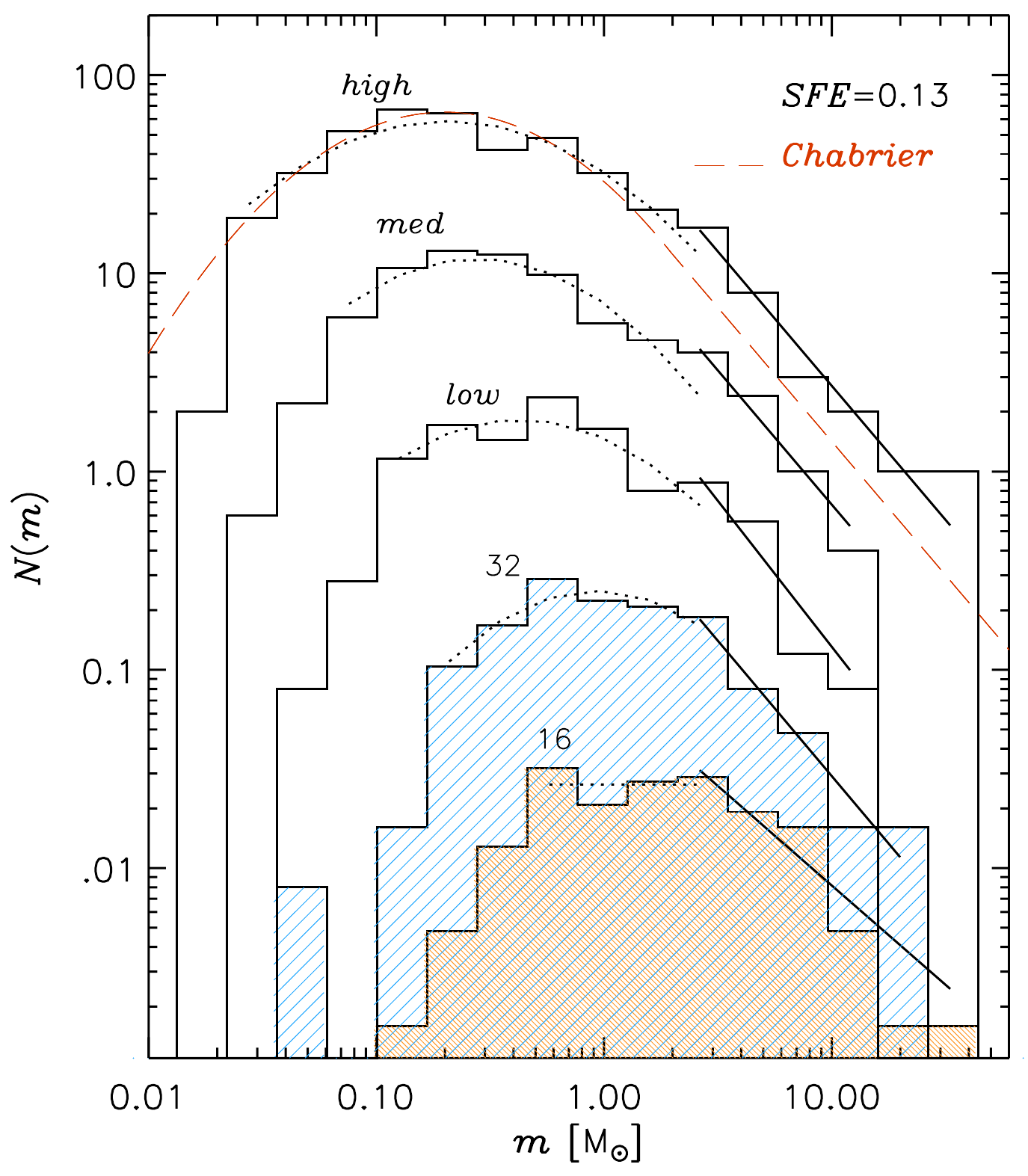}
\caption{Dependence of the IMF on numerical resolution. The five histograms correspond to the IMFs for the runs \emph{16}, \emph{32},
\emph{low}, \emph{med}, and \emph{high} (bottom to top), all sampled at SFE$=0.13$, corresponding to a time of 1.61, 1.91, 2.46, 2.62, and
2,64 Myr, respectively, after the formation of the first star. Except for the top one, the histograms are shifted vertically by a factor of 1/5 (\emph{med}),
1/25 (\emph{low}), 1/125 (\emph{32}) and 1/625 (\emph{16}). The dotted lines are lognormal fits between the smallest mass bin where the
IMF appears to be complete (based on a sharp cutoff at lower masses, more apparent in histograms with narrower bins) and 2 M$_{\odot}$.
The solid lines are power-law fits above 2 M$_{\odot}$. The dashed line corresponds to Chabrier's IMF \citep{Chabrier05} up to 2 M$_{\odot}$,
and Salpeter's IMF \citep{Salpeter55} above that mass.}
\label{fig:imf_peak_conv}
\end{figure}

The star formation efficiency ($SFE$) is the fraction of mass turned into stars at a given time $t$: $\textrm{SFE}(t) = M_{\rm sink}(t) / M_{\rm box}$, where $M_{\rm sink}(t)$
is the total mass in sink particles at the time $t$. The SFR is the time derivative of the SFE and can be expressed in nondimensional units if multiplied
by a characteristic timescale. We follow the convention introduced by \citet{Krumholz+McKee05sfr} to adopt the free-fall time of the mean density as the characteristic
time, so the nondimensional star-formation rate is defined as $\textrm{SFR}_{\rm ff}(t)= d \textrm{SFE}(t) / {\rm d} (t / t_{\rm ff})$. Fig.~\ref{fig:SFE_conv} shows the time evolution
of SFE and SFR$_{\rm ff}$ in the convergence runs. The time dependence is well fitted by a power law, SFR$_{\rm ff} \approx 0.06 ( t / t_{\rm ff})^{0.5}$, a slow increase
of SFR$_{\rm ff}$ from 0.04 at 0.5 Myr (0.4 $t_{\rm ff}$) to 0.08 at 2.4 Myr (2 $t_{\rm ff}$). This increase in SFR$_{\rm ff}$ is probably related to the formation of a dominant
dense cluster toward the end of the run (see Fig.~\ref{fig:coldens}). The increasing density in the cluster-forming region decreases the local virial parameter, increasing
the SFR \citep{Padoan+12sfr,Padoan+17sfr}.

We do not expect this trend of increasing SFR$_{\rm ff}$ to continue for a much longer time under more realistic conditions. To keep the evolution realistic
for a longer timescale, past SFE$=0.13$, external forcing from scales beyond the 4 pc box size would be needed to properly account for the interaction
of larger-scale turbulence with the cluster-forming region, as observed in larger-scale simulations with supernova driving \citep{Padoan+17sfr}. With supernova driving,
the typical disruption time of a MC is $\approx2\,t_{\rm dyn}$ \citep{Padoan+16SN_I}, equivalent to $\sim$2.5 Myr in the simulations of this work, meaning
that external feedback from larger scales should play an important role in the evolution of a 4 pc region after approximately 2 Myr. Protostellar feedback from
the high-mass stars in the dense cluster could also become important as a local feedback agent at that stage, but it is neglected here.

\begin{figure}[t]
\includegraphics[width=\columnwidth]{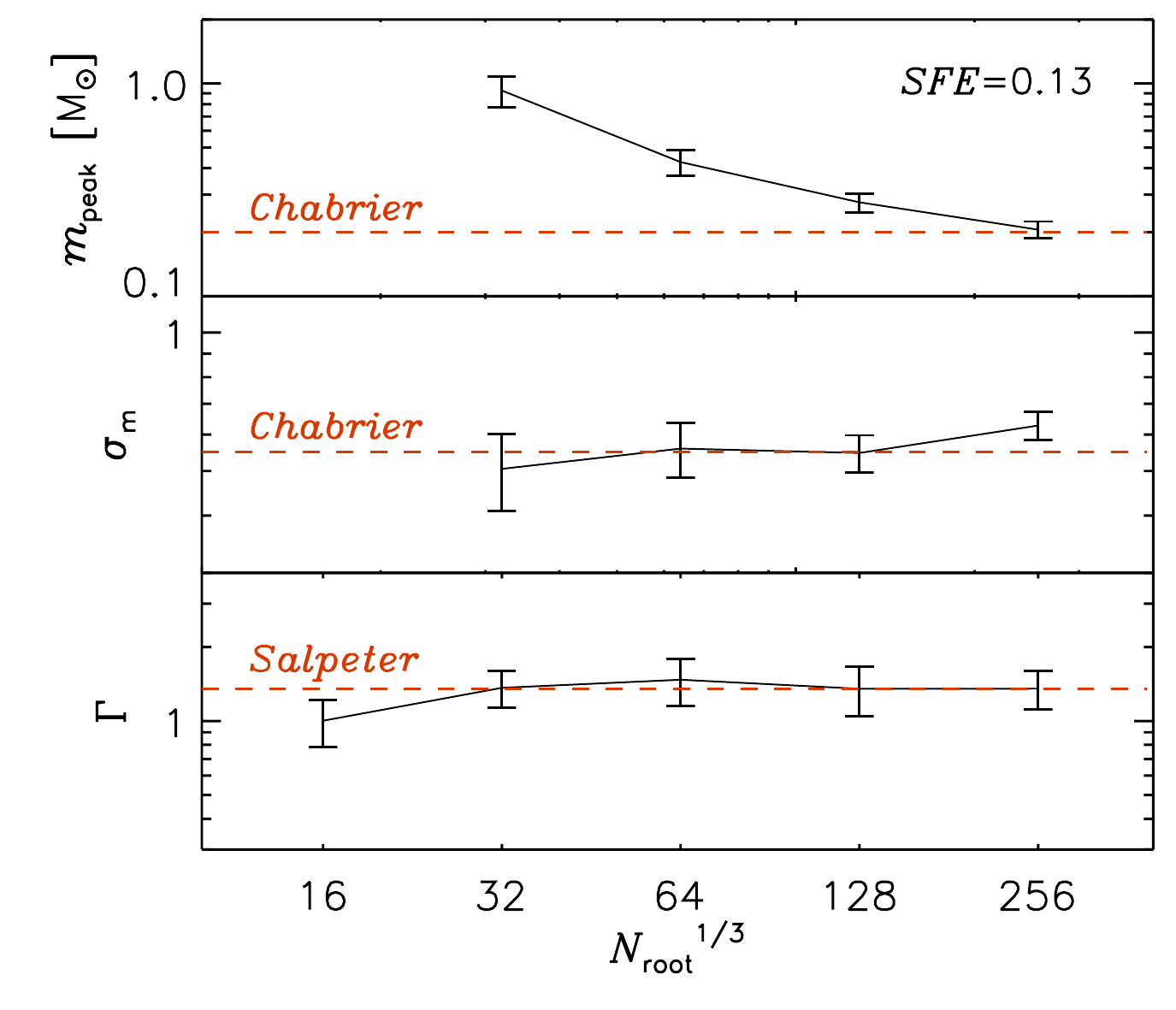}
\caption{Parameters of the lognormal fits (top and middle panels) and the power-law fits (bottom panel) shown in the previous figure, plotted
as a function of the numerical resolution, expressed as the linear size of the root grid in number of computational cells. The error bars show the
1 $\sigma$ uncertainty of the parameters. The top panel shows that the IMF peak tends to converge with resolution, although a full convergence
would probably require an even larger root grid size of at least $512^3$ cells.}
\label{fig:imf_par_conv}
\end{figure}

Compared to the \emph{high} run, SFR$_{\rm ff}$ is converged in the \emph{med} run and nearly converged in the \emph{low} run,
while at lower resolutions the runs are clearly not converged, and possibly also affected by numerical fragmentation, given the low minimum
Jeans number of $L_J=2$. There is a period between 0.7 $t_{\rm ff}$ and 1.2 $t_{\rm ff}$ where the \emph{low} run has a higher SFR$_{\rm ff}$, but otherwise
the three runs \emph{low}, \emph{med}, and \emph{high} show converged SFR$_{\rm ff}$. This result is in agreement with previous studies of the
star-formation rate \citep[e.g.][]{Padoan+Nordlund11sfr,Padoan+12sfr,Federrath+Klessen12}, where it was found that SFR$_{\rm ff}$ converges
at a relatively low numerical resolution, which allowed those works to explore a broad parameter space with many relatively low resolution simulations.
On the other hand, the numerical convergence of the stellar IMF is much more demanding and, in our opinion, has never been convincingly
achieved. It has not even been tried, so far, in the case of isothermal MHD turbulence, which is the main goal of this work.

The convergence test of our numerical IMFs is shown in Figures~\ref{fig:imf_peak_conv} and \ref{fig:imf_par_conv}. It has been carried out for a
single snapshot of each simulation, near the end of each run, to take advantage of the highest value reached by SFE, and thus the larger
statistical sample. All five runs are compared at SFE$=0.13$, corresponding to a time of 1.61, 1.91, 2.46, 2.62, and 2.64 Myr after the formation
of the first star, in order of increasing resolution (SFE$_{\rm ff}$ is higher in the low-resolution runs, so the same value of SFE is reached in a
shorter time). The IMF histograms computed at those times are plotted in Fig.~\ref{fig:imf_peak_conv}. To avoid the confusion generated by
noisy overlapping histograms, the IMFs have been vertically shifted, except in the case of the reference run \emph{high}, where $N(m)$ is indeed
the number of sink particles in each logarithmic mass interval of the histogram.

\begin{figure}[t]
\includegraphics[width=\columnwidth]{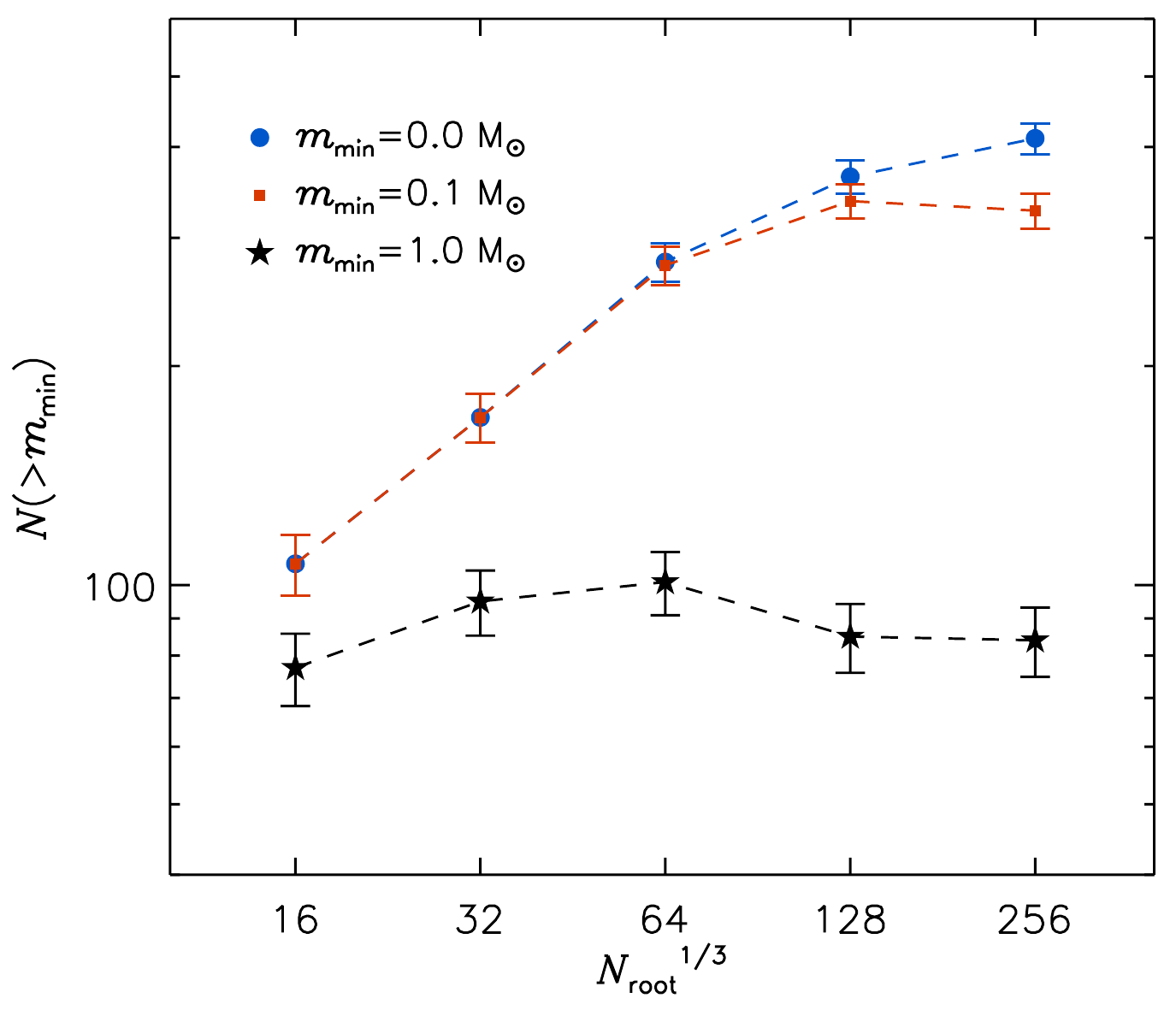}
\caption{Convergence plots for the total number of stars (circles), the number of stars with mass $m> 0.1~{\rm M}_{\odot}$ (squares), and the
number of stars with mass $m> 1.0~{\rm M}_{\odot}$ (stars), at SFE$=0.13$ in all the runs. Within the $\sqrt{N}$ uncertainty shown by the error
bars, all runs have essentially the same number of intermediate- and high-mass stars, while convergence is achieved at the resolution of the run
\emph{med} in the case of $m> 0.1~{\rm M}_{\odot}$. Even the total number of stars ($m_{\rm min}=0~{\rm M}_{\odot}$) shows a clear trend toward
numerical convergence, although it may still slightly increase at a resolution even higher than that of the run \emph{high}.}
\label{fig:imf_par_cum_conv}
\end{figure}

\begin{figure}[t]
\includegraphics[width=\columnwidth]{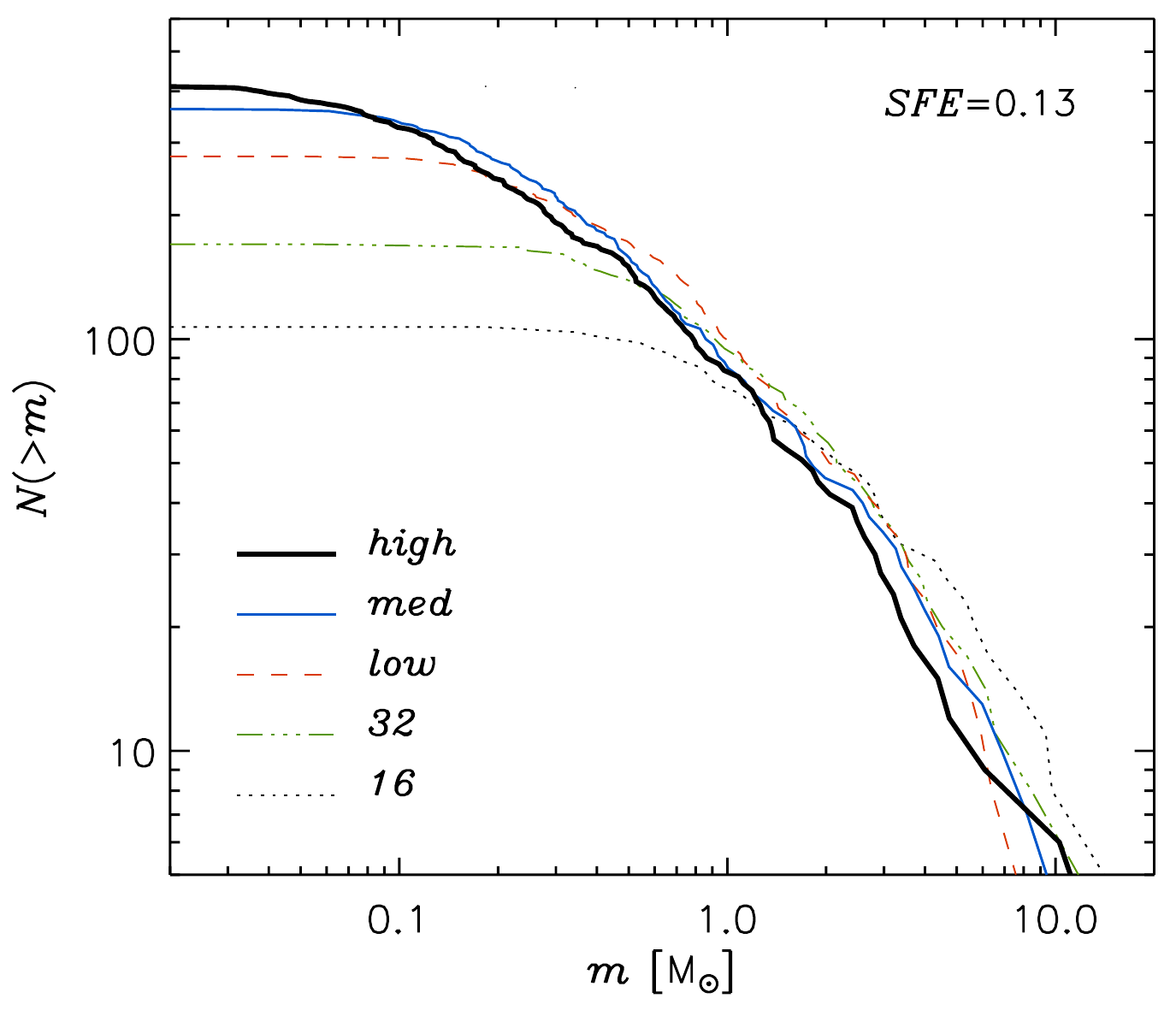}
\caption{Cumulative IMFs of the convergence-test simulations at SFE$=0.13$, as in Fig.~\ref{fig:imf_peak_conv}. The curves show the number of
sink particles above the mass $m$ as a function of $m$. There is a clear tendency toward convergence, with the two highest-resolution runs,
\emph{med} and \emph{high}, having essentially the same cumulative IMF down to a mass of order  $0.1~{\rm M}_{\odot}$, lower than the IMF
peak.}
\label{fig:imf_cum_conv}
\end{figure}

The figure shows a clear shift of $m_{\rm peak}$ toward lower values as the resolution increases, although the shift is quite small between the two
highest-resolution runs. To quantify the result of this comparison, we have fitted all the IMFs with a lognormal function for sink masses
$m\lesssim 2~{\rm M}_{\odot}$ (dotted lines), and with a single power law for masses $m> 2~{\rm M}_{\odot}$ (solid straight lines). Both models
clearly provide very good representations, in their respective mass ranges, of the shape of the IMFs. In the case of the reference run \emph{high},
the IMF is complete over more than three orders of magnitude in mass, with two orders of magnitude covered by the lognormal fit, and over one
order of magnitude by the power law fit. The smallest mass bin where the IMF is assumed to be complete is taken to be the one just above a sharp
drop in $N(m)$, at approximately $0.02~{\rm M}_{\odot}$ in the case of the run \emph{high} (and larger values by approximately a factor of two for
each consecutive step of decreasing resolution). The cutoff is much more apparent when the histograms are computed with narrower mass
bins, which is what we have done in order to determine these approximate completeness limits.

The dashed line in Fig.~\ref{fig:imf_peak_conv} shows the Chabrier IMF for single stars \citep{Chabrier05} below $2~{\rm M}_{\odot}$, continued
by the Salpeter IMF \citep{Salpeter55} at larger masses. It fits almost exactly our highest-resolution IMF, except that the Salpeter range is slightly
higher in the case of the simulation, partly because the turnover region has a slightly larger width than in the Chabrier IMF. The values of the
best-fit parameters are plotted in Fig.~\ref{fig:imf_par_conv}, showing that all IMFs above \emph{16} have a width, $\sigma_{\rm m}$, consistent
with that of the Chabrier IMF, except for the slight increase in the case of the run \emph{high}, and a power-law slope, $\Gamma$, consistent with
the Salpeter value. In the case of the lowest-resolution run, \emph{16}, $\sigma_{\rm m}$ cannot be measured and $\Gamma$ is slightly smaller
than Salpeter's value. The IMF peak, $m_{\rm peak}$, shows a clear dependence on resolution (top panel of Fig.~\ref{fig:imf_par_cum_conv}), but
also a trend toward numerical convergence (which may or may not have been already reached at the resolution of the run \emph{high}).

An alternative way to test for numerical convergence of the IMFs with increasing resolution is to use the cumulative IMFs. Although the value of
$m_{\rm peak}$ is best defined by a lognormal fit to the IMF, as we did above, the procedure has some dependence on the choice of bin
size and location, while the cumulative IMF is immune to such choices. In Fig.~\ref{fig:imf_cum_conv}, we show the cumulative IMFs of the
convergence-test simulations at SFE$=0.13$, expressed as the number of sink particles above the mass $m$ as a function of $m$. The
figure shows a clear tendency toward numerical convergence. The two highest-resolution runs, \emph{med} and \emph{high}, have essentially
the same cumulative IMF down to a mass of order $0.1~{\rm M}_{\odot}$, lower than the IMF peak. The rate of convergence is illustrated in
Fig.~\ref{fig:imf_par_cum_conv}, where we plot the number of stars above a given mass, $m_{\rm min}$, as a function of the root grid
size of the simulation. Even in the case of the total number of stars, $m_{\rm min}=0~{\rm M}_{\odot}$, the cumulative IMFs are clearly
converging with increasing resolution, consistent with the convergence of $m_{\rm peak}$ in the top panel of Fig.~\ref{fig:imf_par_conv}.
For stars with $m> 0.1~{\rm M}_{\odot}$, the convergence is achieved at the resolution of the run \emph{med}, and all runs have essentially
the same number of intermediate- and high-mass stars ($m_{\rm min}=1.0~{\rm M}_{\odot}$).

Based on the above convergence tests, we can conclude that we have found, for the first time, clear evidence of the convergence of the
IMF turnover (with a nearly converged value of $m_{\rm peak}$ in agreement with the observations) in the case of isothermal MHD turbulence,
as predicted by the turbulent fragmentation models of the IMF
\citep[PN02,][]{Padoan+97imf,Hennebelle+Chabrier08imf,Hennebelle+Chabrier09imf,Padoan+Nordlund11imf,Hopkins12imf}.

\section{IMF Variability} \label{sec_var}

The universality of the stellar IMF is hotly debated. While most works emphasize the apparent invariance of the IMF 
\citep[e.g.][]{Chabrier05,Bastian+10,Massey11,Offner+14ppvi,Weisz+15}, some stress compelling evidence of IMF variability 
\citep[e.g.][]{Kroupa01,Dib+10,Marks+12,Scholz+13,Kroupa+13,Dib14,Dib+17}. Observational determinations of the stellar IMF have been
approximated with different models, such as a multicomponent power law function \citep{Kroupa01,Kroupa02}, a tapered power law function \citep{Parravano+11},
or a lognormal function below 1 M$_\odot$ continued by a power law at larger masses \citep{Chabrier05}. The slope of the power law at large masses is
usually assumed to be $\Gamma=1.35$, as first estimated by Salpeter \citep{Salpeter55}, or slightly shallower (e.g. $\Gamma=1.3$ in Kroupa's and Chabrier's IMFs).
However, in the most complete and homogeneous study to date, based on 85 resolved clusters in M31, the IMF is actually found to be somewhat steeper, with
$\Gamma=1.45^{+0.03}_{-0.06}$ \citep{Weisz+15}. Besides this well-defined mean value, the value of $\Gamma$ exhibits variations from cluster
to cluster that are found to be within the error bars, with only a few outliers \citep[e.g.][]{Weisz+15}, or interpreted as indications of intrinsic IMF
variations \citep[e.g.][]{Dib+17}. The IMF peak is found to be at approximately 0.2 M$_\odot$, but statistically significant variations
from cluster to cluster are suggested by \cite{Dib14}. These variations, if confirmed, may reflect both the environment and the age of the 
observed stellar populations.

Physical models for the origin of the IMF predict a dependence on the average physical parameters of the star-forming environment,
which may in principle result in a larger IMF variability than observed. One may explore extra processes that would keep
the predicted IMF invariant, such as the radiative feedback from the accretion luminosity of protostars  \citep{Krumholz+16imf_peak}, which
appears to be crucial in simulations neglecting the magnetic field \citep[e.g.][]{Bate12,Krumholz+12imf}. However, it is also possible that the
variability predicted by the theory does not violate the observational constraints. In this section, we consider the turbulent fragmentation model by
PN02 and use our simulations to test its prediction for the dependence of the IMF turnover on physical
parameters. We then apply the model to the physical parameters of MCs derived from large MC surveys, showing that the predicted IMF variations
are within the observational constraints. The PN02 model also implies an early time evolution of the IMF, which we show to be qualitatively
confirmed by the simulations.

\subsection{The IMF turnover from turbulent fragmentation}

The origin of the characteristic stellar mass, essentially the turnover and peak of the IMF, is arguably the most fundamental question in star formation. 
\citet{Padoan+97imf} proposed that the IMF turnover is the direct result of the pdf of gas density in a turbulent MC
and modeled the turnover as a probability distribution of Jeans masses in the isothermal gas with a lognormal density pdf. Although that early
model did not account for the power-law tail of the IMF at large masses, its original explanation for the origin of the IMF turnover has been essentially
retained in following turbulent fragmentation models that can predict the full IMF \citep[PN02,][]{Hennebelle+Chabrier08imf,Hennebelle+Chabrier09imf,
Hopkins12imf}.

A numerical derivation of the turnover mass from the PN02 model yields $m_{\rm peak}\sim M_{\rm BE,0}{\mathcal M_{\rm s}}^{-1.1}$, where $M_{\rm BE,0}$ is the
Bonnor--Ebert (BE) mass with the external density equal to the average density of the star-forming region and ${\mathcal M_{\rm s}}$ is the rms Mach number of the
turbulent flow (equation (7) in \citet{Padoan+07imf}). This result can be easily derived as the characteristic BE mass in the turbulent flow, meaning the BE
mass with external density equal to the characteristic post-shock density, or, equivalently, the BE mass with external pressure equal to the characteristic
dynamic pressure of the turbulent flow. The standard BE mass confined by a thermal pressure $P_{\rm th,0}$ is given by
\begin{equation}
M_{\rm BE} = 1.182 \frac{\sigma^4_{\rm th}}{G^{3/2} P_{\rm th,0}^{1/2}}.
\label{mbe}
\end{equation}
Including the dynamic pressure of the turbulence, the external pressure is given by
$P_0=P_{\rm th,0}+P_{\rm dyn,0} =  P_{\rm th,0} (1+ {\mathcal M_{\rm s}}^2)$. Substituting into the
previous expression of $M_{\rm BE}$, we get a modified turbulent BE mass:
\begin{equation}
M_{\rm BE,t} \approx \frac{1.182\,\sigma^4_{\rm th}}{G^{3/2} P_0^{1/2}}= \frac{M_{\rm BE,0}} {(1+{\mathcal M_{\rm s}}^2)^{1/2}} \approx \frac{M_{\rm BE,0}}{{\mathcal M_{\rm s}}},
\label{mbe2}
\end{equation}
which is a good approximation to the turnover mass in the turbulent fragmentation models mentioned above, providing an intuitive
explanation of the origin of the IMF peak. To test the validity of this prediction, we express the IMF peak as
\begin{equation}
m_{\rm peak}\equiv \epsilon_{\rm BE} \, M_{\rm BE,t},
\label{mpeak}
\end{equation}
where $\epsilon_{\rm BE}$ is a local efficiency parameter analogous to {\eout} in the sink particle accretion model, and use the
simulations to verify whether it provides a good fit to the numerical IMFs.

\begin{figure}[t]
\includegraphics[width=\columnwidth]{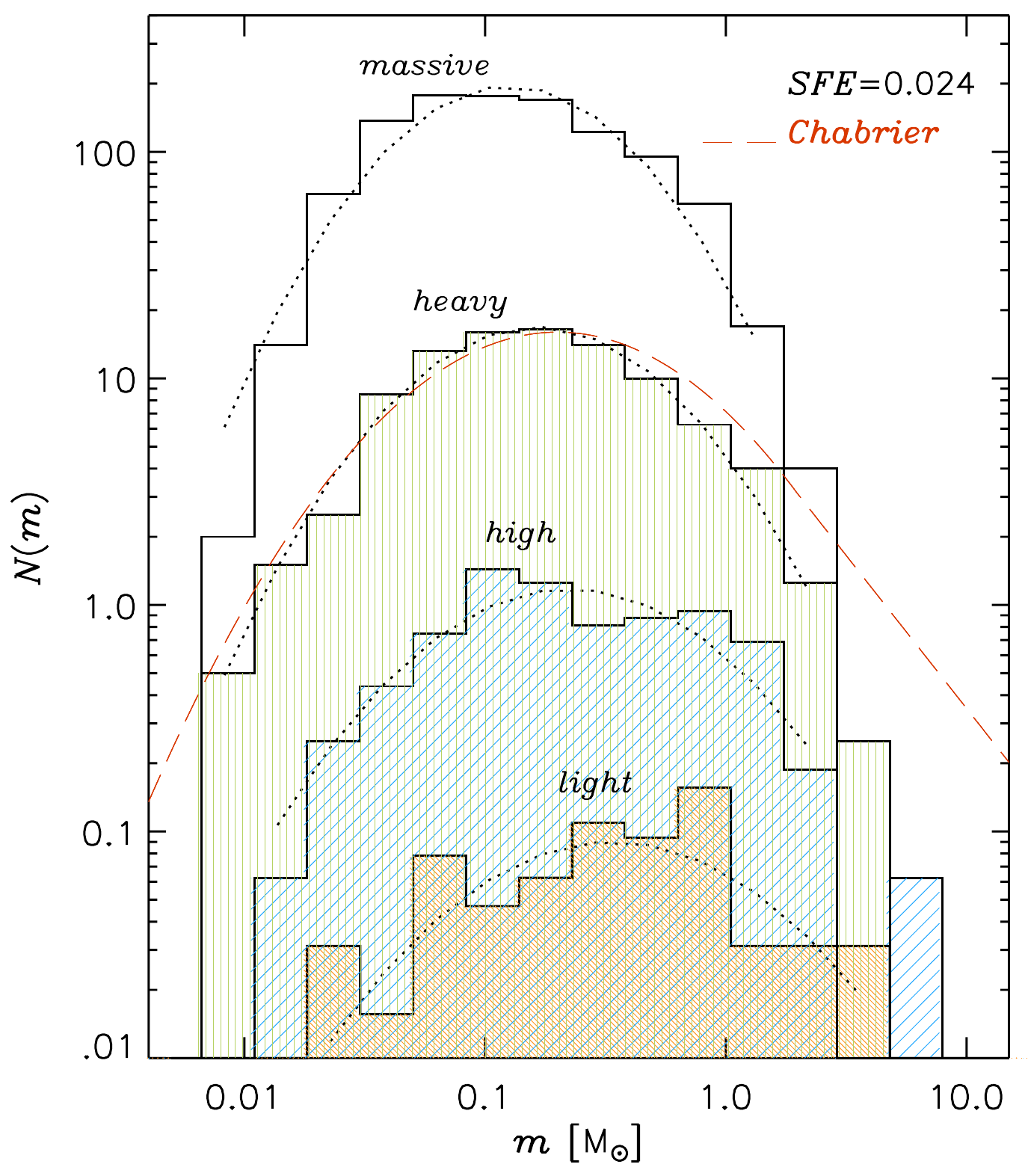}
\caption{Dependence of the IMF turnover on virial parameter (or mean density, equivalently), from the four simulations with a $256^3$ root grid,
\emph{light}, \emph{high}, \emph{heavy} and \emph{massive}, from bottom to top. The IMFs are all sampled at SFE$=0.024$, corresponding to
a time of 2.07, 0.83, 0.46, and 0.23 Myr, respectively, after the formation of the first star. Except for the top one, the histograms are shifted vertically
by a factor of 1/4 (\emph{heavy}), 1/16 (\emph{high}) and  1/64 (\emph{light}). The dotted lines are lognormal fits between the smallest mass bin
where the IMF appears to be complete and approximately $10\times m_{\rm peak}$. The IMF peak clearly shifts toward smaller values as the mean
density increases.}
\label{fig:imf_peak_virial}
\end{figure}

\begin{figure}[t]
\includegraphics[width=\columnwidth]{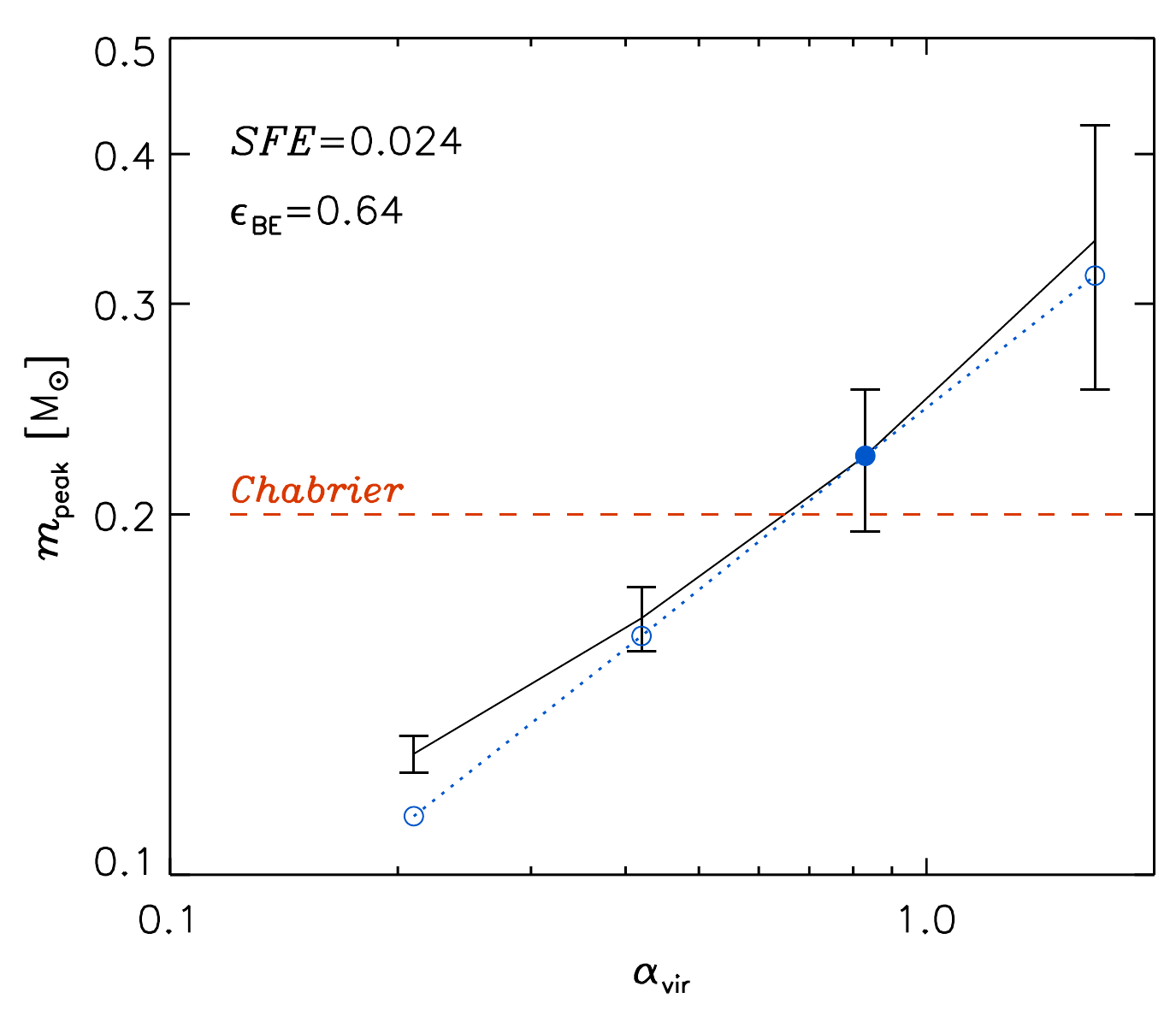}
\caption{Values of the IMF peak, $m_{\rm peak}$, from the lognormal fits of the previous figure, plotted as a function of the virial parameter of each
simulation (a proxy for the inverse of the mean gas density at constant rms velocity and size). The filled circle shows the value predicted by equation
\ref{mpeak} for the simulation \emph{high}, assuming an efficiency factor $\epsilon_{\rm BE}=0.64$, in order to match exactly $m_{\rm peak}$ measured from
the simulation. Assuming this fixed value of $\epsilon_{\rm BE}$, the open circles show the prediction of equation \ref{mpeak} for the other three simulations.
The measured value for the highest-density run is larger than the prediction, possibly because of a decreasing numerical convergence of the value of
$m_{\rm peak}$ as this becomes smaller with increasing mean density.}
\label{fig:imf_par_virial}
\end{figure}

For this purpose, we use the four simulations \emph{light}, \emph{high}, \emph{heavy}, and \emph{massive} with a root grid of $256^3$ cells and
six AMR levels, with four different values of the virial parameter (see Table~\ref{tc}). The virial parameter is varied by leaving the rms velocity constant
and increasing or decreasing the mean density (total mass) in the computational volume by a factor of two or four relative to the reference run \emph{high}
(see \S~\ref{sec_model} and Table~\ref{tc}). The overdensity threshold at which the root grid is refined is changed from $\rho_{\rm ref}=10 \langle\rho\rangle$
in run \emph{high} to $\rho_{\rm ref}=20\langle\rho\rangle,\,5\langle\rho\rangle{\rm, and}\,2.5 \langle\rho\rangle$ in \emph{light}, \emph{heavy}, and
\emph{massive}, respectively, to keep the minimum Jeans number constant, at 14.4. 
The IMFs from these four simulations are shown in Figure \ref{fig:imf_peak_virial}, where the histograms
are shifted vertically by a factor of four between consecutive runs, except for the top histogram, to minimize the confusion of overlapping plots.
The IMFs are all sampled at SFE$=0.024$, corresponding to a time of 2.07, 0.83, 0.46, and 0.23 Myr after the formation of the first star, for the runs
\emph{light}, \emph{high}, \emph{heavy}, and \emph{massive}, respectively. We have chosen a rather low SFE for this comparison because the
run \emph{light} has a very low SFR$_{\rm ff}$, such that to reach a much higher SFE the simulation should be integrated for much longer than 2~Myr.
As commented above, on a scale of 4~pc the influence of larger-scale feedbacks should become quite significant after approximately 2~Myr, making
this idealized setup driven by a random force somewhat questionable at later times. Despite the short timescale of the higher $\alpha_{\rm vir}$ runs
at SFE$=0.024$, we have found that the value of $m_{\rm peak}$ (and the ratios of its values from different runs) is already reasonably stable to
allow this comparison.

The dotted lines in Fig.~\ref{fig:imf_peak_virial} are lognormal fits of the IMFs (the power law fit at larger masses is not possible in this case because
the high-mass tail is not developed yet at this early time). The lowest-mass bin for the fit is based on the approximate IMF completeness limit judged as
in the numerical convergence test, while the highest-mass bin is approximately $10\times m_{\rm peak}$, assuming that the beginning of the power law
tail is also shifted to higher masses as the mean density decreases. The IMF peak clearly shifts toward smaller values as the mean density increases,
as predicted by the isothermal turbulent fragmentation model of the IMF. The best-fit values of the lognormal peaks are shown in Fig.~\ref{fig:imf_par_virial},
plotted as a function of the $\alpha_{\rm vir}$ value of each run (a proxy for the inverse of the mean density at a fixed rms velocity and size). The prediction of
equation~\ref{mpeak} is shown by the open circles, after normalizing the relation by the measured value of $m_{\rm peak}$ in the run \emph{high}. The
normalization corresponds to the choice $\epsilon_{\rm BE}=0.64$, quite close to the related local efficiency parameter set in the sink particle accretion
model, $\eout=0.5$.

Fig.~\ref{fig:imf_par_virial} shows that the measured variation of $m_{\rm peak}$ with the mean density is approximately consistent with the prediction
of equation~\ref{mpeak}. Although the slight discrepancy in the case of the run \emph{massive} may seem significant, it is not significant if one takes into
account the uncertainty in the measured value for the run \emph{high}. Furthermore, because we have established that the value of $m_{\rm peak}$ in
the run \emph{high} may not be fully converged (see Fig.~\ref{fig:imf_par_conv}), it is also possible that the value of $m_{\rm peak}$ in
the run \emph{massive} is even less converged, as the total mass in this run is larger and the peak smaller than in the run \emph{high}. The increasingly
higher lack of numerical convergence with increasing mean density could then explain the observed deviation from the prediction of equation~\ref{mpeak}.

We set the system rms velocity assuming a temperature of 10 K and the system size (or total mass) based on a standard Larson velocity-size relation
(see \S~\ref{sec_model}). If we chose not to follow the observed Larson velocity--size relation, both the rms velocity and the size (or total mass) of the system could be
rescaled, as long as the nondimensional parameters of the simulation, ${\mathcal M_{\rm s}}$ and $\alpha_{\rm vir}$, were not changed. Thus, one may suspect that the predicted
IMF peak is consistent with the numerical IMFs only for specific values of gas temperature or system size, but it can be easily shown that this agreement
is immune to the rescaling of the simulation. The virial parameter can be expressed as:
\begin{equation}
\alpha_{\rm vir} \propto {\mathcal M_{\rm s}}^2 T M_{\rm tot}^{-2/3}\rho_0^{-1/3}.
\label{alpha_vir}
\end{equation}
Because both ${\mathcal M_{\rm s}}$ and $\alpha_{\rm vir}$ are fixed in the simulation, the mass can only be scaled according to
$M_{\rm tot}\propto T^{3/2} \rho_0^{-1/2} \propto M_{\rm BE,0}$. This shows that imposing a value for both ${\mathcal M_{\rm s}}$ and $\alpha_{\rm vir}$
in the simulation  implies a fixed value of the ratio $M_{\rm tot}/M_{\rm BE,0}$, and thus a fixed value of $M_{\rm tot}/M_{\rm BE,t}$. Thus, rescaling
the temperature or size (or total mass) of the system does not affect our comparison of the predicted IMF peak with the IMF peak from the simulations.

To fully test the prediction of the turbulent fragmentation model with respect to the IMF turnover (and the width of the IMF as well), we should also
consider the dependence of $m_{\rm peak}$ on the sonic and Alfv\'{e}nic rms Mach number. Because all the simulations of this work have the
same Mach number, this important test will be addressed in a separate study.

\subsection{Variability of the IMF turnover with environment} \label{sec_env}

\begin{figure}[t]
\includegraphics[width=0.45\textwidth]{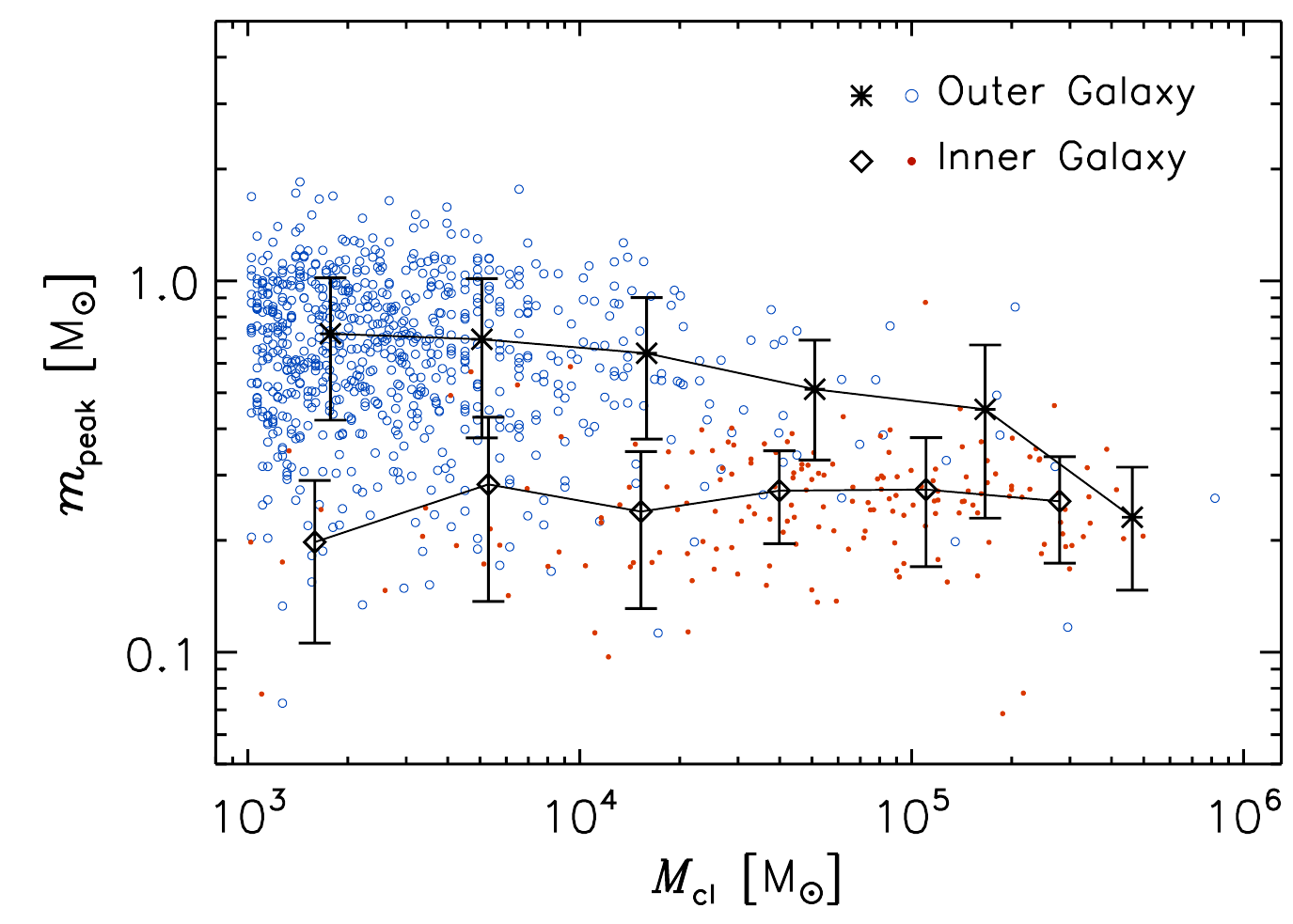}
\caption{Predicted IMF peak according to equation (\ref{mpeak}) versus cloud mass, for Outer Galaxy Survey clouds from \citet{Heyer+98} and the Galactic Ring
Survey clouds from \citet{Roman-Duval+10}, more massive than $10^3$ M$_\odot$ (see main text for details about the cloud selection). The error bars give the
mean and standard deviation of $m_{\rm peak}$ in six logarithmic bins of $M_{\rm cl}$.}
\label{fig:mbet}
\end{figure}

The theoretical and numerical prediction that the IMF peak scales with $M_{\rm BE,t}$ implies an environmental dependence of the IMF. In previous works,
we have already stressed that if the virial parameter does not vary significantly in star forming regions, and assuming standard velocity--size and mass--size
relations, the predicted IMF peak should have only mild variations \citep{Padoan+07imf}. Here, we try to quantify the expected scatter of $m_{\rm peak}$
based on the scatter in the observed properties of star-forming regions. We can express $m_{\rm peak}$ as a function of the nondimensional
parameters of the simulation and the total mass:
\begin{equation}
m_{\rm peak}\approx 1.124 \, M_{\rm tot} {\mathcal M_{\rm s}}^{-4} \alpha_{\rm vir}^{3/2},
\label{mpeak2}
\end{equation}
which shows that for constant $\alpha_{\rm vir}$ and for standard Larson relations, $M_{\rm tot} \propto L^2$ and $\sigma_{\rm v}\propto L^{1/2}$, $m_{\rm peak}$
is constant. However, observed MCs have a range of values of $\alpha_{\rm vir}$ and yield Larson relations with a significant scatter and with exponents
in general different from those standard values. Thus, our IMF model should predict non-negligible IMF peak variations from cloud to cloud.

In order to quantify the observational scatter in $m_{\rm peak}$ predicted by the model, we consider two of the largest Galactic MC samples available:
the MC catalog by \citet{Heyer+01}, extracted from a decomposition of the $^{12}$CO FCRAO Outer Galaxy Survey \citep{Heyer+98}, and the MC
catalog by \citet{Roman-Duval+10}, extracted from the UB--FCRAO Galactic Ring Survey \citep{Jackson+06}. To limit the distance and mass uncertainties,
\citet{Heyer+01} consider only MCs with circular velocity $<-20$ km s$^{-1}$, which yield a sample of 3901 clouds. \citet{Roman-Duval+10} provide
an estimate of the error in the mass determination of each of the 750 MCs in their catalog. We select a subsample of their clouds with a mass error
$< 20\%$, in order to minimize the scatter in cloud properties due to observational errors instead of intrinsic cloud differences. Finally, we retain
only MCs with mass $> 10^3$ M$_{\odot}$ (smaller clouds would not yield a well-sampled IMF), resulting in 720 MCs from the Outer Galaxy Survey
and 174 MCs from the Galactic Ring Survey.

Figure \ref{fig:mbet} shows the estimated value of $m_{\rm peak}$ for the clouds in the two observational samples. In the case of the Outer Galaxy,
$m_{\rm peak}$ shows a tendency to decrease with increasing cloud mass, while $m_{\rm peak}$ is essentially independent of cloud mass in the case of the
Galactic Ring. On the average, the expected IMF peak is more than twice larger for clouds in the Outer Galaxy than for those in the Galactic Ring, because of
the larger values of $\alpha_{\rm vir}$ in the Outer Galaxy clouds.  For the most massive clouds (few$\times 10^5$ M$_{\odot}$), where $\alpha_{\rm vir}$ is relatively
low also in the case of the Outer Galaxy Survey, the two samples give approximately the same value, $m_{\rm peak}\approx 0.25$, consistent with the
peak of the Chabrier IMF. In order to estimate a characteristic value of the peak, we consider the clouds with $\alpha_{\rm vir} < 3.0$, because of the strong
suppression of star formation at larger values of the virial parameter \citep[e.g.][]{Padoan+Nordlund11sfr,Padoan+12sfr,Padoan+17sfr}, and with mass
$M_{\rm cl} > 10^4$ M$_{\odot}$, because most of the mass is in the most massive clouds, based on the cloud mass distribution. With these subsets of
clouds from the two surveys, the mean and standard deviations are $m_{\rm peak}=0.6\pm0.25$ M$_\odot$ and $m_{\rm peak}=0.26\pm0.09$ M$_\odot$
for the outer and inner Galaxy respectively, with over 90\% of these star-forming clouds yielding values in the range $0.1 <  m_{\rm peak} < 1.0$ M$_\odot$.

This scatter in the peak of the stellar IMF predicted for different MCs is the consequence of the scatter in the velocity--size and mass--size relations, or, equivalently,
the scatter in the relation between virial parameter and mass (see Figures 31, 33, 34, and 35 in \citet{Padoan+16SN_I} and Figures 5, 6 and 7 in \citet{Padoan+16SN_III}).
We have recently shown that supernova driven turbulence generates MCs with properties consistent with the observations \citep{Padoan+16SN_I,Pan+16,Padoan+16SN_III}.
Because of this successful comparison between MCs selected from our simulation and the observations, we can use the simulation to infer that most of the scatter in the
observational Larson relations may originate from true physical variations from cloud to cloud, rather than be dominated by statistical uncertainties in the observational
measurements. Thus, we conclude that the predicted variations of the IMF peak from cloud to cloud, illustrated by Figure \ref{fig:mbet}, are realistic. This result is consistent
with the recent finding that the IMF of young nearby stellar clusters show significant variations from region to region. Using a Bayesian analysis of the IMFs of
eight young Galactic clusters, \citet{Dib14} has demonstrated that the posterior probability distribution functions of the IMF parameters of different clusters do not
generally overlap within the 1$\sigma$ uncertainty level. In the case of the Chabrier plus power-law fit, he derives IMF peak values in the range 0.29-0.69 M$_\odot$;
in the case of the fit with Parravano's tapered power law, the range is even larger, 0.14-0.80 M$_\odot$.\footnote{We are neglecting the case of NGC 2024
that has only 69 stars.} These observed IMF peak values are consistent with the ones predicted by our model applied to the
star-formation conditions of typical Galactic MCs. Thus, their scatter is consistent with that expected as a consequence of cloud-to-cloud variations
in ${\mathcal M_{\rm s}}$ and $\alpha_{\rm vir}$ at fixed cloud mass (essentially the scatter in the Larson relations).

\subsection{Variability of the IMF from time evolution} \label{sec_var_time}

As explained in \citet{Padoan+Nordlund11imf}, the PN02 turbulent fragmentation model implies a time evolution of the IMF, because more massive stars
are the result of converging motions from larger scales in the turbulent flow, requiring larger time to assemble the stellar mass (the turnover time of turbulent eddies
increases with their size) than lower-mass stars. Therefore, at very early times, massive stars are still not fully formed, as they require
a timescale comparable to the turnover time of the largest turbulent scales in the flow, of order of a Myr in typical MCs. This is much longer than the
formation time of 100 kyr in the model of massive star formation of \citet{McKee+Tan02,McKee+Tan03}.

\begin{figure}[t]
\includegraphics[width=\columnwidth]{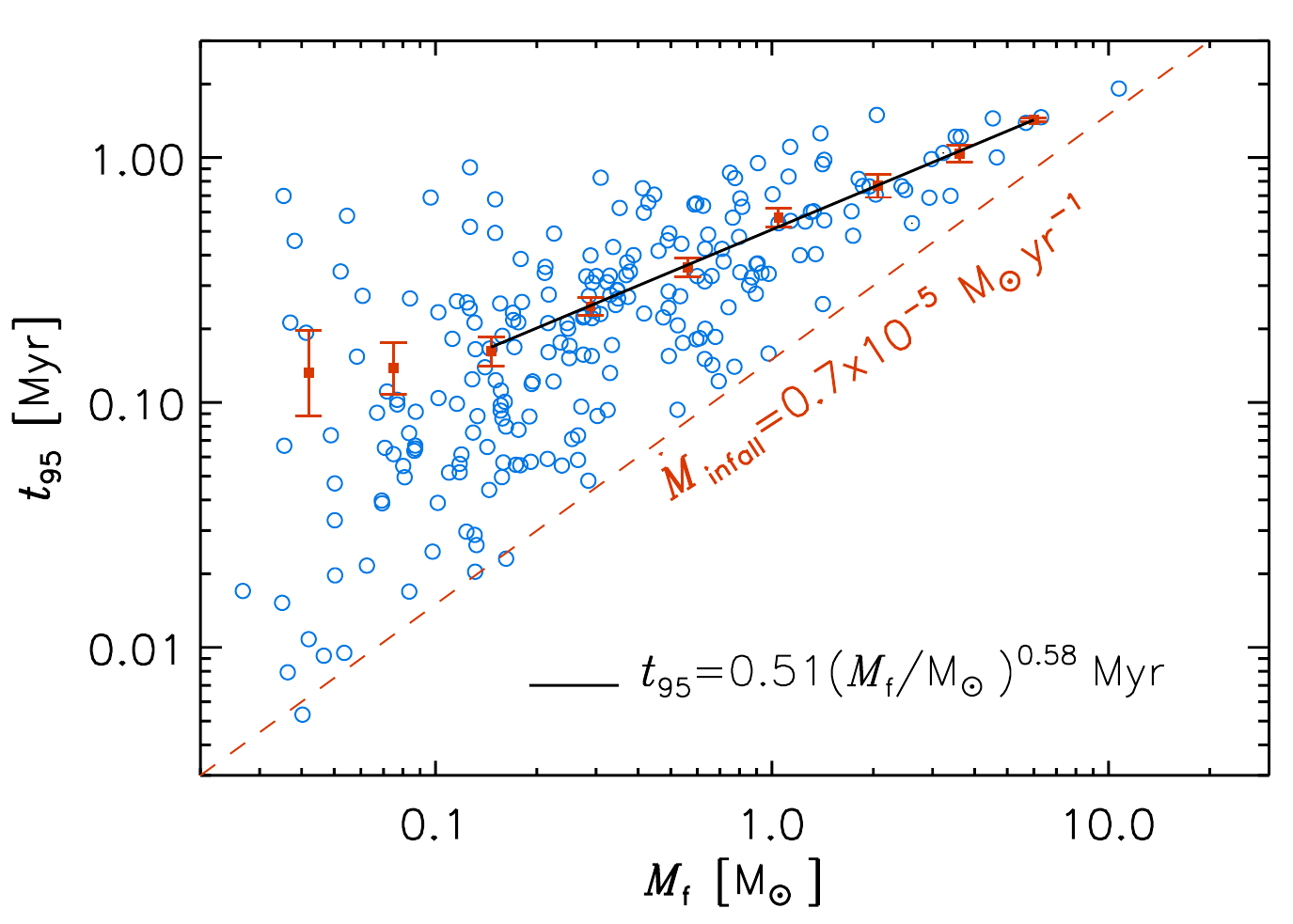}
\caption[]{Formation time of sink particles when 95\% of their final mass has been assembled, versus sink particle final mass, defined as the sink particle mass at
the end of the simulation \emph{high}, at $t=2.5$ Myr. The squared symbols and the error bars show the average and standard deviation of $t_{95}$ computed inside
logarithmic intervals of the final mass. The solid black line is a linear fit to the logarithmic values of $t_{95}$ versus final mass, giving
$t_{95}=0.51\,{\rm Myr} \,(M_{\rm f}/{\rm M}_{\odot})^{0.58}$, and the dashed line is an approximate lower envelope of the plot, corresponding to a
constant infall rate of $0.7	\times 10^{-5}$ M$_{\odot}\,$yr$^{-1}$.}
\label{fig:t95}
\end{figure}

It should be stressed that the mechanism of massive star
formation (and thus the origin of the Salpeter slope of the IMF tail) in the turbulent fragmentation models of \citet{Hennebelle+Chabrier08imf} and \citet{Hopkins12imf}
is quite different than in PN02, and, unlike PN02, may lead to the McKee and Tan scenario of massive star formation. In these models, massive stars originate from
massive cores that manage to exceed their Jeans mass. The reason why a large mass is needed to exceed the Jean mass is that the turbulence is included
as a source of pressure support defining the Jeans mass, despite the fact that such a generalization of the Jeans mass is actually valid only in the case in which the
turbulent outer scale is much smaller than the core size and the turbulent velocity is much smaller than the speed of sound \citep{Chandrasekhar51}, both conditions being
largely violated in the context of these models. The collapse of such a massive core cannot occur until it is fully formed, meaning until it has exceeded this turbulent
Jeans mass. Once that happens, the core collapses and forms a massive star essentially in a free-fall time, similarly to the scenario of the McKee and Tan model.
However, massive prestellar cores as predicted by these turbulent fragmentation models may have too low gas density (too large sizes), on average, compared with
observed cores, or even with the initial conditions of the McKee and Tan model, because they only need to be mild density fluctuations in the turbulent flow, rather
than post-shock regions.

Turbulent pressure support against self-gravity plays no role in PN02, where the turbulence is only viewed as a source of density enhancement through shocks.
Prestellar cores are assumed to emerge in the post-shock gas, where the turbulence has been largely dissipated. The inertial converging flows feeding such
post-shock cores can accumulate enough mass to form a massive star, over a characteristic turnover time on the scale of such flows, much longer than the free-fall
time in the post-shock gas. At the post-shock density, such mass would be many times larger than the Jeans mass (excluding support from turbulent motions that
is not important in the post-shock gas), so the core cannot be supported against collapse for the whole time necessary to gather all the available mass. As soon as
the critical mass for collapse in the post-shock gas has been reached, a protostar of intermediate mass is formed by the collapse of the core, and the rest of the mass
has to be accreted through a circumstellar disk fed by the same converging flows that had assembled the prestellar core. In other words, the stellar mass predicted
by the PN02 model should be seen as the total mass available to form a star, while the actual mass of a prestellar core (prior to its collapse into a protostar) could
be significantly smaller, at least in the case of massive stars (see Figure 1 in \citet{Padoan+Nordlund11imf}). This results in a difference between the prestellar core
mass function (MF) and the stellar IMF, with the prestellar core MF having a steeper high-mass tail than the Salpeter IMF \citep{Padoan+Nordlund11imf}. In the case
of low-mass stars, the stellar mass is not much larger than the characteristic BE mass in the post-shock gas, so most of the core mass is assembled before the core
collapses.

\begin{figure}[t]
\includegraphics[width=\columnwidth]{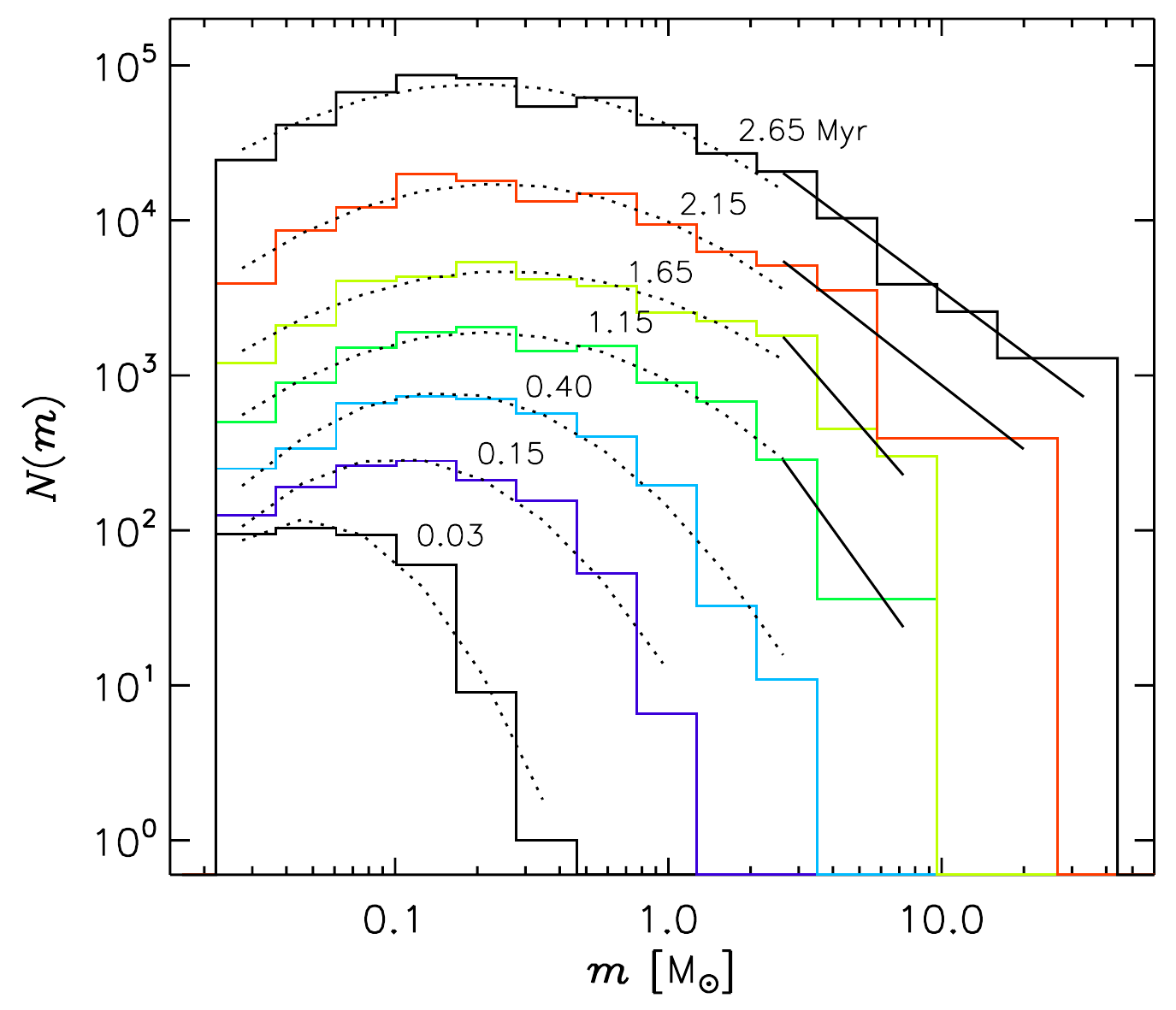}
\caption[]{Time evolution of the mass distribution of sink particles in the reference simulation \emph{high}. The time of each IMF since the formation of the
first sink particle is given next to each histogram.  The histograms are shifted vertically by an arbitrary value, except for the case of 0.03 Myr, which shows the
actual number of stars in each mass bin. The dotted lines are lognormal fits between the smallest mass bin where the IMF appears to be complete
(based on a sharp cutoff at lower masses, more apparent in histograms with narrower bins and corresponding to late times) and 2 M$_{\odot}$. The solid
lines are power-law fits above 2 M$_{\odot}$.}
\label{fig:imf_time}
\end{figure}

Earlier turbulence simulations without self-gravity and sink particles have already demonstrated the post-shock origin of prestellar cores as assumed in PN02
\citep{Padoan+2001cores,Padoan+07imf}, at odds with the scenario of
\citet{Hennebelle+Chabrier08imf} and \citet{Hopkins12imf}. Using a clump-find algorithm (instead of sink particles),
 \citet{Padoan+07imf} identified dense post-shock cores containing many Jeans masses (and not supported against self-gravity
by their turbulent pressure) in very large simulations of supersonic MHD turbulence without self-gravity. They also found that the core mass distribution
was consistent with a power law with the Salpeter slope, proving that such cores could contain the mass reservoir responsible for the formation of massive
stars. Evidently, if self-gravity had been present in the simulations, those massive cores would have collapsed much before gathering their total mass, and
the rest of their mass would have been accumulated over many free-fall times, as indeed shown by more recent simulations with self-gravity and sink particles,
such as in the work by \citet{Padoan+14acc}. In \citet{Padoan+14acc}, using a simulation with almost identical physical and numerical parameters as the model \emph{high}
in this work, we obtained nearly 1300 sink particles over a time of 3.2 Myr, with a mass function closely following a Chabrier IMF at small masses and a Salpeter
IMF at masses larger than 1-2 M$_{\odot}$. We used that simulation to argue that the large-scale infall from the turbulent inertial flows feeding the protostars
through an accretion disk could explain the observed luminosity distribution of protostars. We also showed that, on average, the time to gather 95\% of the final
stellar mass, $t_{95}$, increases with increasing final stellar mass, $M_{\rm f}$, according to
$t_{95} = 0.45\, {\rm Myr} \times (M_{\rm f}/1 {\rm M}_{\odot} )^{0.56}$, so it takes on average over 1 Myr to form a 10 M$_{\odot}$ star
(see Figure 13 in \citet{Padoan+14acc}). However, we did not see an accelerated accretion rate as the stars gain mass, so our results
are at odds with the predictions of the competitive accretion scenario \citep{Bonnell+01comp,Bonnell+Bate06competitive}.

\begin{figure}[t]
\includegraphics[width=\columnwidth]{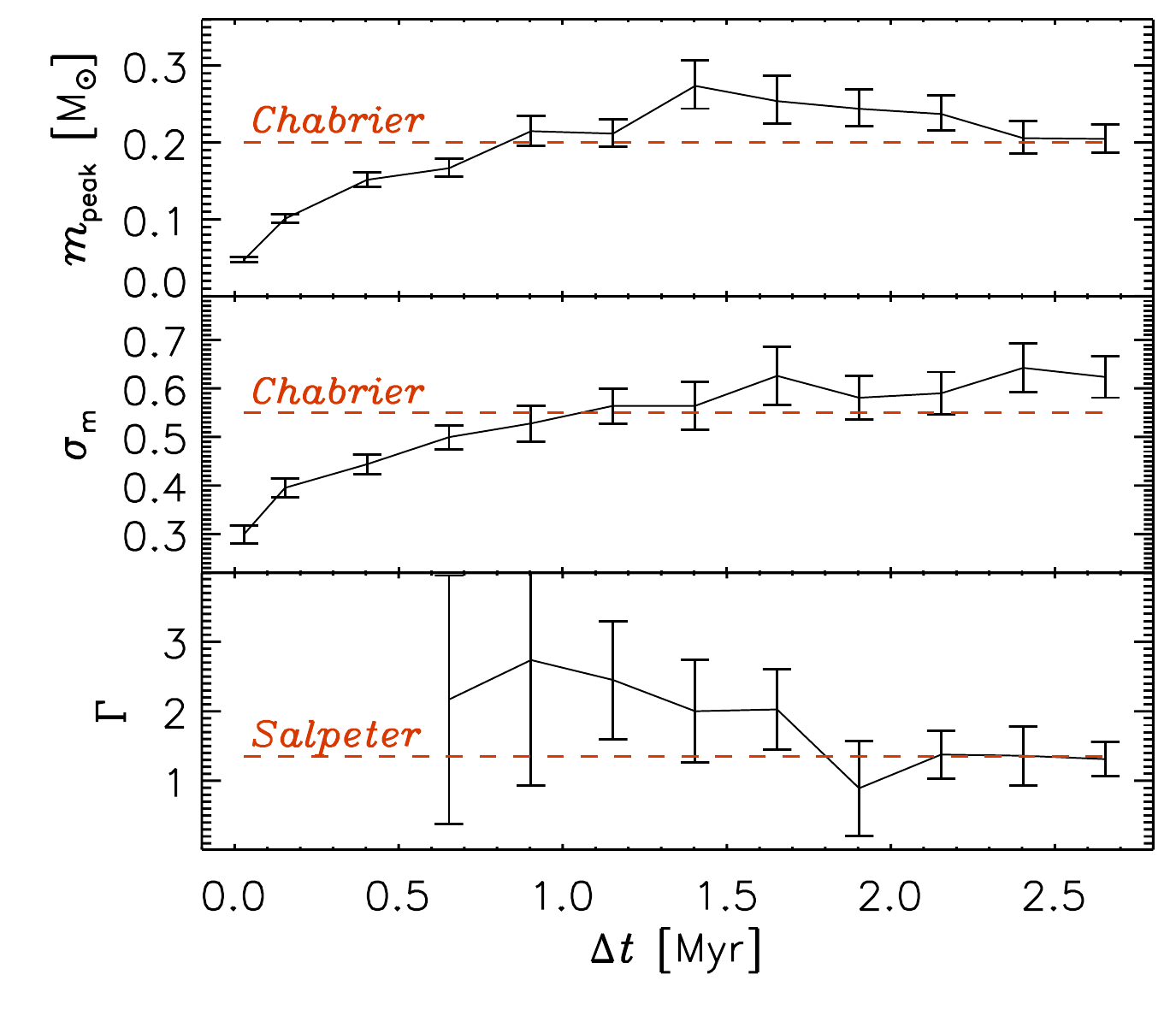}
\caption[]{Time evolution of the IMF parameters derived from the fits shown in the previous figure. The IMF peak is already established after less than 1 Myr
from the creation of the first sink particles, although it is a bit larger around 1.5 Myr (top panel). The power law tail at large masses takes approximately 2 Myr
to develop beyond 10 M$_{\odot}$ and achieve a stable slope, $\Gamma$, consistent with Salpeter's value (bottom panel). The progressive buildup of the tail
and the decreasing value of $\Gamma$ are reflected by a gradual increase in the width of the IMF, $\sigma_{\rm m}$, during the initial 1.7 Myr (middle
panel).}
\label{fig:imf_par}
\end{figure}

The dependence of the formation time on the final stellar mass
is confirmed by the simulations of this work. Figure \ref{fig:t95} shows the dependence of $t_{95}$ on $M_{\rm f}$, for
our reference simulation \emph{high} with $256^3$ root grid. Only stars that have practically stopped accreting by the end of the simulation are
included in the plot, to ensure that the value of $t_{95}$ is not artificially truncated by the finite integration time. This is enforced by selecting
only the sink particles whose accretion rate averaged over the final 100 kyr of the simulation is less than 10\% of their accretion rate averaged from their birth
time to the time they reach 95\% of their final mass. This selection retains 72.5\% of the sink particles. We have verified that a much more stringent
selection, where the accretion rate in the last 100 kyr has dropped to less than 0.005\% of its lifetime average, retains only 28\% of the sink particles
but yields the exact same power-law fit given below, albeit with larger uncertainties of the slope and intercept. Even the power-law fit obtained by retaining
all the sink particles yields the same parameters as in equation \ref{t95} below, within the 1 $\sigma$ uncertainties, as long as the two stars more massive
than 20 M$_{\odot}$ (and still actively accreting) are excluded from the fit. The apparent insensitivity of the relation between $t_{95}$ and $M_{\rm f}$ is
due to the long integration time of the simulation relative to the $t_{95}$ values of even the most massive stars.

Figure \ref{fig:t95} shows that the values of $t_{95}$ increase with increasing $M_{\rm f}$, with
a lower envelope approximately consistent with a constant infall rate of approximately $0.7\times 10^{-5}$ M$_{\odot}\,$yr$^{-1}$.
A power-law fit of the average values of $t_{95}$ in logarithmic bins of $M_{\rm f}$  gives
\begin{equation}
t_{95} = 0.51\pm0.02 \, {\rm Myr} \times (M_{\rm f}/1 {\rm M}_{\odot} )^{0.58\pm0.02},
\label{t95}
\end{equation}
consistent with our previous result in \citet{Padoan+14acc}. Although the fit would be barely affected by including the two lowest-mass bins, those
bins were excluded because the relation seems to flatten at the lowest masses. Furthermore, we do not expect $t_{95}$, as we measure it, to scale with $M_{\rm f}$
for masses below the IMF peak, as such stars (mainly brown dwarfs) do not result from characteristic turbulent compressions at a certain (small) scale,
but from very rare compression events biased toward very large density
\citep[necessary for the BE mass to reach brown-dwarf values;][]{Padoan+Nordlund04bd}.

The relatively long formation timescale should result in a time dependence of the IMF. Using the same simulation \emph{high}, we quantify this time evolution
through a lognormal fit of the IMF between 0.03 and 1.5 M$_{\odot}$ and a power-law fit of the IMF above 1.5 M$_{\odot}$. Figure \ref{fig:imf_time} shows
the IMF and the two fits at seven select times.
Every time includes stars of ages between 0 and $\Delta t$, so it is equivalent to observe a star-forming region of age $\Delta t$. At very early times,
only a few stars have been formed, so the IMF would be poorly defined. To increase the statistical sample, for every value of $\Delta t$ we consider all the
adjacent time intervals of size $\Delta t$ included between the beginning and the end of the star formation process (2.5 Myr in total) and measure the mass
of all the sink particles formed in each of those intervals (hence with age between 0 and $\Delta t$). In this way, the stellar sample is always of the order
of the total number of sink particles at the end of the simulation, for every value of $\Delta t$. Under the reasonable assumption that the average physical
conditions in the simulated star-forming region are stationary, this procedure should provide the correct IMF also for very small values of $\Delta t$.  For
clarity, the IMF histograms in Figure \ref{fig:imf_time} have been shifted vertically by an arbitrary value, except for the case of 0.03 Myr, which shows the
actual number of stars in each mass bin.

Figure \ref{fig:imf_time} shows a clear time evolution of the IMF, with the lognormal part at low masses becoming gradually broader over the first
megayear of evolution, and a gradual buildup of the power law tail. The time dependences of the best-fit parameters are plotted in Figure \ref{fig:imf_par},
where the horizontal dashed lines mark the value of mean and standard deviation of the observational IMF from \citet{Chabrier05} (middle and top
panels) and the Salpeter IMF slope (bottom panel). Both the peak and the width, $\sigma_{\rm m}$, of the IMF grow with time and settle after approximately 1 Myr.

While $\sigma_{\rm m}$ settles at a value
indistinguishable from the observational one, $m_{\rm peak}$ remains a bit larger than Chabrier's value. The time-averaged value of the IMF peak after
the first 1 Myr is  $\langle m_{\rm peak}\rangle=0.28\pm0.02$. The exponent of the power-law tail has a rather large uncertainty, due to the limited
number of intermediate- and high-mass stars in the simulation. However, it is clearly decreasing over time as the high-mass tail of the IMF develops. The gradual
buildup of the IMF tail takes nearly 2 Myr, after which its power-law slope, $\Gamma$, is essentially constant and consistent with Salpeter's value.

This time dependence of the IMF is very hard to constrain observationally, because a very young star-forming region, say less than 0.5 Myr, would
typically have very few stars and thus a poorly defined IMF (we cannot generate many $\Delta t$ intervals as we have done with the simulation). 
Furthermore, the variations of $m_{\rm peak}$ due to variations in the physical parameters
of the star-forming regions according to equation (\ref{mpeak2}) may be difficult to disentangle from the effect of time evolution, particularly if the physical
parameters of a given region also vary with time.

\section{Conclusions}

We have presented results from a large set of simulations of randomly driven supersonic MHD turbulence, meant to represent a characteristic
MC region of 4 pc and 3000 M$_{\odot}$. To match the assumptions of the turbulent fragmentation model of the IMF, we have adopted an isothermal
equation of state, neglecting both radiative and mechanical feedbacks from protostars. Our main results are as follows:

\begin{enumerate}

\item Thanks to a large number of test runs, we have found a definitive set of optimal parameters of our sink particle model:
number of cells per Jeans' length at the sink particle creation density $L_{\rm J,s}=2$, exclusion radius for creation $r_{\rm ex}=8 \,\Delta x$, 
ratio of accretion threshold density to sink creation density $\rho_{\rm th}/\rho_{\rm s}= 2$ and accretion radius $r_{\rm acc}=4\, \Delta x$.

\item We have found clear trends toward a numerical convergence of the IMF, and have shown that isothermal MHD turbulence can 
reproduce both the IMF turnover, consistent with the Chabrier IMF, and the power-law tail at high masses, consistent with the Salpeter slope.

\item The dependence of the IMF peak on the mean gas density predicted by the turbulent fragmentation model has been confirmed using three simulations
with different values of the mean density.

\item We have estimated the expected variations of the IMF peak due to variations of virial parameter and Mach number in MCs, based on observed
cloud properties derived from large Galactic MC surveys. The predicted variations are in line with the observational evidence of IMF variations in
young stellar clusters.

\item The timescale of star formation increases with stellar mass, with 10 M$_{\odot}$ stars requiring on average over 1 Myr to reach 95\% of their final mass.
This results in a time dependence of the IMF. It takes of order 1 Myr for the width and peak of the IMF to grow to their final values, and nearly 2 Myr for the power-law 
high-mass tail to get established and for the slope to settle to the Salpeter value. 

\item The time evolution of the IMF is expected from the turbulent fragmentation model of PN02, but not necessarily from other turbulent fragmentation models. It is also 
in contradiction with the idea that massive stars form from the rapid collapse of massive cores, as well as with the idea that the growth of massive stars is
controlled primarily by their own mass through competitive accretion. 

\end{enumerate}

We conclude that isothermal MHD turbulence yields a stellar IMF consistent with the observations. Other physical processes, such as radiation and outflows
from protostars, competitive accretion, or dynamical interactions, may not play a dominant role. Because we have focused on typical star-forming conditions in
Galactic MCs, it is still possible that these other processes become dominant in more extreme environments, such as near the Galactic center or in starburst
galaxies. The important implications for the formation of massive stars mentioned in this work will be addressed in detail in a separate work.

\acknowledgements
TH is supported by a Sapere Aude Starting Grant from the Danish Council for Independent Research.
PP is supported by the Spanish MINECO under project AYA2014-57134-P. 
{\AA}N is supported by a large framework grant from the Danish Council for Independent Research.
Research at the Centre for Star and Planet Formation is funded by the Danish National Research Foundation (DNRF97).
Computing resources for this work were partly provided by the NASA High-End Computing (HEC) Program through the NASA Advanced 
Supercomputing (NAS) Division at Ames Research Center. We acknowledge PRACE for awarding us access to
the computing resource JUQUEEN based in Germany at Forschungszentrum J\"ulich and access to the computing resource CURIE based in France at CEA,
where part of the simulations where carried out. The astrophysics HPC facility at the University of Copenhagen, which is supported by a research
grant (VKR023406) from VILLUM FONDEN, was used for carrying out part of the simulations, the analysis, and long-term storage of the results.

\appendix

\section{MHD in highly supersonic flows}\label{A:MHD}
Performing simulations of magnetized and highly supersonic flows is a challenge for Godunov solvers, which is only amplified when coupled with
self-gravity, sink particles, and fine-coarse transitions of an adaptive mesh. The average Mach number in the presented simulations is 10, but at exceptional
points in the flow the combined advection and fast-mode speed can be more than 500 times the sound speed. To stabilize the solver in these extreme conditions,
we have made several changes to the public MHD solver included in {\sc ramses}. First of all, we have contributed a small bug fix to how the magnetic fluxes
are updated at fine--coarse interfaces that occasionally would result in nonphysical solutions. Second, we have made a modification of how states in the
HLLD solver are computed when in principle a ``0/0'' situation arises \citep{hlld} in the computation of the transverse velocities and magnetic fields in the
Riemann fan. The equations for the intermediate states are

\begin{align}
e^* &= \rho (S - v)(S-v^*) - B_\parallel^2 \\
v_\perp^* &= v_\perp - B_\parallel B_\perp (v^* - v) / e^* \\
B_\perp^* &= B_\perp e / e^*\,, \\
\end{align}
where $\rho$, $v$, $B$, $e$, and $S$ are the density, velocity, magnetic field, internal energy, and combined advection and fast-mode speed, respectively.
Here $\parallel$ and $\perp$
indicate the parallel and perpendicular components, respectively, with respect to the normal of the interface over which the Riemann problem is solved. This is indeterminate when $e^*=0$.
Instead, the limiting expressions \citep{hlld}
\begin{align}
e^* &= \rho (S - v)(S-v^*) - B_\parallel^2 = 0 \\
v_\perp^* &= v_\perp \\
B_\perp^* &= B_\perp \\
\end{align}
should be used. In practice, numerical round-off introduce large errors when $e^*$ is close to 0, compared to the quantities involved. We have therefore modified the
condition, such that we use the limiting expressions already when $e^* < 10^{-4} B_\parallel^2$. This has removed cases where the solver crashed immediately from one
time step to the next. Both fixes have been propagated to the public version of {\sc ramses}.

In a finite-volume code, unphysical ringing and striping in the solution can arise near strong shocks. The principle mechanism to counteract the problem is
using a slope-limiting procedure that lowers the order of interpolation from centered to interface values in the vicinity of the shocks introducing diffusion.
If the slope-limiting procedure is only based on variations in variables across the interface, in a dimensionally split MHD code plane-parallel shock fronts can
trigger the so-called carbuncle instability \citep{Sutherland+03}. To counteract this, we have introduced a new three-dimensional slope-limiting procedure, which
has the added benefit that it introduces diffusion isotropically close to shocks irrespective of the direction of propagation maintaining e.g. spherical symmetry
in supernova driven shock fronts. Letting $f(i,j,k)$ be the variable that is to be slope limited, we then compute all 26 slopes
\begin{equation}
\Delta f_{\alpha\beta\gamma}(i,j,k) = f(i+\alpha,j+\beta,k+\gamma) - f(i,j,k)\,
\end{equation}
where $\alpha,\beta,\gamma=-1,0,+1$, and also the face-aligned slopes
\begin{align}
df_x(i,j,k) &= \frac{1}{2}(f(i+1,j,k) - f(i-1,j,k)) \\
df_y(i,j,k) &= \frac{1}{2}(f(i,j+1,k) - f(i,j-1,k)) \\
df_z(i,j,k) &= \frac{1}{2}(f(i,j,k+1) - f(i,j,k-1))\,
\end{align}
This is used to construct an isotropic slope limiter
\begin{equation}
{\rm lim} = \min\left(1, \delta\frac{\min(|\min(\Delta f)|,|\max(\Delta f)|)}{\sqrt{df_x^2 + df_y^2 + df_z^2}}\right)
\end{equation}
where $1\le\delta\le2$ is an input parameter that smoothly interpolates the slope limiter between behaving
as a min-mod and a monotonized-central limiter. The limiter is then applied to the final slopes equally in all
directions: $df_{x,y,z}(i,j,k) \to {\rm lim}\,\times\, df_{x,y,z}(i,j,k)$. We have typically used a value of $\delta=1.5$
for the simulations presented in this paper.

With the above modifications the HLLD solver in {\sc ramses} has turned out to be exceptionally stable. Even though the
solution is stable, sometimes the time step will decrease dramatically owing to cells with strong magnetic fields and
low densities typically in the vicinity of sink particles. To finish the runs in a reasonable time, we calculate the
combined advection and fast-mode speed, $c_B$, at all cell centers. If $c_B > c_B^{\rm max}$ we use a
diffusive Local Lax-Friedrichs (LLF) solver on all connecting edges and interfaces of the cell to calculate the Riemann
problem. This is effective in diffusing the solution sufficiently for the time step to stabilize at
reasonable values. Typically we have used $c_B^{\rm max}=500$, which activates LLF  in approximately
1 out of $10^4$ of the cells.

\section{Gravity}\label{A:gravity}
In {\sc ramses} the Poisson equation is solved on the adaptive mesh using a combination of multigrid and conjugate gradient methods
\citep{Teyssier+11grav}. The code can use not only spatial but also temporal subcycling, with a factor of two resolution
difference between each AMR level. The latter is required when using 10 or more AMR levels for efficient evolution of the model.
The code is structured such that the gravitational potential is calculated on the coarse levels first, with a recursive call midway
through the time step to the time evolution routine on the finer level. The time evolution of the cells is done after this recursive
call, resulting in the finest AMR levels to be evolved forward first resulting in a W-like evolution cycle for the different AMR levels.
Therefore, the boundary conditions on individual patches in the AMR hierarchy, computed by interpolating the potential at coarser levels
to boundary cells constructed on the fly, can be
out of temporal sync at the second step in a subcycle. If two sink particles are placed inside individual patches with coarser grid
cells in between, this will lead to a delayed transmission of the gravitational force and a complete decay of, e.g.,~binary
systems in less than 10 orbits. In a periodic box, tests with a few particles show that the delayed transmission of gravity through lower
levels subjects moving particles to self-interaction, and will slow them down on relatively short timescales. Notice that self-interaction is in
practice not a problem in our simulation since a single particle will not dominate the potential. While these effects are most obvious
for the particles (both sink and dark matter), similar effects happen for the gas dynamics. Fortunately, the gravitational
potential is the second integral of the density, and as such in general smooth, both spatially and in time. In the case of
time subcycling we have implemented extrapolation forward in time in the boundary cells between different levels. Letting $\Delta t_c$ be
the past time step at the coarse level and $\Delta t_f$ be the current time step at the fine level, then the gravitational potential at
the boundary then becomes $\phi_{t+\Delta t / 2} = (1+ \Delta t_f / \Delta t_c) \phi_t - (\Delta t_f / \Delta t_c) \phi_{t-\Delta t}$. This
extrapolation has effectively removed the decaying orbits and
self-interaction induced by time subcycling and is now also integrated in the public version of the {\sc ramses} code.

\citet{Federrath+10sinks} use a combination of a grid-based potential for the gas and a direct computation of the gravitational
forces of the sink particles, which has a cost that is proportional to $(N_{\rm cell} + N_{\rm sink}) \times N_{\rm sink}$. They also subcycle the sink
particles in time to resolve the orbits of individual sink orbits. While this method is arguably very precise in evolving in particular multiple
stars in close orbits, it rapidly becomes prohibitively expensive when considering large-scale simulations, such as those presented in this paper,
with up to 1000 sink particles. \citet{Ostriker+13} present an alternative implementation where the combined gravitational potential
of the gas and sink particles is computed on the grid, and the force on individual particles is found through interpolation of the forces
on the nearest grid cell centers. While this method is scalable, it is not precise at the subcell scale, and long-term stable orbits
require binaries to be separated by at least 10 grid cells at all times, and even then the orbits are prone to precession and some
level of decay.

To couple gas and sink particles gravitationally, we have developed two methods for evaluating the force
on individual particles. We use either CIC (8 neighbors) or TSC  (27 neighbors) interpolation from the sink particle position to
deposit the mass on the grid. The combined density of the gas and the sink particles is used as input to the Poisson equation, and
the resulting gravitational potential $\phi_{\rm sink+gas}$ is used for finding the gravitational acceleration of the gas.

In the first method to calculate the gravitational acceleration on the sink particles we use a matching CIC or TSC interpolation of the gradients
of $\phi_{\rm sink+gas}$ to the sink particle position. This corresponds closely to the method described in \citet{Ostriker+13}.
In the second method, which is the default for the runs presented here, only the gas is updated by calculating the gravitational acceleration from
$\phi_{\rm sink+gas}$. To find the gravitational acceleration of the sink particles, we solve the Poisson equation again, but this time for
the gas only. The gravitational acceleration of the gas on the sink particles can then be computed from $\phi_{\rm gas}$,
while the gravitational force of the sink particles instead is evaluated by direct summation of the forces from individual
particles. We use a cubic spline softening of the gravitational force \citep{Federrath+10sinks} with a typical softening
length of $0.3\Delta x$ when directly computing the forces between different sink particles.

Both of the methods are scalable to thousands of sink particles, and the second method gives as
precise orbits in multiple-star systems as the method of \citet{Federrath+10sinks} at the added cost of requiring an additional solution
to the Poisson equation.

The smaller-scale $64^3$ root grid test models in Appendix~\ref{A:sinks}, \emph{RUN1}--\emph{RUN27}, are using
the first method, while all other runs have been evolved with the second method. We have performed test runs at $\mathcal{M}_s=10$ with both
the first and the second method. Even though individual binary systems are much better described by the second method, the overall impact on
the IMF in the test runs is negligible and significantly smaller than the Poisson scatter in the IMF distribution.

To move the sink particles through the volume, we use a simple leapfrog Kick-Drift-Kick algorithm, identical to the method used
to integrate the orbits of dark matter particles in {\sc ramses}. The global Courant condition is modified to take into account both the
velocity $v_{x,y,z}$ and acceleration $a_{x,y,z}$ of the sink particles. Let $\Delta x^s$ be the minimum of the distances between sink particles and the softening length
\begin{equation}
\Delta x^s = \min[\mathcal{S} \Delta x, {\rm sink-sink\,distances}]\,,
\end{equation}
where $\Delta x = 2^{-\ell}$ is the cell size at refinement level $\ell$ and $\mathcal{S}=0.3$ is the softening number.
Then the limit on the time step imposed by each sink particle $i$ is
\begin{equation}
dt_s = \min \left[ \frac{C_{\Delta t} \Delta x^s}{2 \max[|v_x^i|,|v_y^i|,|v_z^i|]},
\left(\frac{C_{\Delta t} \Delta x^s}{|a^i|}\right)^{1/2} \right]\,,
\end{equation}
where $C_{\Delta t}$ is the Courant factor, which is set to 0.5 in the runs.

\begin{table*}[t]
\setlength{\tabcolsep}{3pt}
{\centering
\caption{Numerical parameters of the test runs}
\begin{tabular}{l|cccccccr|cccc|ccccc}
\hline\hline \\[-2ex]
 & \multicolumn{8}{c|}{Run Parameters}  & \multicolumn{4}{c|}{Creation of Sinks} & \multicolumn{5}{c}{Accretion to Sinks}  \\
Run & $Root$ & $N_{\rm AMR}$ &$\Delta x$  & $L_{\rm J}$ & Mass & $\alpha_{\rm vir}$ & $t_{\rm end}$ & SFE & $L_{\rm J,s}$  &  $\rho_{\rm s}$ & $\rho_{\rm s}$ & $r_{\rm ex}$ & $\rho_{\rm acc}$ & $\rho_{\rm acc}$ & $N_{\rm body}$ & $r_{\rm acc}$& {\eout} \\
 & Grid &  & (au) & & (M$_\odot$) & & (Myr) & & & (cm$^{-3}$) & ($\langle \rho \rangle$)  & ($\Delta x$) &(cm$^{-3}$) & ($\langle \rho \rangle$) & & ($\Delta x$) &\\ [0.8ex]
\hline \hline \\[-2ex]
\emph{RUN1}   & 64$^3$ & \textbf{6} & 200 & 8.0 & 3000 & 0.83 & 1.5 & 25\% & \textbf{8} & $6.6\times10^6$ & $8.3\times10^3$ & \textbf{16} & $3.3\times10^6$ & $4.1\times10^3$ & no & 2 & 1.0 \\  
\emph{RUN2}   & 64$^3$ & \textbf{6} & 200 & 8.0 & 3000 & 0.83 & 1.6 & 25\% & \textbf{8} & $6.6\times10^6$ & $8.3\times10^3$ & \textbf{32} & $3.3\times10^6$ & $4.1\times10^3$ & no & 2 & 1.0 \\
\emph{RUN3}   & 64$^3$ & \textbf{6} & 200 & 8.0 & 3000 & 0.83 & 1.6 & 25\% & \textbf{8} & $6.6\times10^6$ & $8.3\times10^3$ & \textbf{64} & $3.3\times10^6$ & $4.1\times10^3$ & no & 2 & 1.0 \\
\emph{RUN4}   & 64$^3$ & \textbf{6} & 200 & 8.0 & 3000 & 0.83 & 1.5 & 25\% & \textbf{4} & $2.6\times10^7$ & $3.3\times10^4$ & \textbf{32} & $1.3\times10^7$ & $1.6\times10^4$ & no & 2 & 1.0 \\  
\emph{RUN5}   & 64$^3$ & \textbf{6} & 200 & 8.0 & 3000 & 0.83 & 1.5 & 25\% & \textbf{4} & $2.6\times10^7$ & $3.3\times10^4$ & \textbf{32} & $1.3\times10^7$ & $1.6\times10^4$ & no & 2 & 1.0 \\
\emph{RUN6}   & 64$^3$ & \textbf{6} & 200 & 8.0 & 3000 & 0.83 & 1.6 & 25\% & \textbf{4} & $2.6\times10^7$ & $3.3\times10^4$ & \textbf{64} & $1.3\times10^7$ & $1.6\times10^4$ & no & 2 & 1.0 \\
\emph{RUN7}   & 64$^3$ & \textbf{6} & 200 & 8.0 & 3000 & 0.83 & 1.5 & 25\% & \textbf{2} & $1.0\times10^8$ & $1.3\times10^5$ & \textbf{16} & $5.2\times10^7$ & $6.6\times10^4$ & no & 2 & 1.0 \\ 
\emph{RUN8}   & 64$^3$ & \textbf{6} & 200 & 8.0 & 3000 & 0.83 & 1.5 & 25\% & \textbf{2} & $1.0\times10^8$ & $1.3\times10^5$ & \textbf{32} & $5.2\times10^7$ & $6.6\times10^4$ & no & 2 & 1.0 \\ 
\emph{RUN9}   & 64$^3$ & \textbf{6} & 200 & 8.0 & 3000 & 0.83 & 1.5 & 25\% & \textbf{2} & $1.0\times10^8$ & $1.3\times10^5$ & \textbf{64} & $5.2\times10^7$ & $6.6\times10^4$ & no & 2 & 1.0 \\
\hline \\[-2ex]
\emph{RUN10}  & 64$^3$ & \textbf{7} & 100 & 8.0 & 3000 & 0.83 & 1.5 & 25\% & \textbf{8} & $2.6\times10^7$ & $3.3\times10^4$ & \textbf{16} & $1.3\times10^7$ & $1.6\times10^4$ & no & 2 & 1.0 \\
\emph{RUN11}  & 64$^3$ & \textbf{7} & 100 & 8.0 & 3000 & 0.83 & 1.6 & 25\% & \textbf{8} & $2.6\times10^7$ & $3.3\times10^4$ & \textbf{32} & $1.3\times10^7$ & $1.6\times10^4$ & no & 2 & 1.0 \\ 
\emph{RUN12}  & 64$^3$ & \textbf{7} & 100 & 8.0 & 3000 & 0.83 & 1.5 & 25\% & \textbf{8} & $2.6\times10^7$ & $3.3\times10^4$ & \textbf{64} & $1.3\times10^7$ & $1.6\times10^4$ & no & 2 & 1.0 \\
\emph{RUN13}  & 64$^3$ & \textbf{7} & 100 & 8.0 & 3000 & 0.83 & 1.6 & 25\% & \textbf{4} & $1.0\times10^8$ & $1.3\times10^5$ & \textbf{16} & $5.2\times10^7$ & $6.6\times10^4$ & no & 2 & 1.0 \\
\emph{RUN14}  & 64$^3$ & \textbf{7} & 100 & 8.0 & 3000 & 0.83 & 1.6 & 25\% & \textbf{4} & $1.0\times10^8$ & $1.3\times10^5$ & \textbf{32} & $5.2\times10^7$ & $6.6\times10^4$ & no & 2 & 1.0 \\
\emph{RUN15}  & 64$^3$ & \textbf{7} & 100 & 8.0 & 3000 & 0.83 & 1.5 & 25\% & \textbf{4} & $1.0\times10^8$ & $1.3\times10^5$ & \textbf{64} & $5.2\times10^7$ & $6.6\times10^4$ & no & 2 & 1.0 \\
\emph{RUN16}  & 64$^3$ & \textbf{7} & 100 & 8.0 & 3000 & 0.83 & 1.5 & 25\% & \textbf{2} & $4.2\times10^8$ & $5.2\times10^5$ & \textbf{16} & $2.1\times10^8$ & $2.6\times10^5$ & no & 2 & 1.0 \\ 
\emph{RUN17}  & 64$^3$ & \textbf{7} & 100 & 8.0 & 3000 & 0.83 & 1.5 & 25\% & \textbf{2} & $4.2\times10^8$ & $5.2\times10^5$ & \textbf{32} & $2.1\times10^8$ & $2.6\times10^5$ & no & 2 & 1.0 \\
\emph{RUN18}  & 64$^3$ & \textbf{7} & 100 & 8.0 & 3000 & 0.83 & 1.5 & 25\% & \textbf{2} & $4.2\times10^8$ & $5.2\times10^5$ & \textbf{64} & $2.1\times10^8$ & $2.6\times10^5$ & no & 2 & 1.0 \\[0.8ex]
\hline \\[-2ex]
\emph{RUN19}  & 64$^3$ & \textbf{8} & 50 & 8.0 & 3000 & 0.83 & 1.6 & 25\% & \textbf{8} & $1.0\times10^8$ & $1.3\times10^5$ & \textbf{16} & $5.2\times10^7$ & $6.6\times10^4$ & no & 2 & 1.0 \\ 
\emph{RUN20}  & 64$^3$ & \textbf{8} & 50 & 8.0 & 3000 & 0.83 & 1.6 & 25\% & \textbf{8} & $1.0\times10^8$ & $1.3\times10^5$ & \textbf{32} & $5.2\times10^7$ & $6.6\times10^4$ & no & 2 & 1.0 \\ 
\emph{RUN21}  & 64$^3$ & \textbf{8} & 50 & 8.0 & 3000 & 0.83 & 1.6 & 25\% & \textbf{8} & $1.0\times10^8$ & $1.3\times10^5$ & \textbf{64} & $5.2\times10^7$ & $6.6\times10^4$ & no & 2 & 1.0 \\
\emph{RUN22}  & 64$^3$ & \textbf{8} & 50 & 8.0 & 3000 & 0.83 & 1.6 & 25\% & \textbf{4} & $4.2\times10^8$ & $5.2\times10^5$ & \textbf{16} & $2.1\times10^8$ & $2.6\times10^5$ & no & 2 & 1.0 \\
\emph{RUN23}  & 64$^3$ & \textbf{8} & 50 & 8.0 & 3000 & 0.83 & 1.6 & 25\% & \textbf{4} & $4.2\times10^8$ & $5.2\times10^5$ & \textbf{32} & $2.1\times10^8$ & $2.6\times10^5$ & no & 2 & 1.0 \\
\emph{RUN24}  & 64$^3$ & \textbf{8} & 50 & 8.0 & 3000 & 0.83 & 1.6 & 25\% & \textbf{4} & $4.2\times10^8$ & $5.2\times10^5$ & \textbf{64} & $2.1\times10^8$ & $2.6\times10^5$ & no & 2 & 1.0 \\
\emph{RUN25}  & 64$^3$ & \textbf{8} & 50 & 8.0 & 3000 & 0.83 & 1.6 & 25\% & \textbf{2} & $1.7\times10^9$ & $2.1\times10^6$ & \textbf{16} & $8.3\times10^8$ & $1.0\times10^6$ & no & 2 & 1.0 \\ 
\emph{RUN26}  & 64$^3$ & \textbf{8} & 50 & 8.0 & 3000 & 0.83 & 1.5 & 25\% & \textbf{2} & $1.7\times10^9$ & $2.1\times10^6$ & \textbf{32} & $8.3\times10^8$ & $1.0\times10^6$ & no & 2 & 1.0 \\
\emph{RUN27}  & 64$^3$ & \textbf{8} & 50 & 8.0 & 3000 & 0.83 & 1.6 & 25\% & \textbf{2} & $1.7\times10^9$ & $2.1\times10^6$ & \textbf{64} & $8.3\times10^8$ & $1.0\times10^6$ & no & 2 & 1.0 \\[0.8ex]
\hline \hline \\[-2ex]
\emph{RUN28}  & \textbf{512$\mathbf{^3}$} & \textbf{3} & 200 & 8.0 & 3000 & 0.83 & 2.0 & 25\% & \textbf{1} & $4.2\times10^8$ & $5.2\times10^5$ & \textbf{8} & $2.1\times10^8$ & $2.6\times10^5$ & yes & 2 & 1.0 \\ 
\emph{RUN29}  & \textbf{512$\mathbf{^3}$} & \textbf{3} & 200 & 8.0 & 3000 & 0.83 & 2.0 & 25\% & \textbf{2} & $1.0\times10^8$ & $1.3\times10^5$ & \textbf{4} & $5.2\times10^7$ & $6.6\times10^4$ & yes & 2 & 1.0 \\ 
\emph{RUN30}  & \textbf{512$\mathbf{^3}$} & \textbf{3} & 200 & 8.0 & 3000 & 0.83 & 1.9 & 25\% & \textbf{2} & $1.0\times10^8$ & $1.3\times10^5$ & \textbf{8} & $5.2\times10^7$ & $6.6\times10^4$ & yes & 2 & 1.0 \\ 
\emph{RUN31}  & \textbf{512$\mathbf{^3}$} & \textbf{3} & 200 & 8.0 & 3000 & 0.83 & 2.0 & 25\% & \textbf{2} & $1.0\times10^8$ & $1.3\times10^5$ &\textbf{16} & $5.2\times10^7$ & $6.6\times10^4$ & yes & 2 & 1.0 \\ 
\emph{RUN32}  & \textbf{512$\mathbf{^3}$} & \textbf{3} & 200 & 8.0 & 3000 & 0.83 & 2.0 & 25\% & \textbf{4} & $2.6\times10^7$ & $3.3\times10^4$ & \textbf{8} & $1.3\times10^7$ & $1.6\times10^4$ & yes & 2 & 1.0 \\[0.8ex] 
\hline \\[-2ex]
\emph{RUN33} & \textbf{64$\mathbf{^3}$} & \textbf{6} & 200 & 8.0 & 3000 & 0.83 & 1.5 & 25\% & \textbf{2} & $1.0\times10^8$ & $1.3\times10^5$ & \textbf{8} & $5.2\times10^7$ & $6.6\times10^4$ & yes & 2 & 1.0 \\ 
\emph{RUN34}  & \textbf{1024$\mathbf{^3}$} & \textbf{2} & 200 & 8.0 & 3000 & 0.83 & 2.1 & 25\% & \textbf{2} & $1.0\times10^8$ & $1.3\times10^5$ & \textbf{4} & $5.2\times10^7$ & $6.6\times10^4$ & yes & 2 & 1.0 \\[0.8ex] 
\hline \hline \\[-2ex]
\emph{test}  & 128$^3$ & 6 &100 &  7.2 & 3000 & 0.83 & 2.5 & 13\% & 2 & $4.2\times10^8$ & $5.3\times10^5$ & 8 & $\mathbf{2.1\times10^8}$ & $2.6\times10^5$ & yes & 4 & \textbf{1.0} \\ 
\emph{acc}  & 128$^3$ & 6 &100 &  7.2 & 3000 & 0.83 & 2.5 & 13\% & 2 & $4.2\times10^8$ & $5.3\times10^5$ & 8 & $\mathbf{4227}$ & $5.3$ & yes & 4 & \textbf{1.0} \\ 
\end{tabular}
\vspace{1ex}

\raggedright{{\bf Note.} the \emph{test} and \emph{acc} runs are identical to the \emph{med} run, except that the \emph{acc} run is using an accretion efficiency $\eout=1$ and that the \emph{test} run, in addition to that, is only using a density threshold for accretion like the rest of the test runs. We have set in bold face the main parameters that are explored in each set of runs.}
\label{t2}
}
\end{table*}

\section{Testing the sink particle algorithm}\label{A:sinks}
The sink particle algorithm contains a number of density and distance thresholds for creation and accretion of gas to the sink
particles, and it is important to understand the impact of different choices, and in particular the impact on the SFR, and the IMF.
To test the algorithm, we have created a large grid of models that enables us to study the
dependence of the IMF on all parameters. The test runs are prepared using the same physical parameters and driving as
the convergence runs and are listed in table \ref{t2}. Our grid of models mainly affects three aspects of the simulation.

\subsection{Creation}
The creation of sink particles is controlled by threshold density and the exclusion radius. The higher the
density threshold, the more probable it is that we have identified a genuinely collapsing region. It is not clear, though, how much
numerical fragmentation due to underresolving of the Jeans length on the grid will affect the creation of sink particles, given that
fragments very close to a newborn sink particle can end up being accreted.
\citet{Ostriker+13} find that using a threshold density $\rhosink$, where the
Jeans length is only resolved by a single cell ($L_{\rm J}\sim1$), gives consistent results compared to using lower-density floors.
In the smallest test runs with a $64^3$ root grid, where we can make a systematic set of models, our runs have $L_{\rm J}\ge$8 everywhere until
the final AMR level. On the final level we use three different $\rhosink$, corresponding to Jeans lengths of $L_{\rm J, s}=$2, 4, and 8.
The exclusion radius $\rexcl$ controls how close we allow a new sink particle to be formed to existing sinks. Very close to the sink particle the
gas flow is disturbed by the sink particle itself, accretion, and the accumulation of magnetic fields. It is not a priori clear how well the MHD
dynamics is described in the vicinity of a sink particle, and pileup of gas can lead to the artificial creation of sink particles very close to $\rexcl$.
To test the effect of $\rexcl$, we use $\rexcl$ between 4 and 64 in our models, measured in units of $\Delta x$ of the finest AMR level.

\subsection{Accretion}
A protostar is initially formed in a gravitational free-fall collapse of the gas, while afterward the star grows through accretion.
As discussed in \S~\ref{sec:methods} we use two different accretion methods: either a combination of velocity and density thresholds that
makes it possible to describe accretion at low densities, or only a density threshold.
Except for the \emph{acc} model, which is used to compare the two methods, we use the second method for all test runs and let the star accrete
when the density reaches $\rhoacc={\rhosink}/2$, and the cell is closer than the accretion radius to a sink particle. The accretion radius is kept at
$\racc=2 \Delta x$ for runs using a pure density threshold, while when using a combination of velocity and density threshold methods it is possible to relax this,
due to a tapering function (equation \ref{eq:tapering}), and use a larger accretion radius. We have run a number of control experiments to probe the impact of having
larger accretion radius and accretion density, and we do not find significant differences as long as $\rhoacc<\rhosink$, and $\racc<\rexcl/2$. Even though a smaller $\rhoacc$ and
larger $\racc$ effectively lower the gas density around the sink particles, avoiding kinks in the gas density distribution close to the star and thus
artificial creation of sink particles, it also disturbs the gas dynamics at larger distances from the sink particle. We find the above compromise,
$\rhoacc=\rhosink/2$ and $\racc=2 \Delta x$, to be a good solution for the test runs, while for the main runs with a combined velocity and density criterion
we use $\racc=\rexcl/2$ \citep[see also][for a discussion of the accretion parameters]{Kuffmeier+17}.

\subsection{Numerical resolution}
The bulk of the models are evolved with a 64$^3$ root grid. To ensure $L_J\ge8$ everywhere, the root grid is refined
at an overdensity of 2, while subsequently the adaptive mesh is refined where the density has grown with a factor of four, to have the same minimum
Jeans length at all AMR levels, except possibly at the finest level of refinement. This Truelove refinement is different from Lagrangian codes, such
as SPH, and corresponds to having increasing mass resolution in the finer cells. To test the impact of numerical resolution, we also run a number of models
with a $512^3$ and $1024^3$ root grid. This has a significant impact on the resolution of the fragmentation, due to the turbulent cascade
(the sonic point is expected to be at $\sim{\mathcal M_{\rm s}}^{-2} L_{\rm box} = 0.01 L_{\rm box}$).

Besides studying the effect of increasing the root grid size, we also vary the resolution by changing the number of AMR levels above the root
grid, $N_{\rm AMR}$. The supersonic turbulence creates a filamentary self-similar gas structure with a lognormal density distribution. To avoid the
creation of spurious sink particles, it is important to achieve a significant separation (at least by a factor of 14, which is the density contrast of a
critical BE sphere) between the density threshold for sink creation, \rhosink{},
and the highest density reached purely through the turbulent compression of the gas. To test the effect of $N_{\rm AMR}$, we evolve models
with $N_{\rm AMR}=$ 6, 7, and 8.

Varying $\rhosink$, \rexcl{}, and $N_{\rm AMR}$ gives a regular grid of 27 models with a $64^3$ root grid. To further test the impact of resolving
the turbulence, we have also evolved five models with a $512^3$ root grid and a single model with a $1024^3$ root grid. Finally, two runs that
are similar to the \emph{med} convergence run are included to compare the different settings for the gas accretion.

In table \ref{t2}, we list 36 models, all simulating the same physical problem and thoroughly testing numerical aspects of our model,
and in particular the impact on the star formation rate and the stellar IMF. In addition, more than 20 other models with variations of root grid resolution,
Jeans length, and auxiliary parameters, such as the hydro solver, the coupling of gravity for the sink particles, and the relative size of $\racc$ compared to
$\rexcl$ have been run during the course of this project, but to keep the table manageable, we only include runs that are directly referenced and used in the paper.

\begin{figure}[t]
\includegraphics[width=0.44\textwidth]{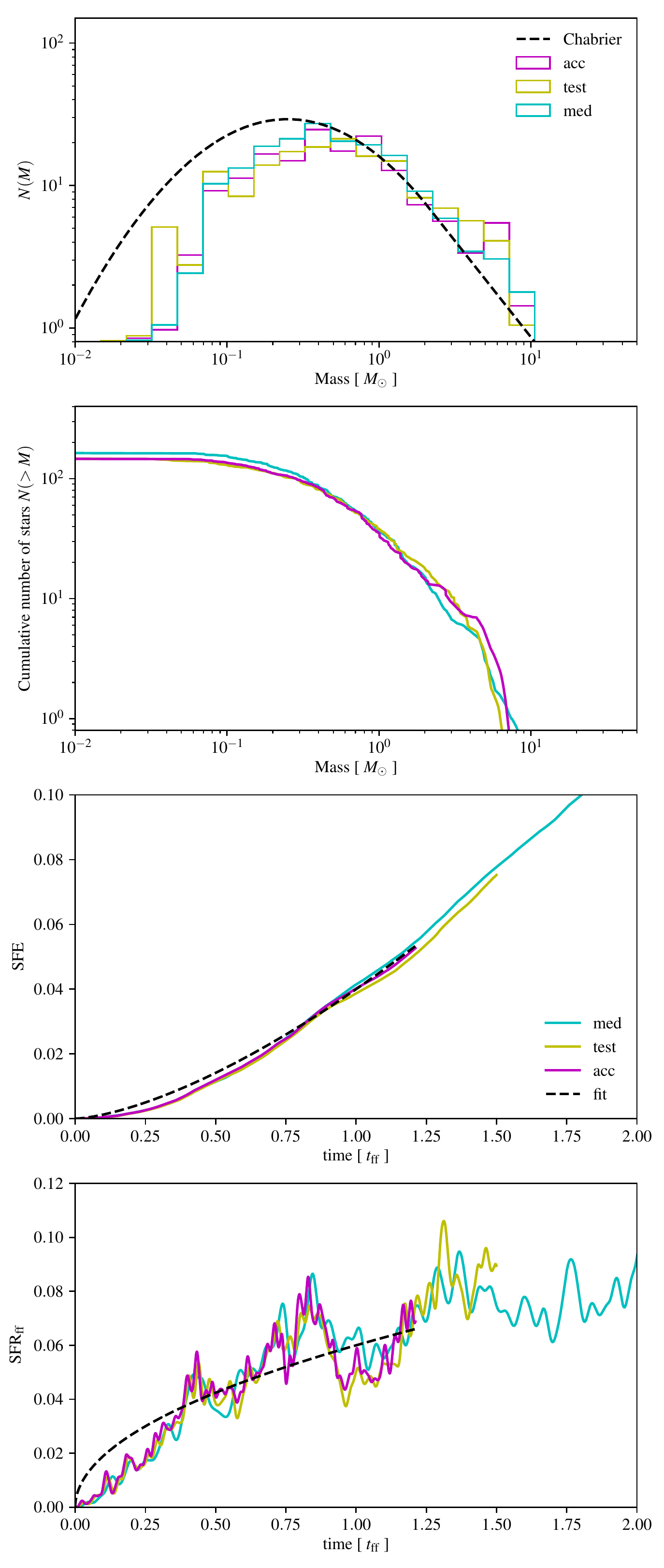}
\caption{Comparing different {\eout} and accretion algorithms. This compares
\emph{med}, \emph{test}, and \emph{acc}. For the first two panels the runs are sampled at an SFE of 4.5\%, 10.8\% and 10.8\% respectively.
The first panel shows the IMF, and the second panel the cumulative IMF. In the third and fourth panels are shown the SFE and SFR per free-fall
time as a function of time. In all panels the masses for the second and third run are rescaled by a factor of 2.4 to show the correspondence
between $\eout=0.5$ and  $\eout=1$.}\label{fig:wind-comp}
\end{figure}
\begin{figure*}[t]
\includegraphics[width=6cm]{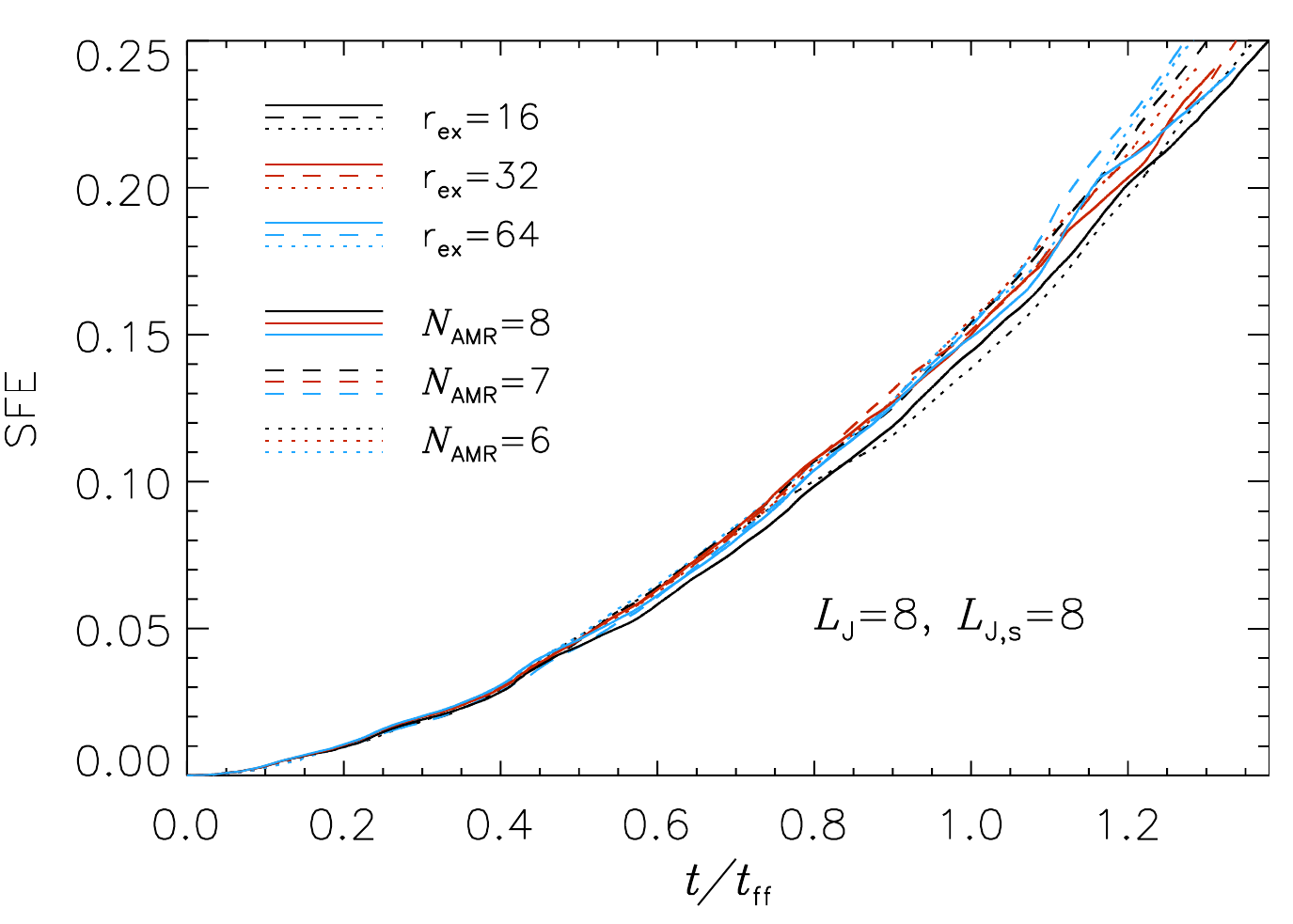}
\includegraphics[width=6cm]{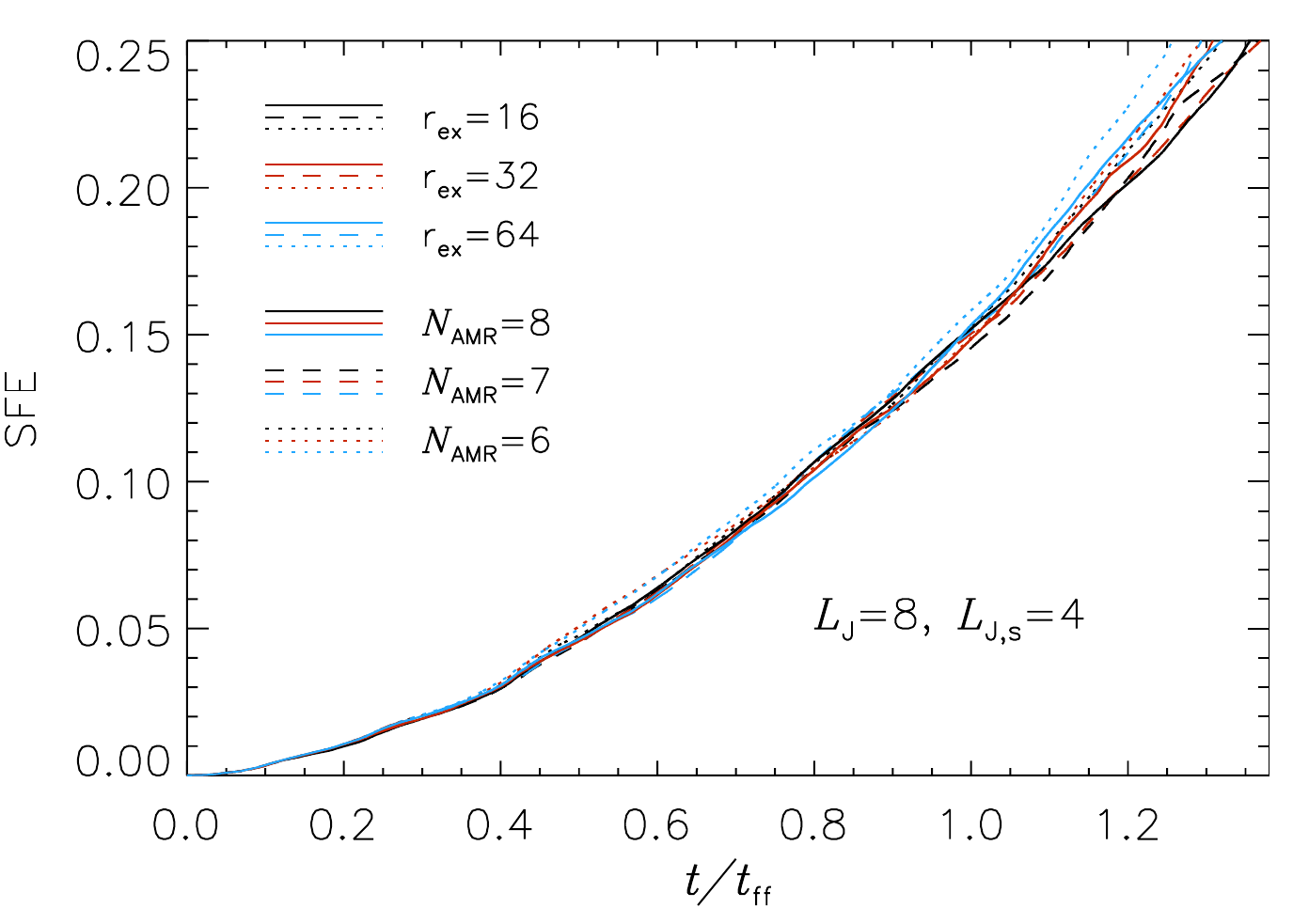}
\includegraphics[width=6cm]{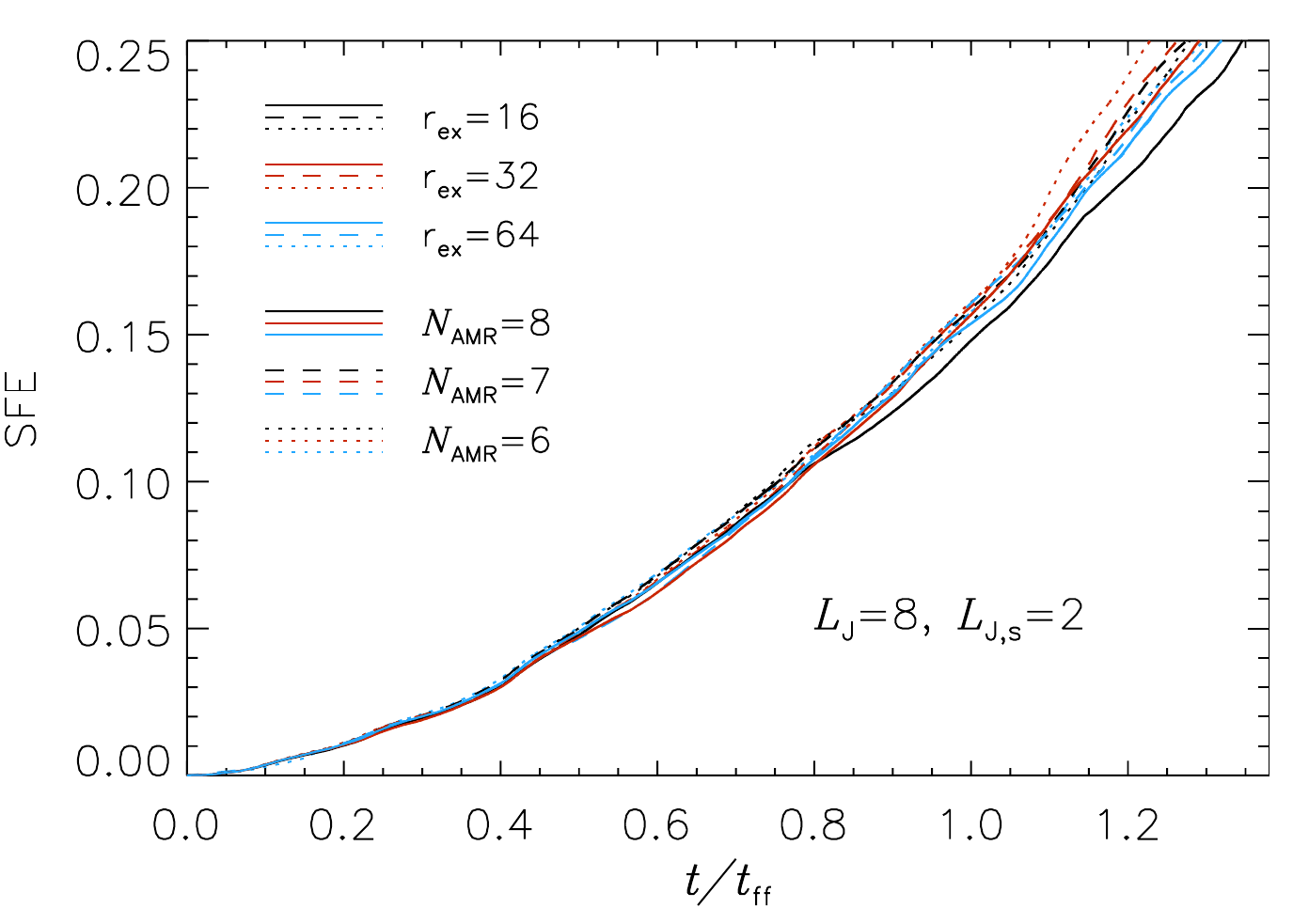}
\caption{SFE versus time from all the test runs with $64^3$ root grid, and $L_{\rm J}=$8. The three panels corresponds to $L_{\rm J,s}=$8 (left),
$L_{\rm J,s}=$4 (middle), and $L_{\rm J,s}=$2 (right).}
\label{fig:SFE}
\end{figure*}
\begin{figure*}[t]
\includegraphics[width=6cm]{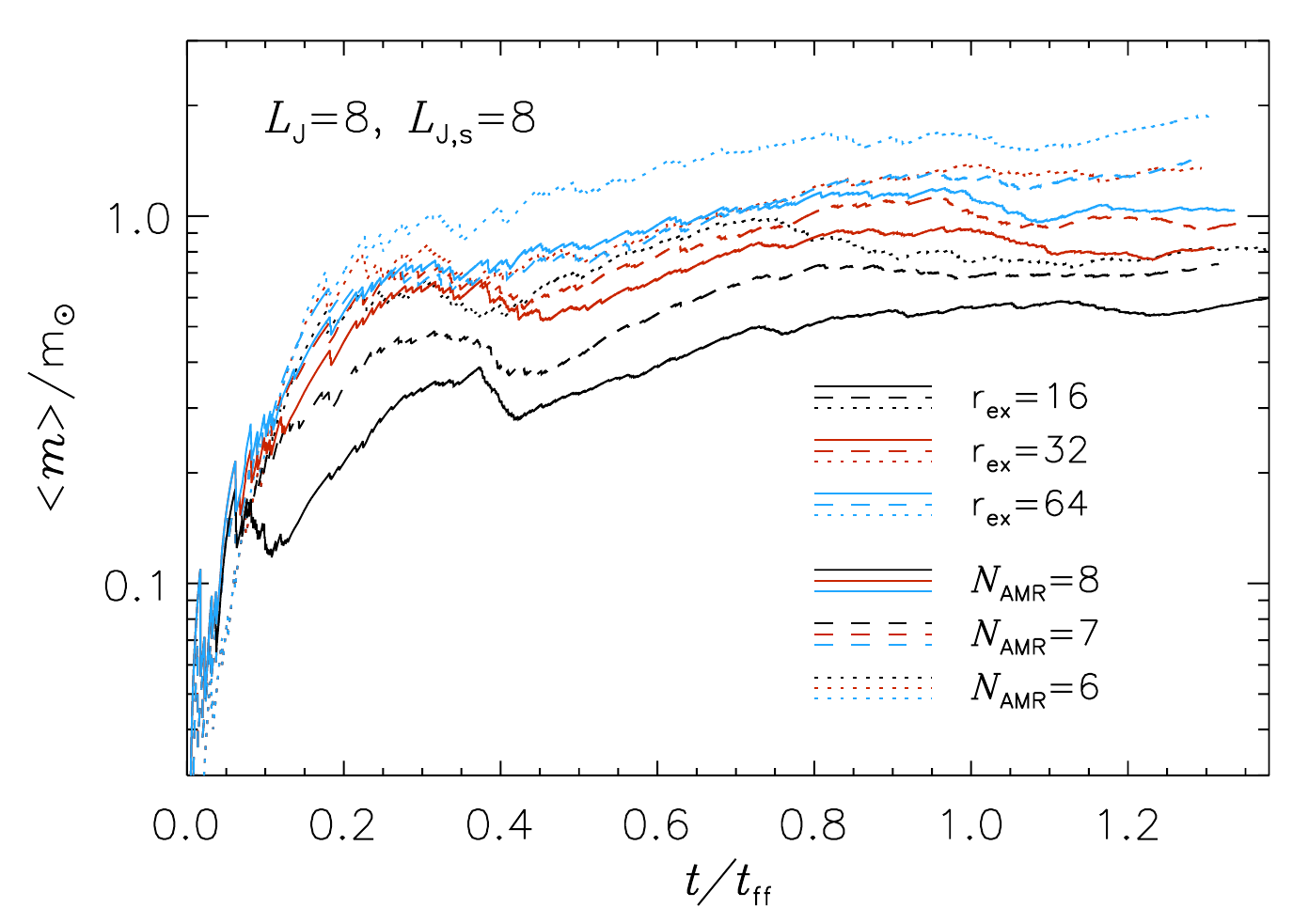}
\includegraphics[width=6cm]{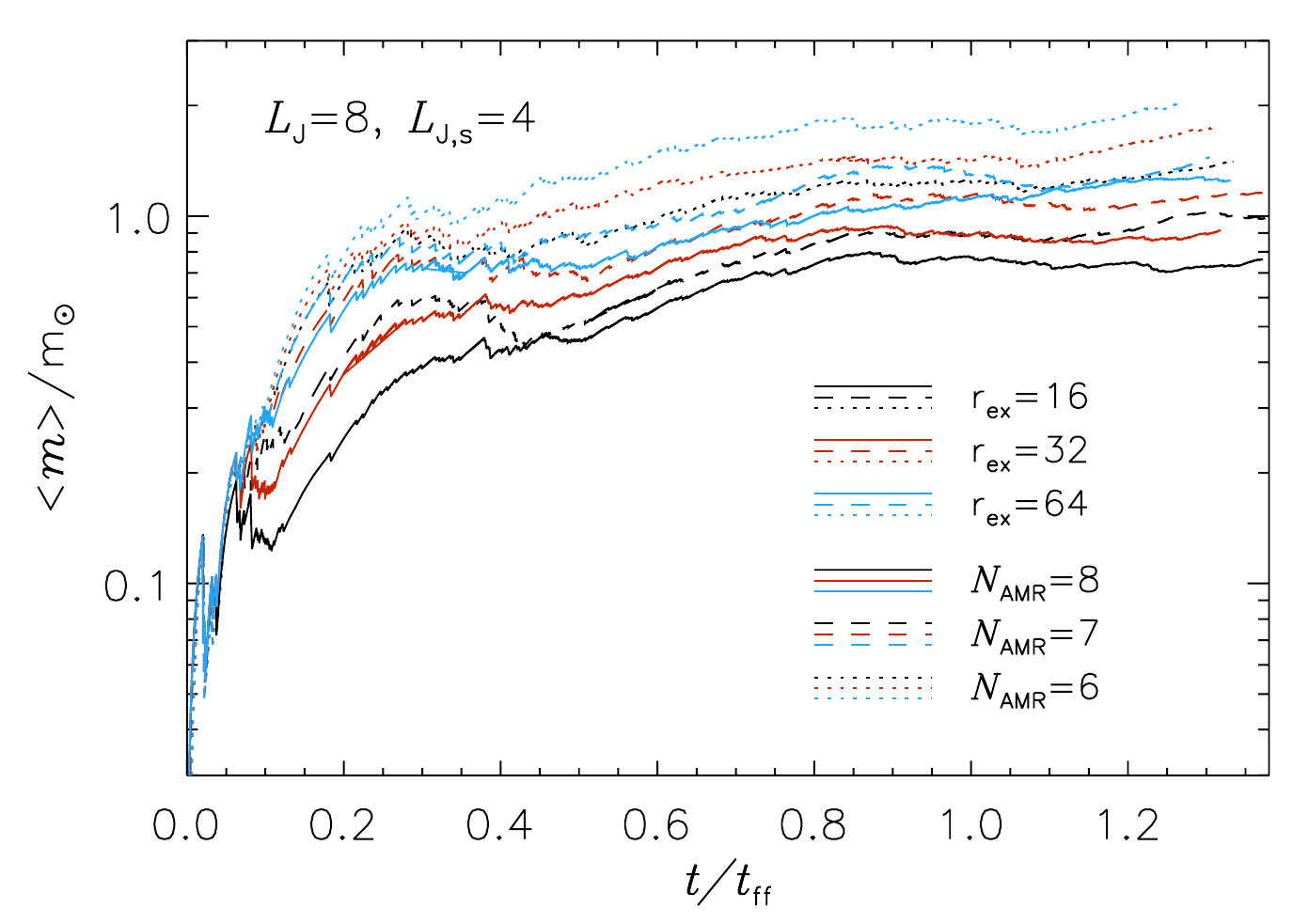}
\includegraphics[width=6cm]{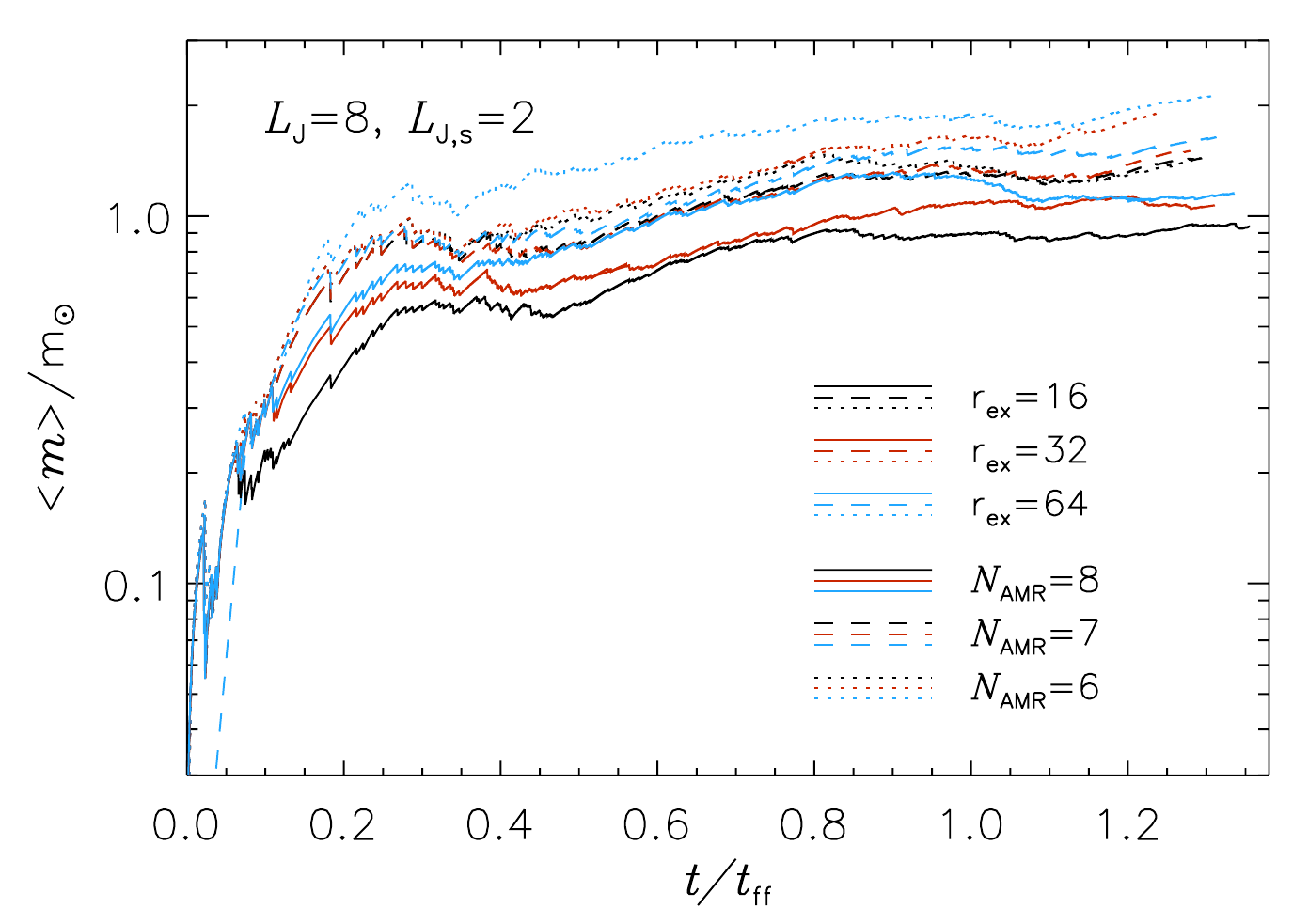}
\caption{Average stellar mass versus time from all the test runs with $64^3$ root grid, and $L_{\rm J}=$8.
The three panels corresponds to $L_{\rm J,s}=$8 (left), $L_{\rm J,s}=$4 (middle), and $L_{\rm J,s}=$2 (right).}
\label{fig:avmass}
\end{figure*}
\subsection{Starformation efficiency and average stellar mass}
The models have run with self-gravity and sink particles for up to 1.5 free-fall times or 1.8 Myr, similar to the lifetimes of nearby star-forming regions.
At the end of the runs, the SFE reaches 25\% Notice that since these models have an accretion efficiency
$\eout=1$, their SFEs have to be rescaled with a factor of 2.4, corresponding to a rescaled SFE of 10\%, to compare them with the convergence runs
as shown in Fig.~\ref{fig:wind-comp} and discussed in \S~\ref{sec:methods}. Fig.~\ref{fig:SFE} shows that the overall SFE is
extremely robust between runs, which is a consequence of gravitational collapse: if gas reaches a certain
threshold density, it will almost unavoidably collapse. Depending on the details of the sink recipe, this will result in
either accretion to an existing sink particle or creation of a new sink particle, but the total accreted mass remains nearly constant.

The average mass of the sink particles produced in each run is shown in Fig.~\ref{fig:avmass}. We have grouped the runs in terms
of the Jeans number at the threshold density for sink particle creation, $L_{\rm J,s}$. There are both common trends and a significant
dependence on the numerical details of the average mass: there is a clear dependence on both the exclusion radius, the number of AMR
levels, and the density threshold at which sink
particles are created. Given that the SFE is very similar for all runs, a smaller average mass corresponds to a higher number of sink
particles and vice versa. A smaller exclusion radius, a higher number of AMR levels, and a lower threshold for sink particle creation all
result in a lower average mass and therefore a higher number of sink particles. The difference between different
models diminishes with increasing threshold for sink creation, showing convergence at the highest resolution. The parameter with the
smallest effect on the average mass is the exclusion radius. Adding a higher number of AMR levels, or changing the sink threshold, both have a larger impact.

As we will see below, the reasons for the significant dependence of the average stellar mass on the sink particle creation parameters
and $N_{\rm AMR}$ may be numerical: with a small root grid of $64^3$ cells, the threshold for sink creation is too low, resulting
in the spurious creation of extra particles. Thus, besides being a clear illustration of the effect of numerical parameters, Fig.~\ref{fig:avmass}
is also a clear indication that the resolution of these test runs is too low to study the stellar IMF, though probably sufficient to study the SFR
(as indicated by Fig.~\ref{fig:SFE}). Although we cannot afford a similar systematic set of test simulations with a much larger root grid size, we expect
that the dependence on numerical parameters shown in Fig.~\ref{fig:avmass} would be significantly weaker in simulations with a root grid size
of $256^3$ cells or larger.  The higher-resolution runs can resolve an
increasing number of close binaries, which should not be confused with the artificial creation of spurious sink particles. Finally, another important
effect that is explored below is how well the turbulent fragmentation and the smallest collapsing cores are resolved.

A common feature of all models is that the average sink particle mass becomes nearly stationary after roughly 0.8 free-fall times. This can
be understood in terms of the time it takes for the most massive stars in a cluster to accrete. Only after $\sim$0.8 free-fall times is the
enough of the high-mass Salpeter slope developed to keep the average mass essentially constant, as demonstrated by Figures
\ref{fig:imf_time} and \ref{fig:imf_par} in \S~\ref{sec_var_time}, which are based on our highest-resolution simulations.

\subsection{Nearest-neighbor separation at formation}\label{A:neighbor}
To investigate the effects of different parameter choices, a higher-order statistics than the SFE or the average stellar mass is needed.
In this section we will introduce the nearest-neighbor separation at formation, and we use it below, together with the IMFs
of the different runs, to better understand the differences between the test runs. The nearest-neighbor separation at formation, $\mathcal{D}$,
or ``neighbor statistics" for brevity, is the histogram of distances from a set of sink particles, at the time of their formation, to the
nearest other particle in the simulation volume.  It is both very simple to extract from a numerical simulation and a powerful statistic
to judge the fidelity of a sink particle distribution, if we have prior knowledge about how it should behave.

In this subsection we will argue that $\mathcal{D}$ should increase as a function of distance for reasonable assumptions about the stellar
distributions. This can then be used to detect when models are polluted by sink particles generated as a consequence of nonphysical
fluctuations close to other sink particles.

Assume that there is no difference between the statistical distribution of newly formed stars and more evolved stars. Then $\mathcal{D}$
can be defined in terms of the nearest-neighbor function of a point-particle set. For a point-particle set with $N$ particles the probability
of finding another point at a distance $r$ of a given point is
\begin{equation}
D^N(r) = - \frac{dP(V(r))}{dr}\,,
\end{equation}
where $P(V(r))$ is the conditional probability (averaged over all points) that there is another point inside the volume $V$ centered on a given point,
and the derivative gives the probability that the nearest neighbor is in the interval $[r,r+dr]$. Since in a star forming region stars are formed one at
a time, then $\mathcal{D}$ can then be found by forming the cumulative sum
\begin{equation}
\mathcal{D}^N(r) = {1 \over N}\sum_{n=1}^{N} D^n(r)
\end{equation}
This is a highly nontrivial distribution related to the $n$-point correlation functions. As an example, if the neighbor diagram has a strong peak close
to the exclusion radius for creation of new particles, $\rexcl$, it could indicate that
\begin{itemize}
\item The overall spatial resolution is so low that clumps that should fragment into many stars are not well resolved.
\item The minimum density for sink formation, $\rhosink$, is low enough that disturbances in the fluid dynamics around other sink particles are sufficient
to induce the formation of new sink particles.
\item The Jeans length is not sufficiently resolved, and therefore numerical fragmentation results in a large number of artificial ``multiple systems".
\end{itemize}

To elucidate the properties of this function, we may consider the simplest theoretical example of a stationary Poisson process with a volume point
density of $n$, where we have
\begin{equation}
D^n(r) = - \frac{d (1 - e^{-n V(r)})}{dr} = 4 \pi r^2 n e^{-n V(r)} .
\end{equation}
If we assume a starting density of one star per unit volume appropriate for our periodic box, we can then write the neighbor statistics for $N$ stars as
\begin{equation}\label{eq:sum}
\mathcal{D}^N(r) = {4 \pi r^2 \over N} \sum_{n=1}^{N} n \exp[-n V(r)]\,.
\end{equation}
To test for numerical artefacts, the solutions at small distances are most relevant, where  ($N V(r) < 1$). There we have $\mathcal{D}(\ln r) \propto r^3$
with a binning in terms of $\ln r$, as done in the histograms from the numerical simulations. 
At larger scales the function has a relatively well-defined peak close to the point where all terms in the sum in Eq.~\ref{eq:sum} still contribute to the total
histogram, i.e.~where $N V(r) \sim 1$, corresponding to the average separation between particles of the final point set.

Define the two-point correlation function, $\xi(r)$, as the excess probability, compared to a homogeneous point set, of finding
another particle at distance $r$ from a given point
\begin{equation}
dP=n(1+\xi(r))dV .
\end{equation}
Assuming for a general distribution that all higher-order correlation functions are negligible, we can then repeat the analysis above to find that
\begin{equation}
\mathcal{D}^N(r) = {4 \pi r^2 [1 + \xi(r)] \over N} \sum_{n=1}^{N} n \exp[-n V_\xi(r)] \,,
\end{equation}
where $V_\xi(r)\equiv\int_V [1+\xi(r)]dV$. For large $N$,
\begin{align}\nonumber
\mathcal{D}^N(r) \approx & {4 \pi r^2 [1 + \xi(r)] \over V_\xi(r)^2} \left\{ (1 + V_\xi(r)) \exp[-V_\xi(r)] \right. \\
 & \quad \left. - (1 + NV_\xi(r)) \exp[-NV_\xi(r)]\right\}
\end{align}
For a clustered distribution such as those observed in young stellar clusters, and assuming a power law $\xi(r)\propto r^\beta$,
we find that at small separations, where $\xi(r)\gg1$,
\begin{equation}
\mathcal{D}(r) \propto r^{3+\beta} d\ln r\,.
\end{equation}

Observationally, the two-point correlation functions in the nearby Taurus, $\rho$ Oph, and Orion regions are found
to be a power law with exponents in the range $-2<\beta<-1$ \citep{Gomez+93,Larson95,Gomez+Lada98}.
This corresponds to a slowly increasing trend for $\mathcal D$ as a function of distance, but with some uncertainty.
In the numerical models we are applying the statistics at the birth of the star, while observationally we see a time-evolved distribution.
Spectroscopic multiple systems formed through e.g.~disk fragmentation may eject stellar companions to the field, and tight multiple stellar systems may
have formed dynamically from systems with larger separations.
In a clustered system, $\mathcal D$ may also have contributions from higher-order correlation functions. Thus, the observational evidence seems
to favor a constant or weakly increasing nearest-neighbor separation at formation as a function of the logarithmic distance binning,
though there is a considerable uncertainty for the innermost scales, below $\sim$1000 au. A decreasing $\mathcal D$ would require a
highly peaked two-point correlation function with a slope less than $\approx$ -3, which seems to be at odds with the observations.

Because $\mathcal{D}$ should increase as a function of distance (up to the mean particle separation), we do not expect it to peak at the smallest 
possible distance in the simulation, that is, at the exclusion radius, $\rexcl$ (see below). Thus, a sharp peak in $\mathcal{D}$ at $\rexcl$ is to be 
interpreted as a clear evidence of spurious sink particles emerging from numerical artifacts (e.g. unphysical density fluctuations near the accretion 
radius).

\begin{figure*}[t]
\includegraphics[width=6cm]{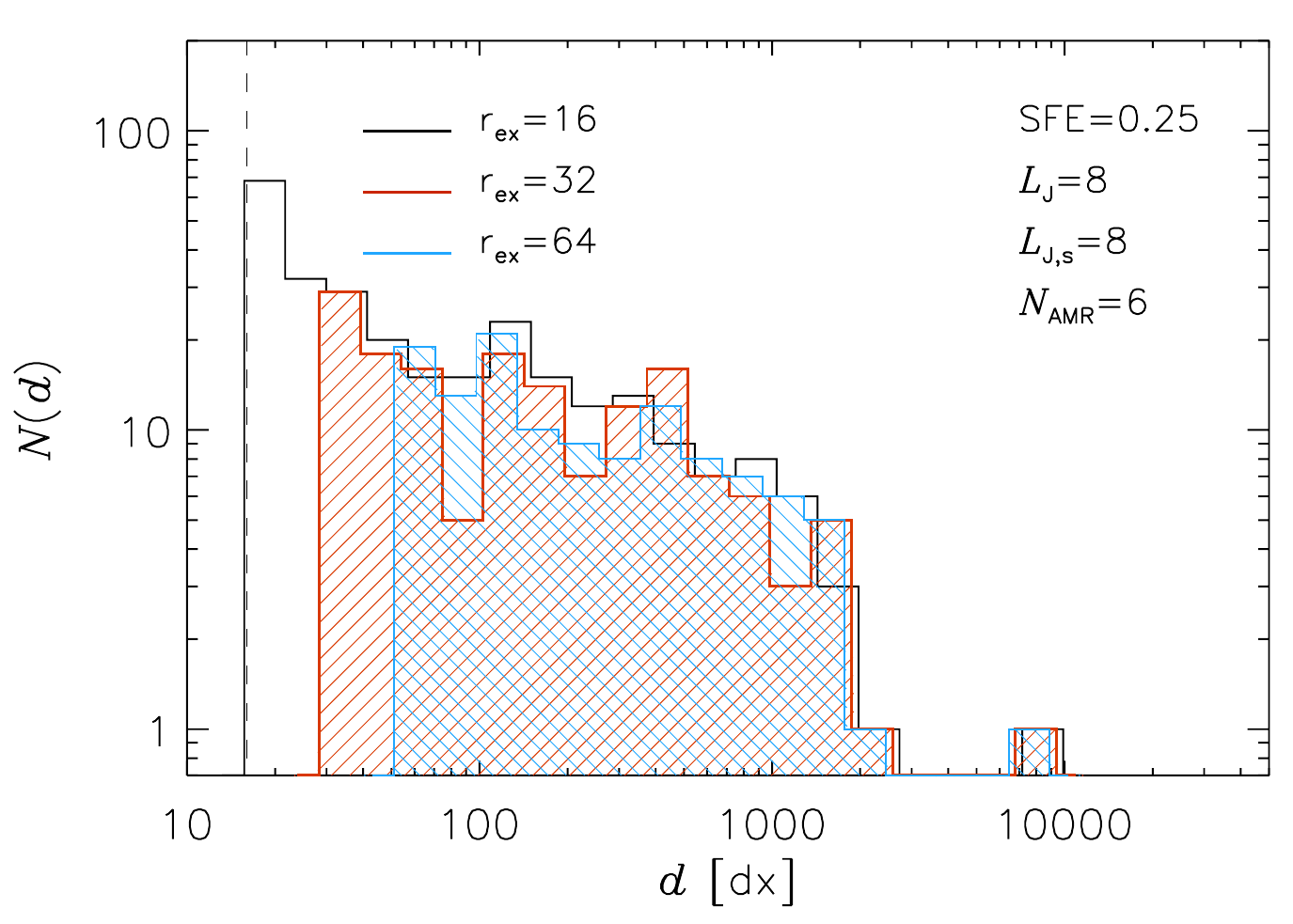}
\includegraphics[width=6cm]{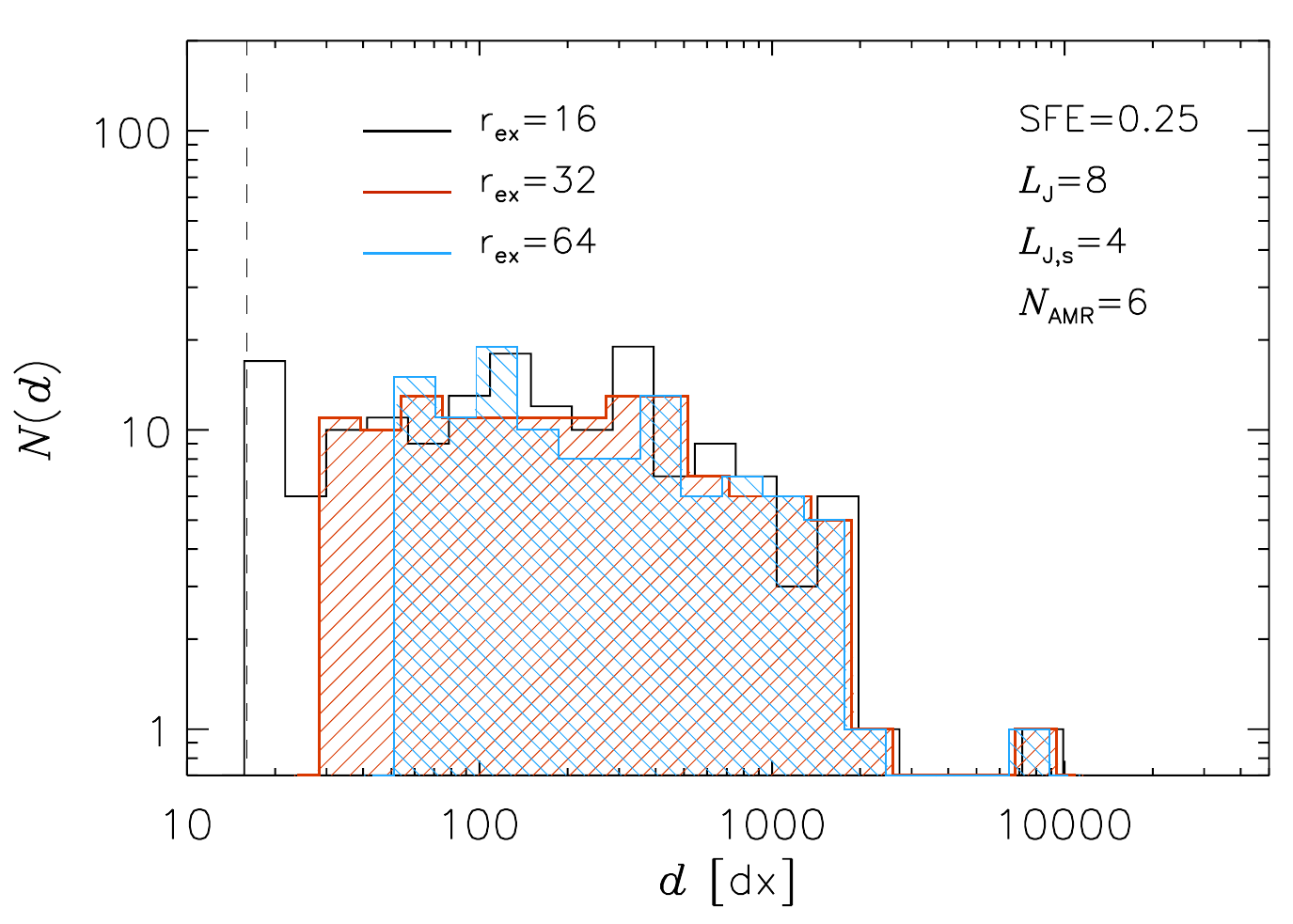}
\includegraphics[width=6cm]{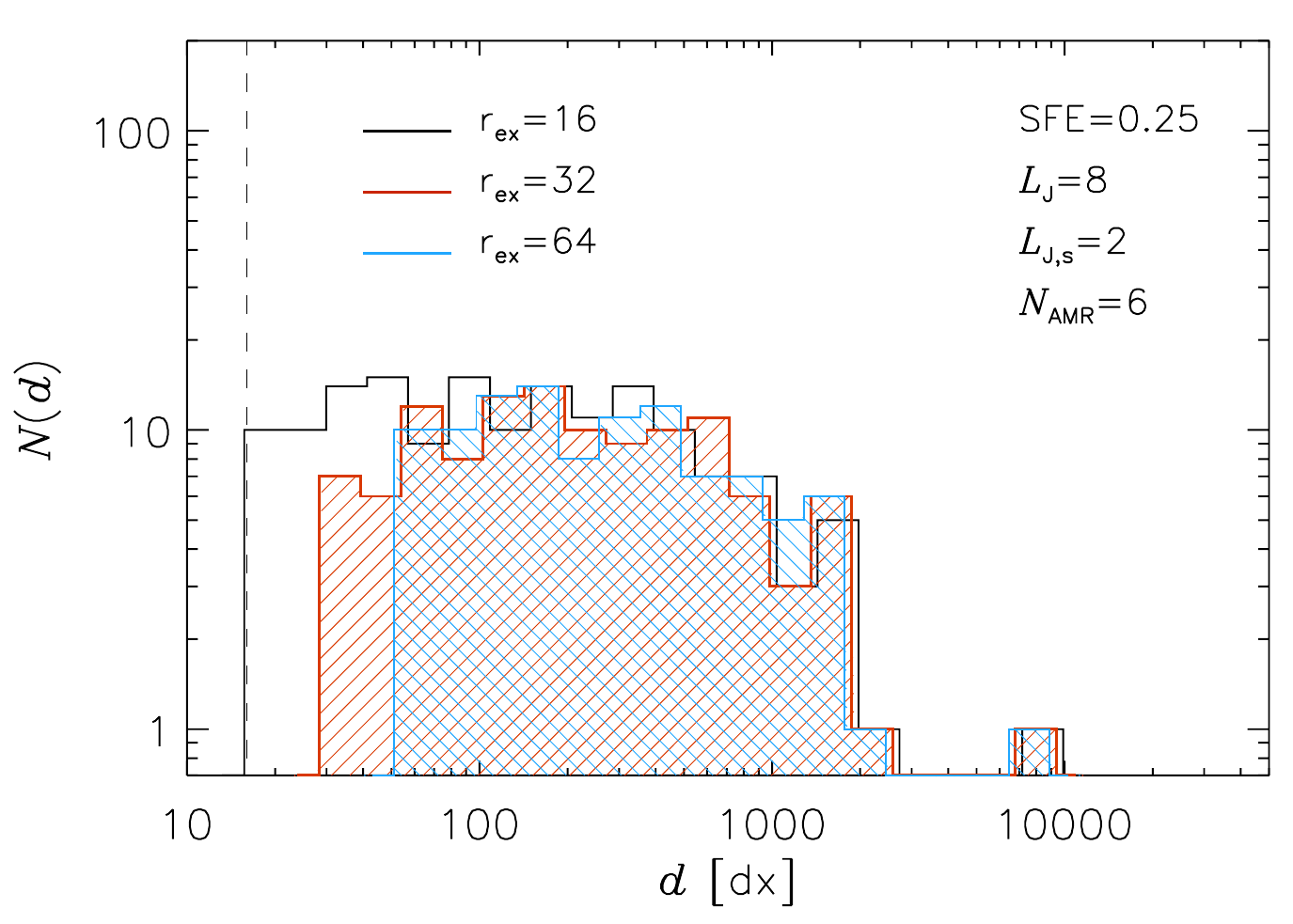}
\includegraphics[width=6cm]{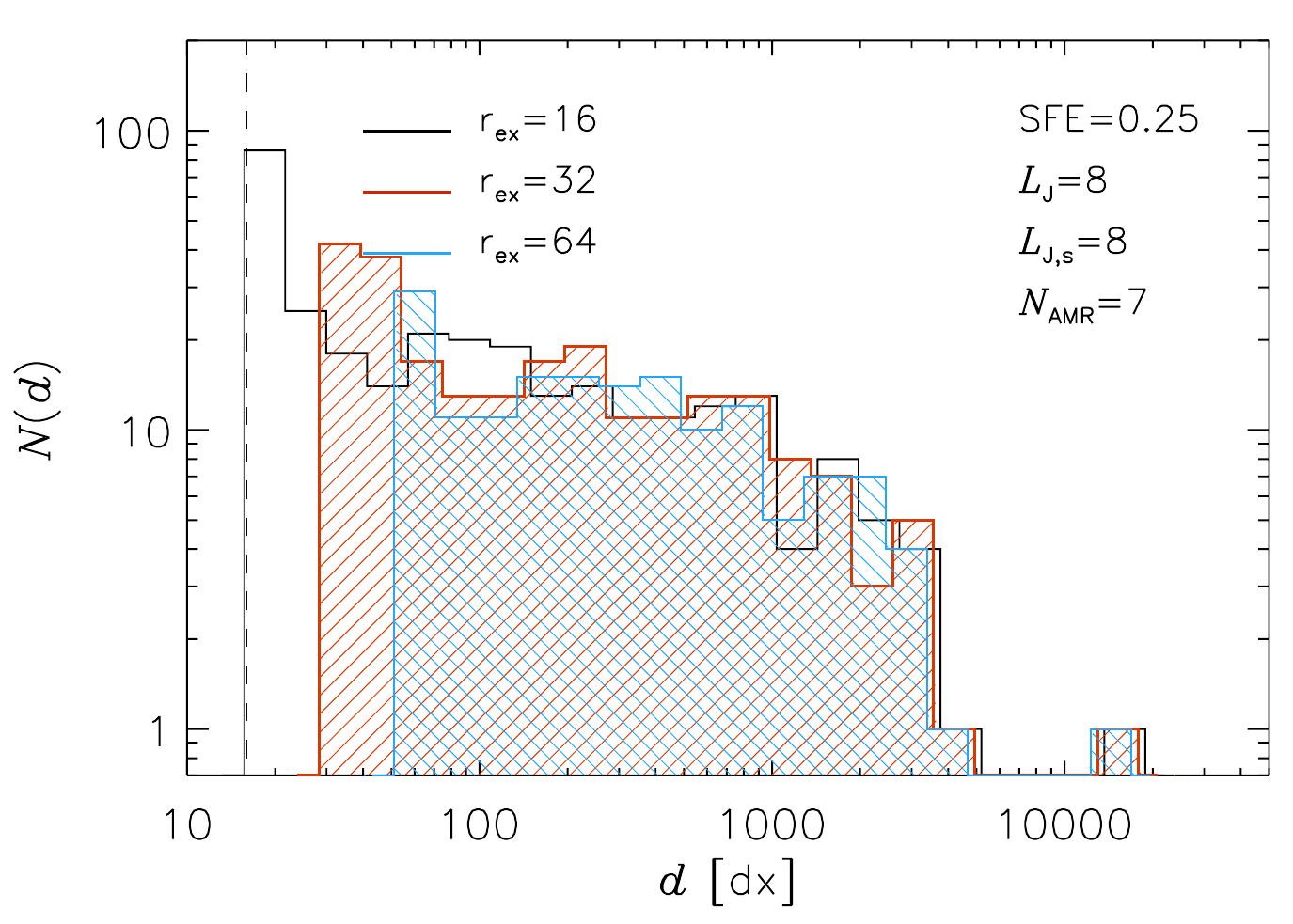}
\includegraphics[width=6cm]{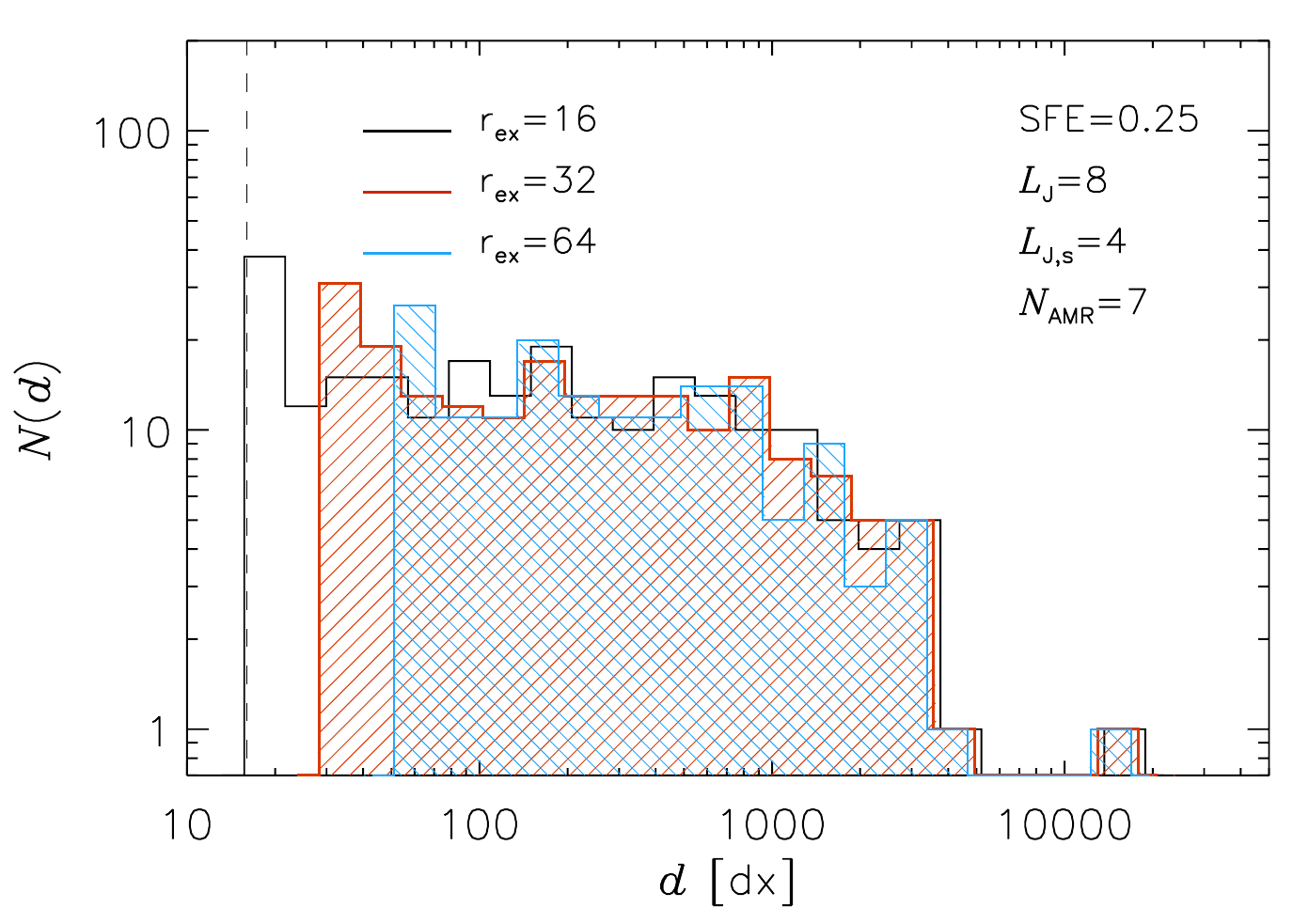}
\includegraphics[width=6cm]{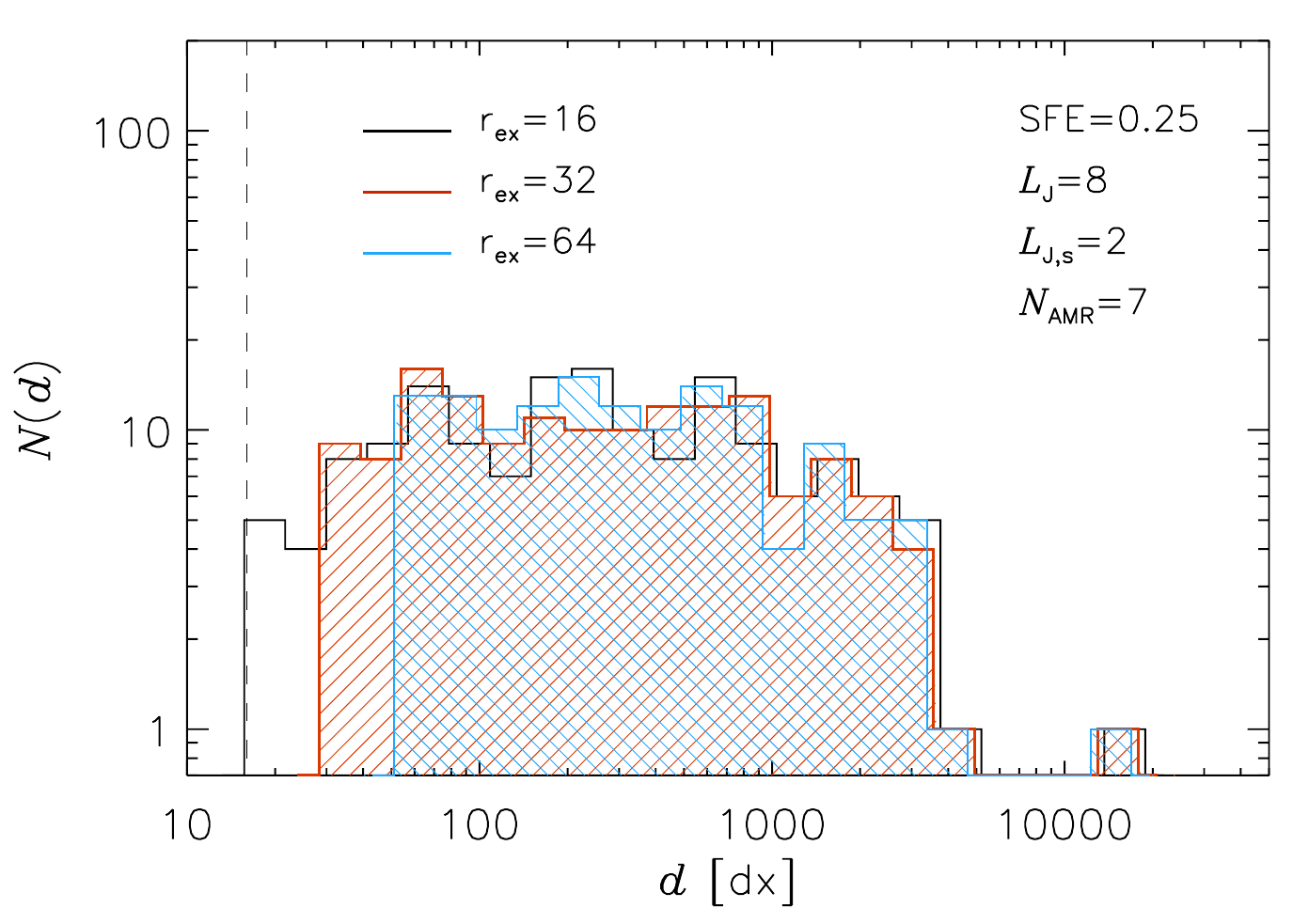}
\includegraphics[width=6cm]{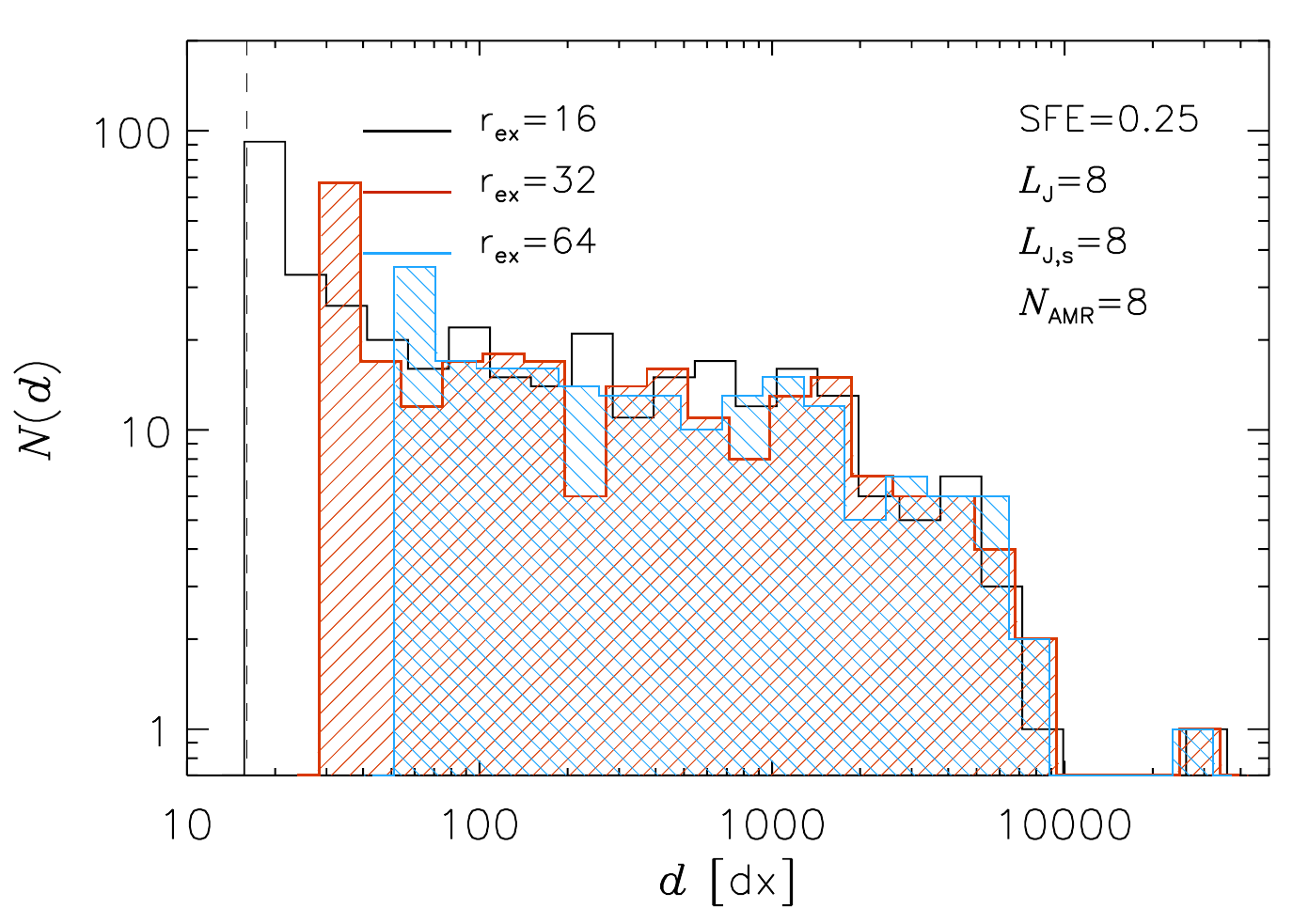}
\includegraphics[width=6cm]{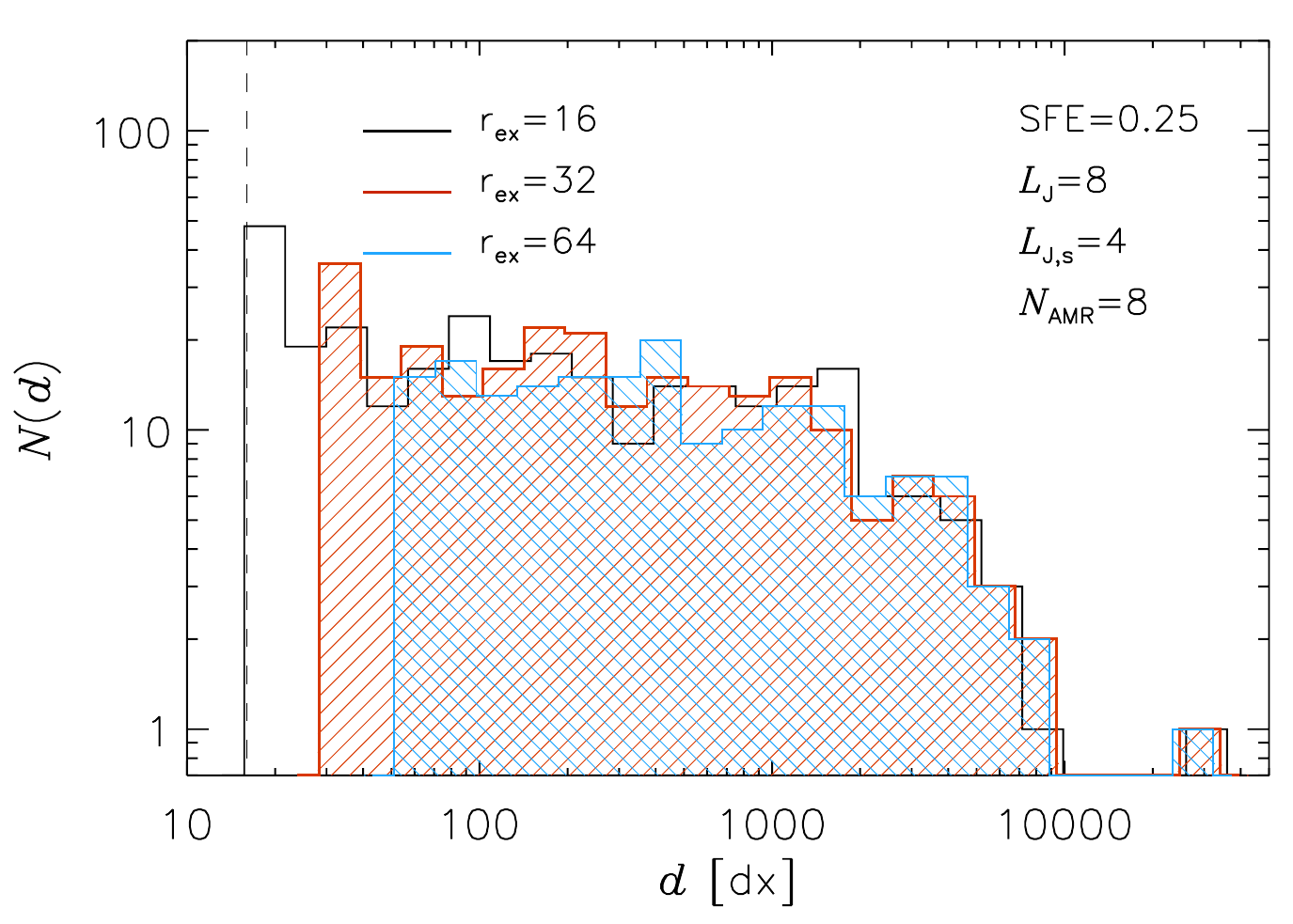}
\includegraphics[width=6cm]{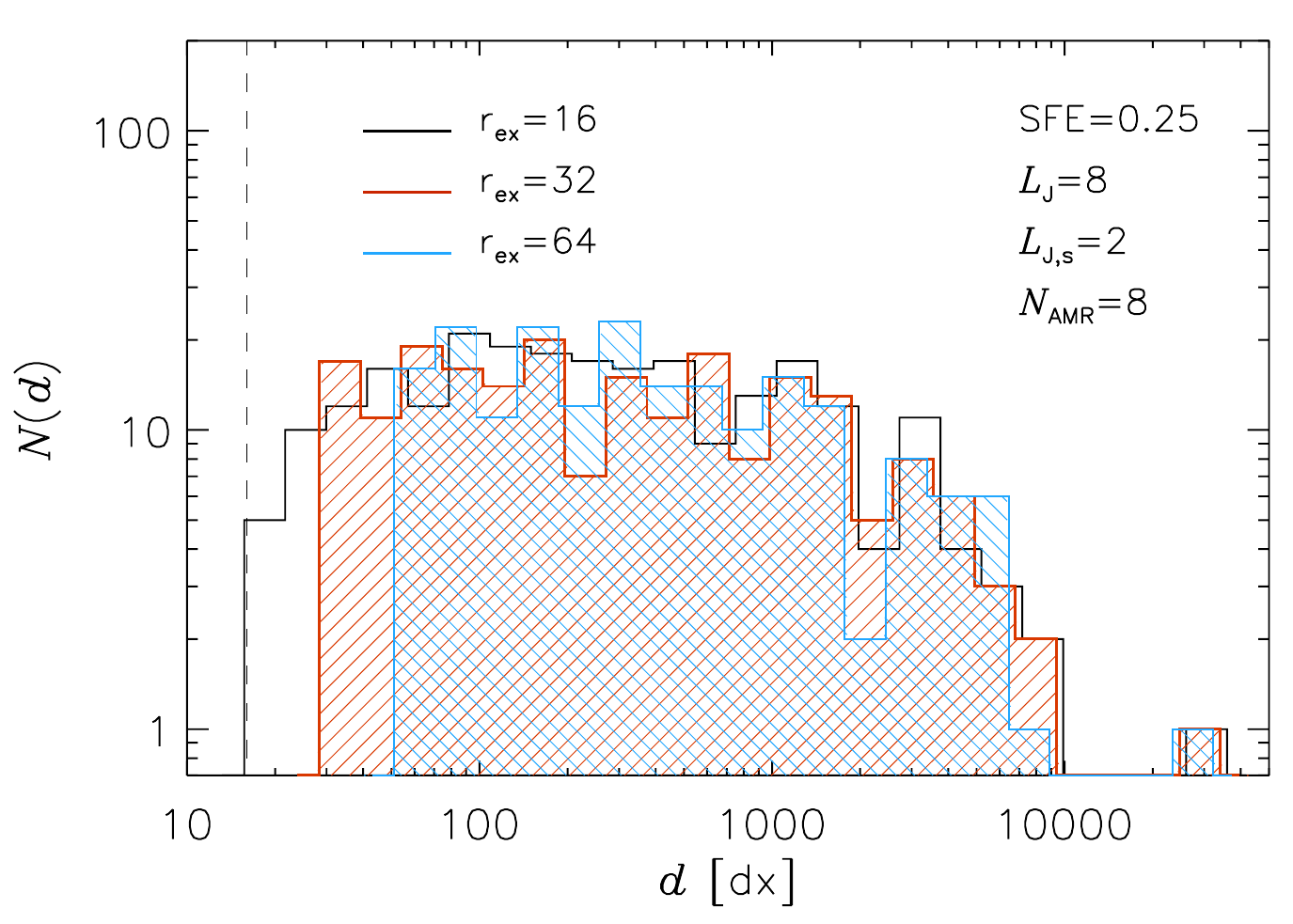}
\caption{Neighbor statistics for \emph{RUN1}--\emph{RUN27} with $64^3$ root grid and $L_{\rm J}=$8. Each row shows the neighbor statistics for 
runs with a constant number of levels of refinement, $N_{\rm AMR}=6$ (\emph{RUN1}--\emph{RUN9}, top), 7 (\emph{RUN10}--\emph{RUN18}, middle), and 8 (\emph{RUN19}--\emph{RUN27}, bottom). From left to right, $L_{\rm J,s}=$8, 4, and 2. Each panel shows three different  values of the exclusion radius, 
$\rexcl=$16$\Delta x$, 32$\Delta x$, and 64$\Delta x$, as black, red, and blue histograms, respectively. The dashed vertical lines mark the distance equal to the smallest $\rexcl$.}
\label{fig:neighbor_64}
\end{figure*}

\begin{figure*}[th]
\includegraphics[width=6cm]{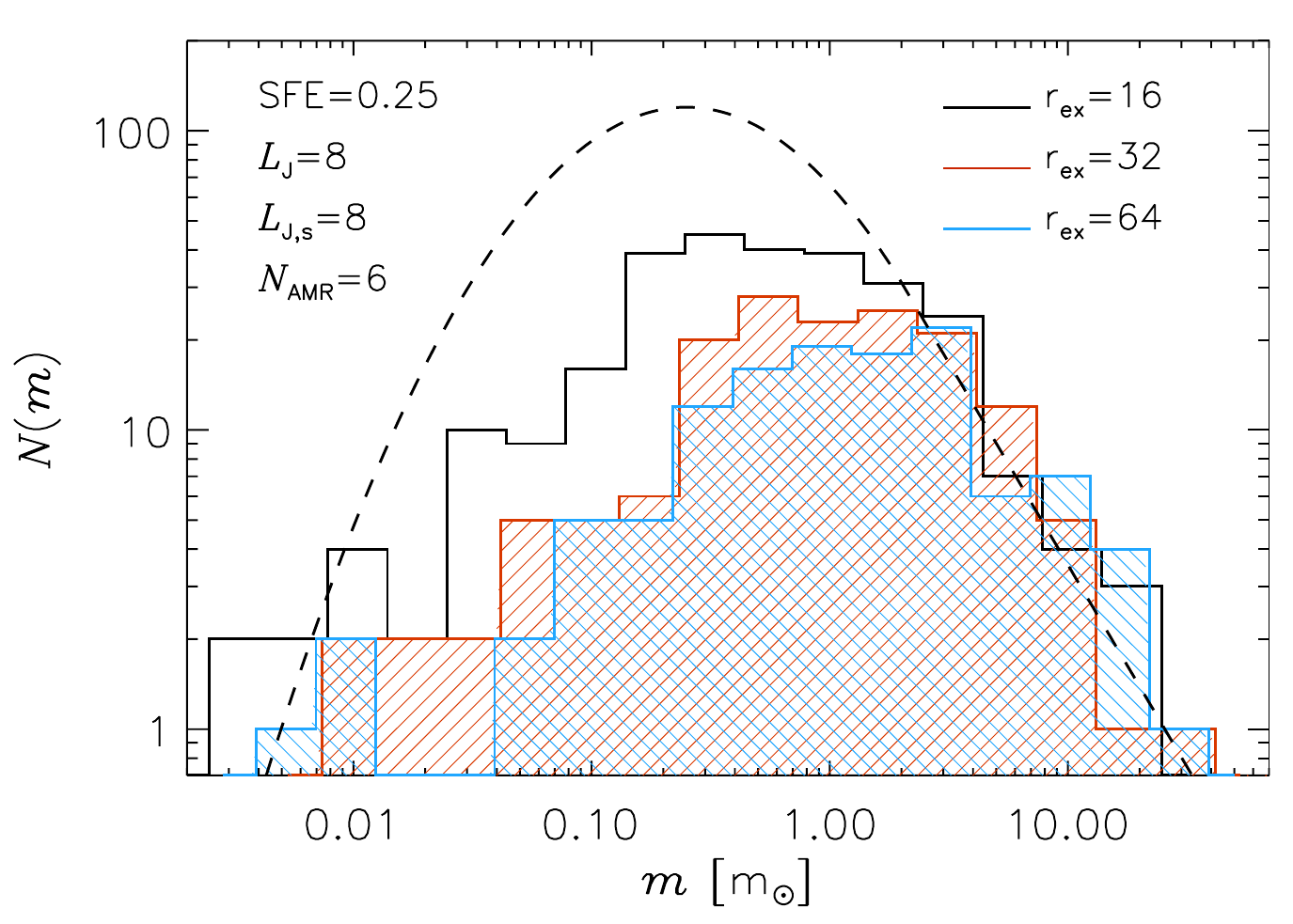}
\includegraphics[width=6cm]{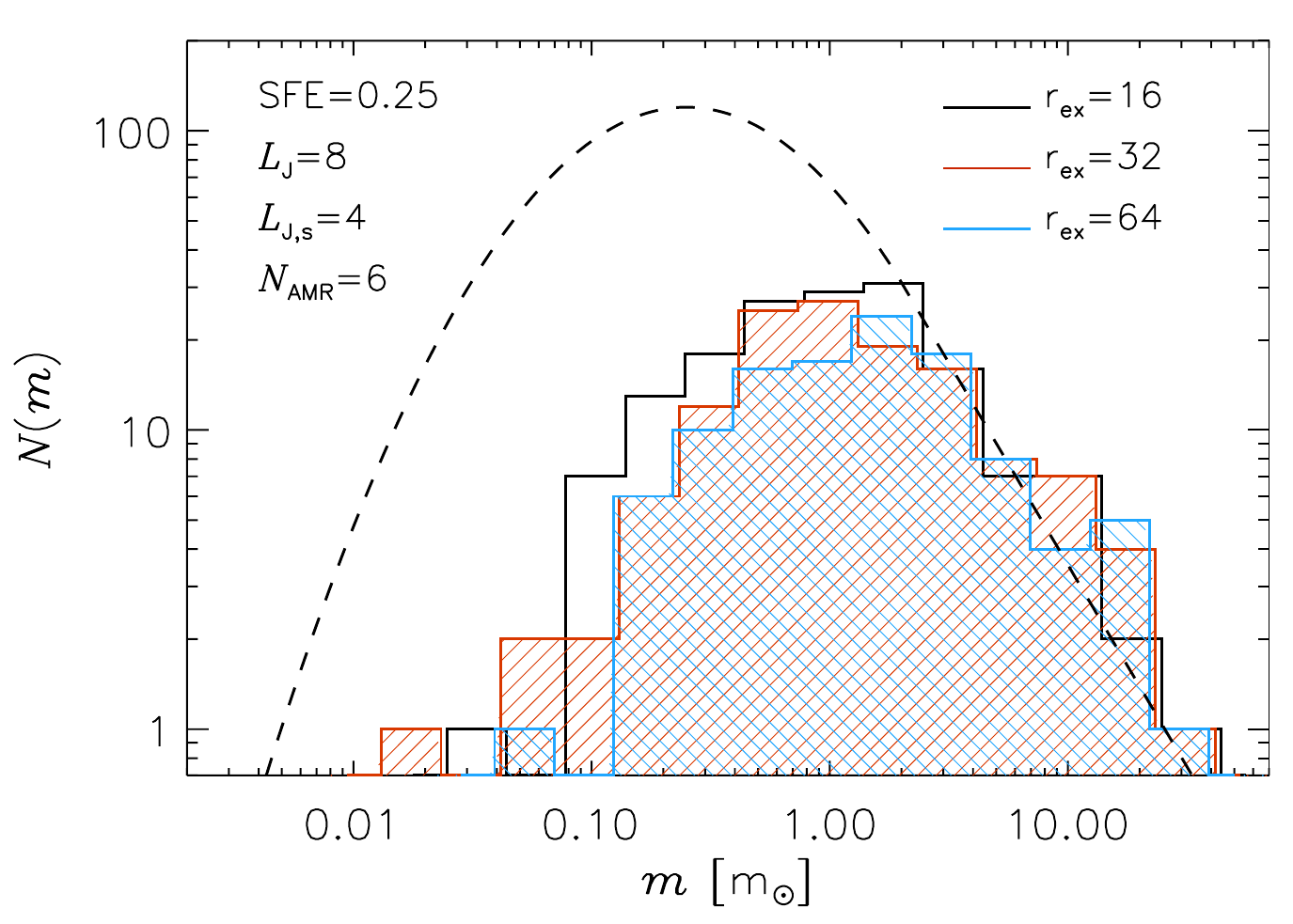}
\includegraphics[width=6cm]{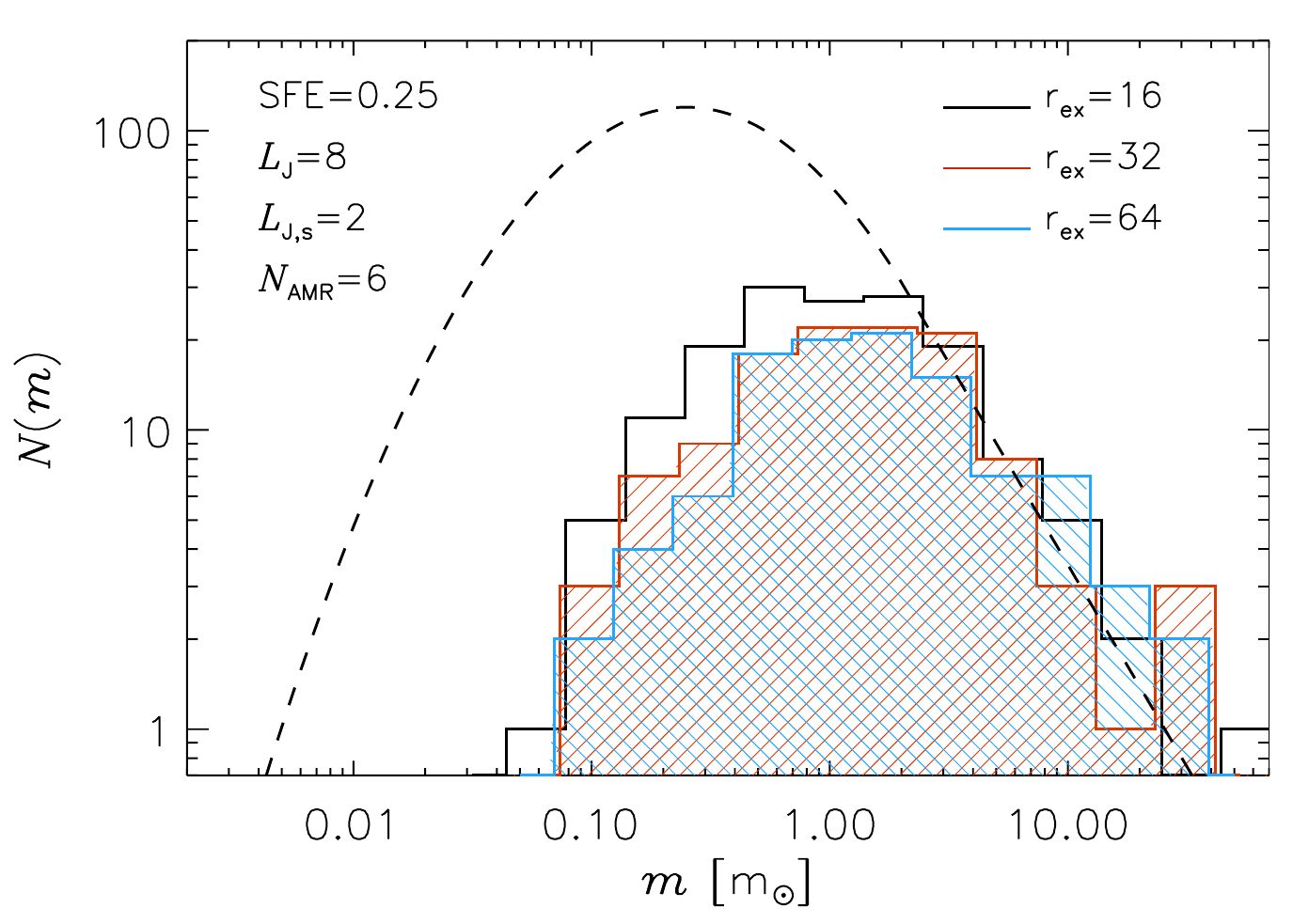}
\includegraphics[width=6cm]{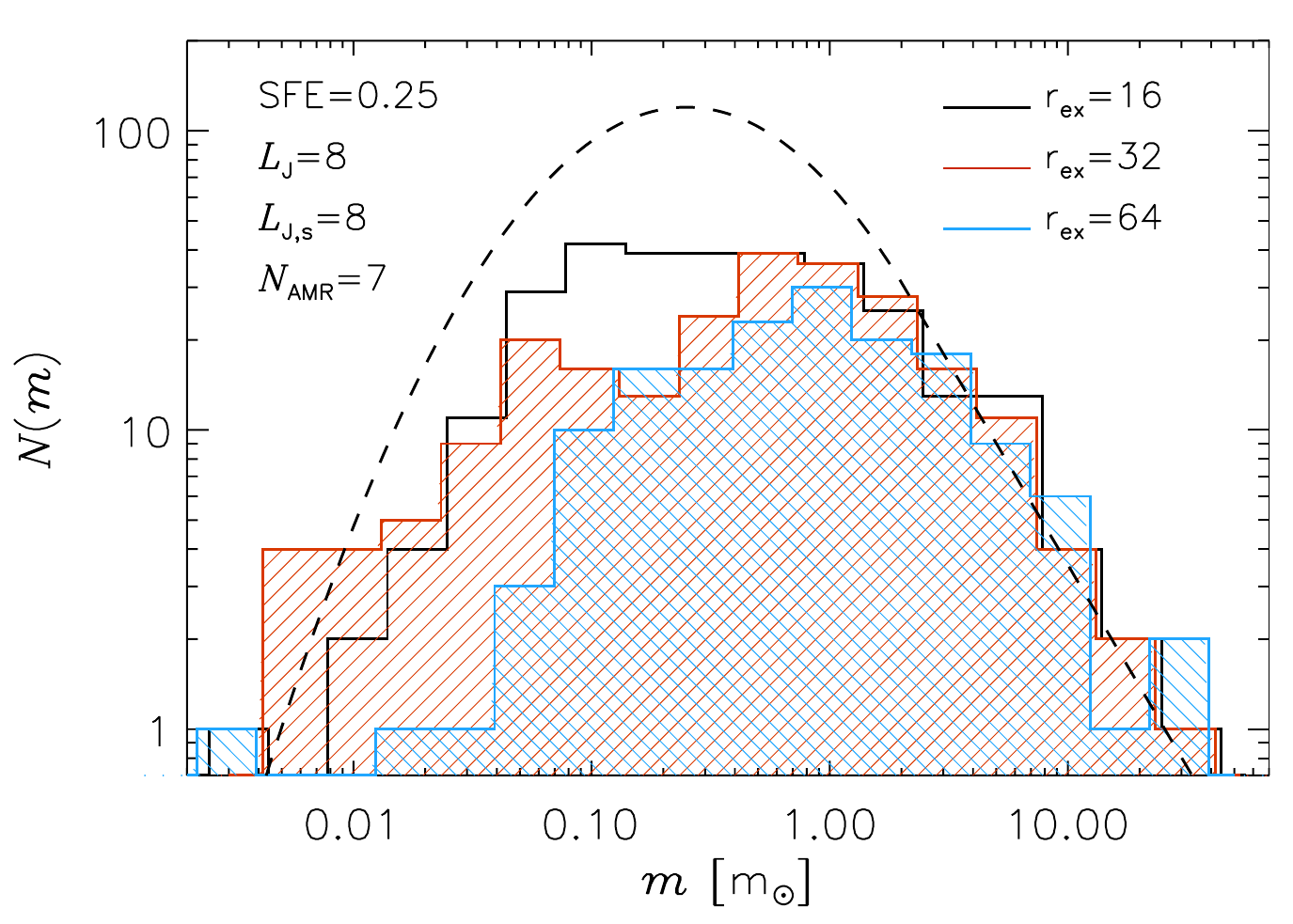}
\includegraphics[width=6cm]{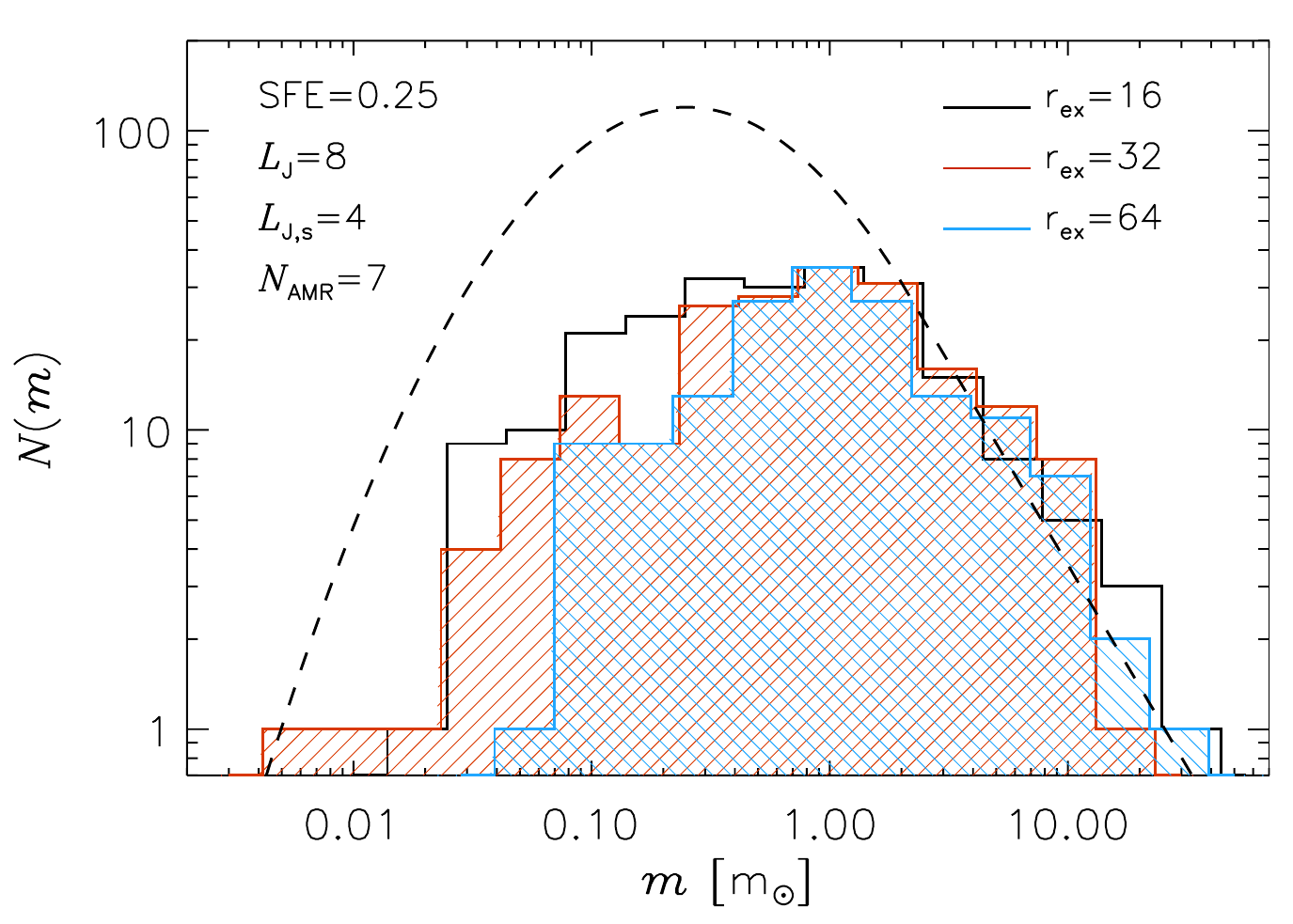}
\includegraphics[width=6cm]{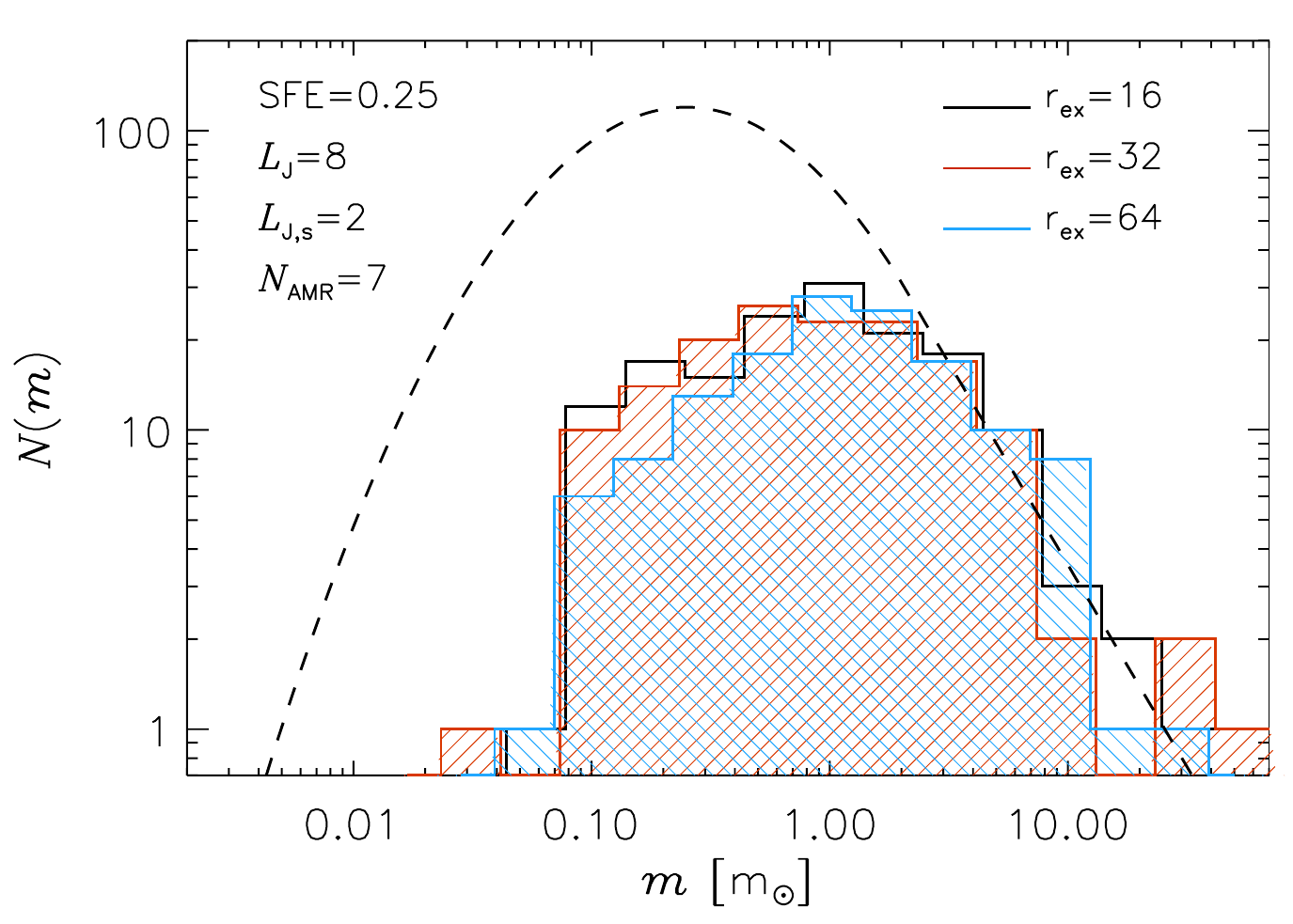}
\includegraphics[width=6cm]{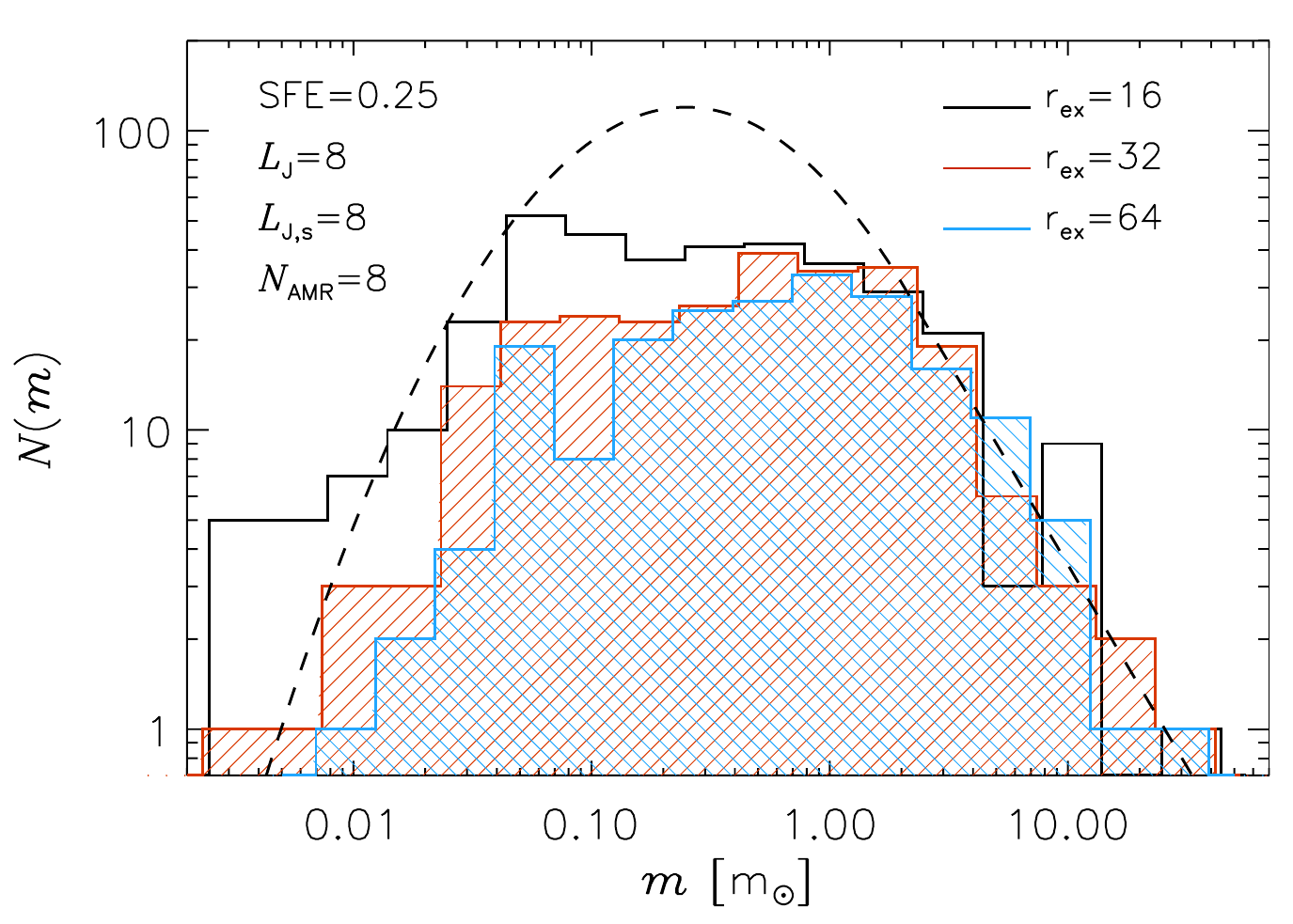}
\includegraphics[width=6cm]{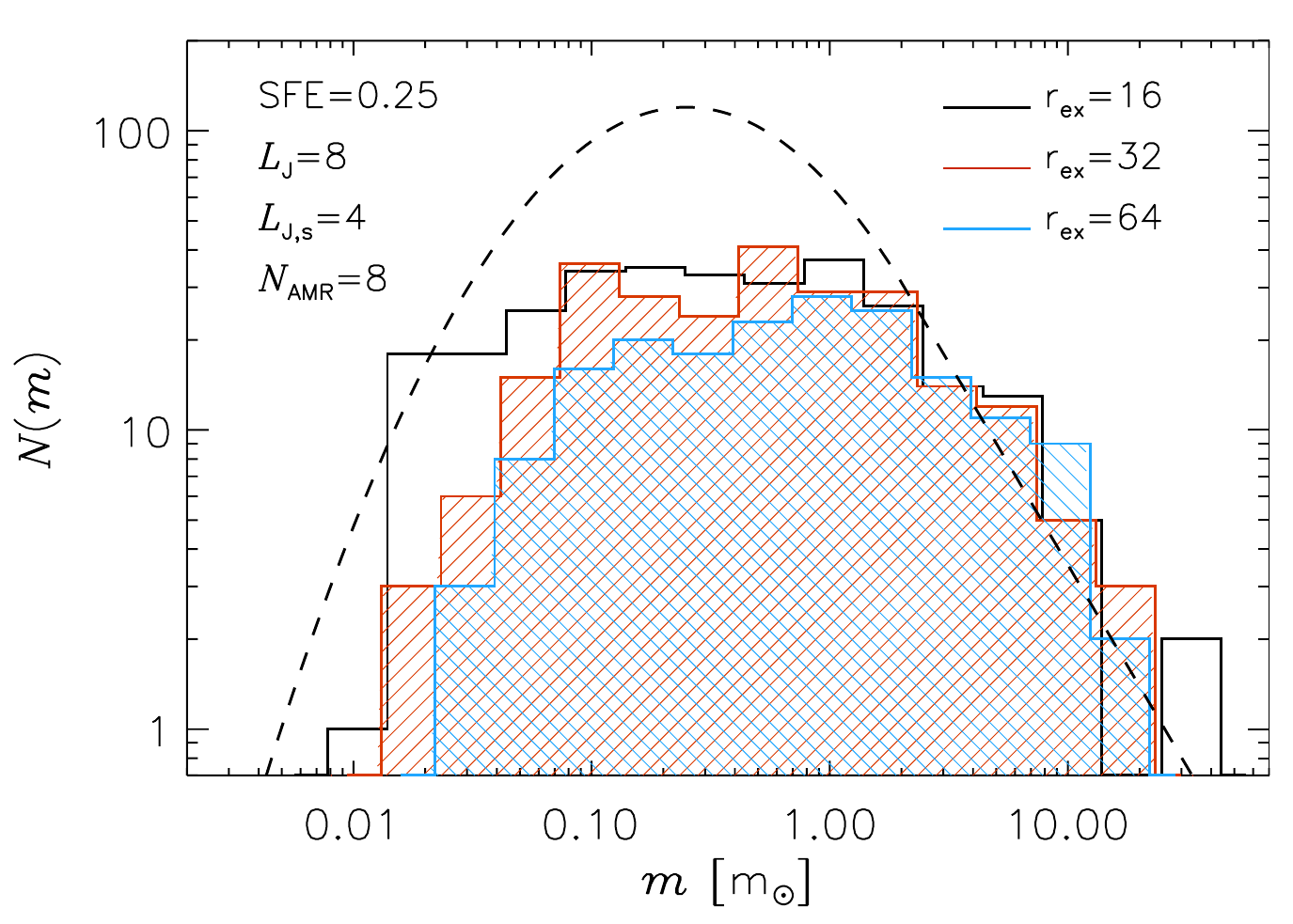}
\includegraphics[width=6cm]{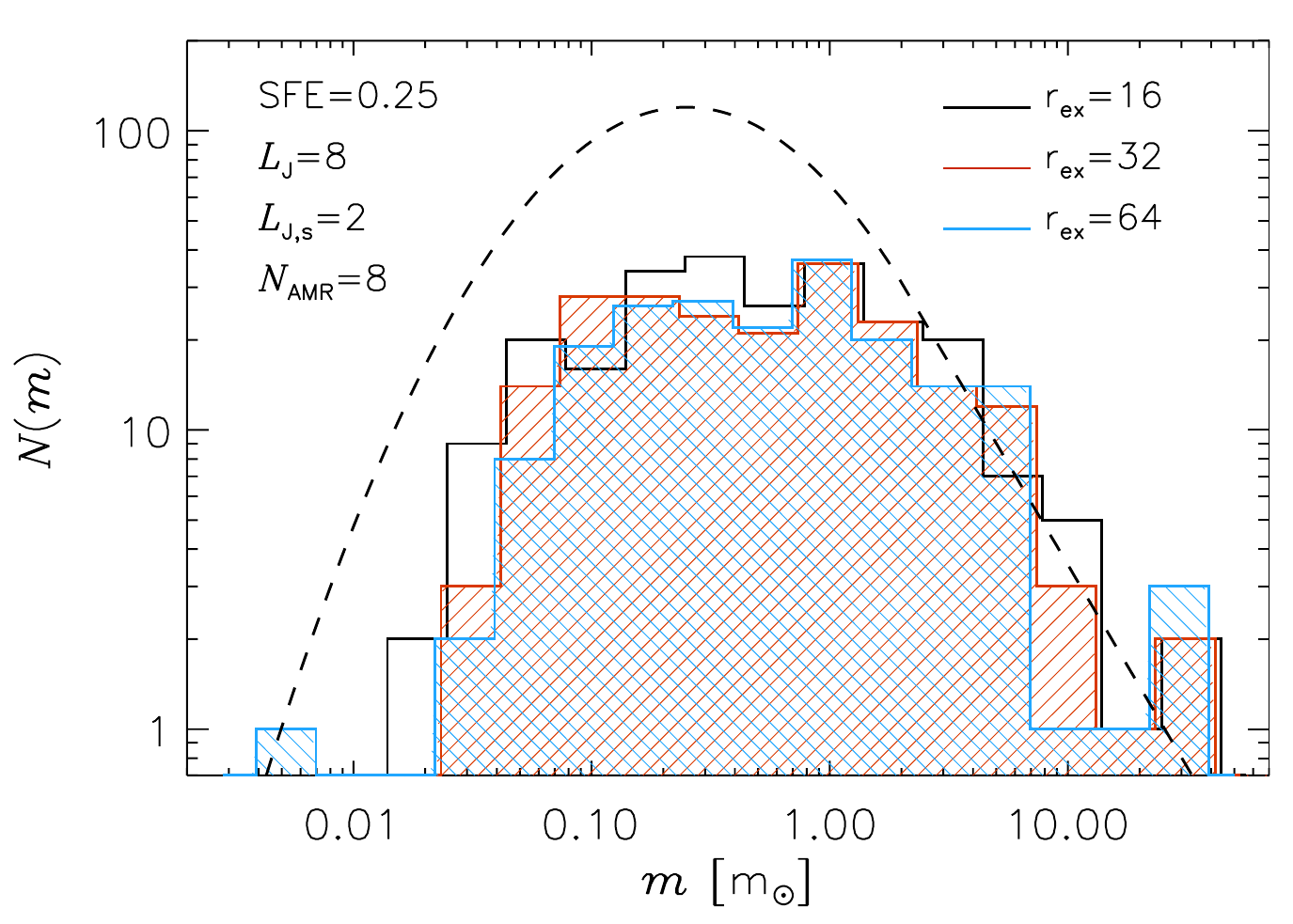}
\caption{IMF for \emph{RUN1}--\emph{RUN27} with $64^3$ root grid and $L_{\rm J}=$8. Each row shows the IMF for runs with a constant number of levels of
refinement $N_{\rm AMR}$ of 6 (\emph{RUN1}--\emph{RUN9}, top), 7 (\emph{RUN10}--\emph{RUN18}, middle), and 8 (\emph{RUN19}--\emph{RUN27}, bottom).
From left to right $L_{\rm J,s}=$8, 4, and 2. Each panel shows three different  values of the exclusion radius, $\rexcl=$16$\Delta x$, 32$\Delta x$, and 64$\Delta x$, as black, red, and blue
histograms, respectively.}
\label{fig:imf_64}
\end{figure*}

\subsection{Impact of sink formation parameters}
In this subsection we discuss how the different numerical parameters in the sink formation model impact the IMF and the neighbor statistics,
$\mathcal{D}$, discussed above.

\emph{Exclusion radius} ($\rexcl$). The purpose of the exclusion volume is to avoid the formation of spurious sink particles near the accretion radius 
of a previously created sink particle. However, secondary density peaks may have a physical origin and lead to the formation of bona fide stellar companions, 
so the exclusion radius should be kept as small as possible in order to resolve as many physical companions as possible. We use the neighbor statistics,
$\mathcal{D}$, as our tool of choice to detect spurious sink particles. As $\rexcl$ is a purely numerical radius, the physics of star formation should not 
yield any special feature in $\mathcal{D}$ at a distance equal to $\rexcl$. A sharp peak at $\rexcl$, for example, is certainly not expected, because $\mathcal{D}$
is predicted to slowly decrease with decreasing distance, as concluded in the previous subsection. On the other hand, the numerical models for sink creation and 
accretion may easily cause a peak in $\mathcal{D}$ at $\rexcl$ (for example, if the gas reservoir within $\racc$ is depleted too rapidly, the density may temporarily 
peak at a distance around $\rexcl$, if $\rexcl\sim\racc$). Thus, we assume that a peak in the neighbor statistics at a distance equal to $\rexcl$ is evidence for 
spurious sink particles and hence too small $\rexcl$; conversely, we assume that if no obvious peak is present at a distance equal to $\rexcl$, the value of $\rexcl$ 
is large enough and the contamination from spurious sink particles is small.  

Fig.~\ref{fig:neighbor_64} shows $\mathcal{D}$ for several test runs with 64$^3$ root grid and six to eight levels of refinement (top to bottom rows of panels). 
Each panel shows the histogram for three different values of $\rexcl$, with the smallest value, $\rexcl=16\Delta x$, indicated by the vertical dashed line. The panels 
show the expected behavior: small values of $\rexcl$ may result in a peak of $\mathcal{D}$ at a distance equal to $\rexcl$, while the peak decreases and 
even disappears as the value of $\rexcl$ is increased. The effect of $\rexcl$ clearly depends on the parameter $L_{\rm J,s}$ and hence on the density threshold where 
sink particles are created, $\rho_{\rm s}$, which is discussed below. One can see that with $L_{\rm J,s}=4$, shown in the middle column of panels of 
Fig.~\ref{fig:neighbor_64}, only the largest value of $\rexcl$ is marginally successful at removing the unphysical peak (one remains at the intermediate resolution 
with seven refinement levels, shown in the central panel), while with $L_{\rm J,s}=2$ the peak is completely removed, also for the smallest value of $\rexcl$. 

The panels in Fig.~\ref{fig:imf_64} show the corresponding IMFs, where it is clear that the spurious sink particles corresponding to the peaks in the neighbor 
statistics are almost exclusively low-mass ones. At masses larger than their peak mass, the IMFs are essentially insensitive to the choice of $\rexcl$.
Furthermore, as for the neighbor histogram, the IMF is practically independent of $\rexcl$ when the lowest value of $L_{\rm J,s}$ is adopted, as shown in 
the right column of panels of Fig.~\ref{fig:imf_64}. 

\begin{figure*}[th]
\center
\includegraphics[width=12cm]{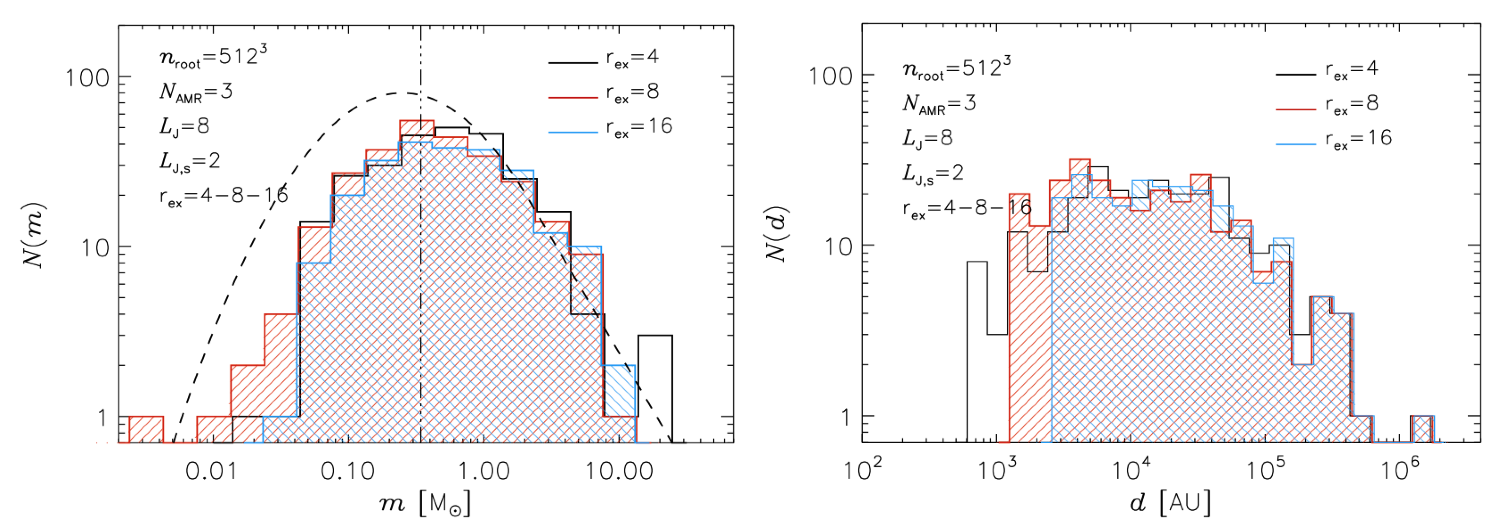}
\caption{IMF and neighbor statistics dependence on the exclusion radius for otherwise well-resolved \emph{RUN29}, \emph{RUN30}, and \emph{RUN31}.
The test runs have a $512^3$ root grid, $L_{\rm J}=8$, and $L_{\rm J,s}=2$. The number of AMR levels
above the root grid is $N_{\rm AMR}=3$. Each panel shows three different values of the exclusion radius,
$\rexcl=$ 4$\Delta x$, 8$\Delta x$, and 16$\Delta x$, as black, red, and blue histograms, respectively. To the left is shown the IMF, and to the right the neighbor statistics.}
\label{fig:exclusion_512}
\end{figure*}

As discussed below, $L_{\rm J,s}=2$ is the optimal value of the numerical Jeans length at the density of sink creation, so we can conclude that, in these test runs,
values of $\rexcl$ as small as $\rexcl=16\Delta x$ do not result in significant numbers of spurious sink particles. In principle, we could adopt an exclusion radius 
as small as $\rexcl=2\, \racc$ (a smaller value, comparable to $\racc$, is guaranteed to generate spurious sinks). To test whether this is acceptable, and also to test whether the 
evidence collected from the low-resolution test runs holds true at higher resolution, we also consider three runs with a 512$^3$ root grid and three levels of 
refinement and three different values of $\rexcl$ ({\it RUN29}, {\it RUN30}, and {\it RUN31}). The dependence of the IMF and $\mathcal{D}$ of these runs on $\rexcl$ 
is shown in Fig.~\ref{fig:exclusion_512}. The neighbor statistics does not show any peak at $\rexcl$, not even at the smallest value, $\rexcl=2\racc=4\Delta x$, 
and the IMFs are barely affected by the choice of $\rexcl$. Thus, we conclude that, as long as $L_{\rm J,s}=2$ (see below), we can adopt the smallest possible 
value of $\rexcl$, that is, $\rexcl=2\racc$. In our convergence-test runs, where $\racc=4\Delta x$, we have adopted $\rexcl=8\Delta x$.

\emph{Density at sink formation} ($L_{\rm J,s}$). The density at sink formation, $\rho_{\rm s}$, has to be as high as possible, so that only bona fide collapsing 
regions are chosen, but not so high that the Jeans length is not resolved at the available spatial resolution. Because the turbulence alone can generate very 
large densities that are not necessarily collapsing regions, $\rho_{\rm s}$ must be at least several times larger than the highest post-shock density in the 
absence of self-gravity at the spatial resolution of the simulation. In our case, a turbulent flow with a sonic Mach number of 10, a characteristic high density 
without self-gravity, is roughly $\sim 10^5$ times the average density (depending on spatial resolution). However, a very large value of $\rho_{\rm s}$ is not 
a sufficient condition to avoid spurious sink particles. In the converging flow in the vicinity of an already-formed sink particle, the gas may reach densities higher 
than $\rho_{\rm s}$ owing to the gravitational potential around the sink, possibly triggering the formation of spurious sinks. To address this concern, we parameterize 
the dependence of the IMF on the density at sink formation through the value of the numerical Jeans length at the density of sink formation, $L_{\rm J,s}$. As 
$\rho_{\rm s}$ is increased (with constant $\Delta x$), the value of $L_{\rm J,s}$ decreases. 

The ideal value of $L_{\rm J,s}$ is a compromise between the requirement of spatially resolving the Jeans length at the end of the AMR hierarchy (high 
$L_{\rm J,s}$), and the need to adopt a large value of $\rho_{\rm s}$ to avoid the creation of spurious sink particles in secondary density peaks that should 
not collapse into separate stars (low $L_{\rm J,s}$). Too low values of $L_{\rm J,s}$ would cause over-merging of density fluctuations, meaning that a single 
star would emerge out of a fragmented region that should have yielded a multiple system (in our experience, this outcome is more likely than the risk of artificial 
fragmentation due to a violation of Truelove's condition). Too high values of $L_{\rm J,s}$ may result in spurious sink particles. 

The dependence of the neighbor statistics and the IMF on $L_{\rm J,s}$ is shown in Figures~\ref{fig:neighbor_64} and \ref{fig:imf_64}. Each column of 
panels in those figures has a constant value of $L_{\rm J,s}$, with $L_{\rm J,s}=8$ on the left, $L_{\rm J,s}=4$ in the middle, and $L_{\rm J,s}=2$ on the 
right. Fig.~\ref{fig:neighbor_64} shows that an artificial peak of spurious sinks at a distance equal to $\rexcl$ is always present for the largest values of
$L_{\rm J,s}$, at any value of $\rexcl$. As $L_{\rm J,s}$ is decreased, the peak decreases, particularly for larger values of $\rexcl$. It is only with $L_{\rm J,s}=2$ 
that the peak is completely gone, independent of $\rexcl$. As mentioned above in relation to the dependence on $\rexcl$, the peak in the neighbor diagram is 
reflected by an increase in the number of low-mass stars, as shown by the IMFs of Fig.~\ref{fig:imf_64}. 

\begin{figure*}[th]
\center
\includegraphics[width=12cm]{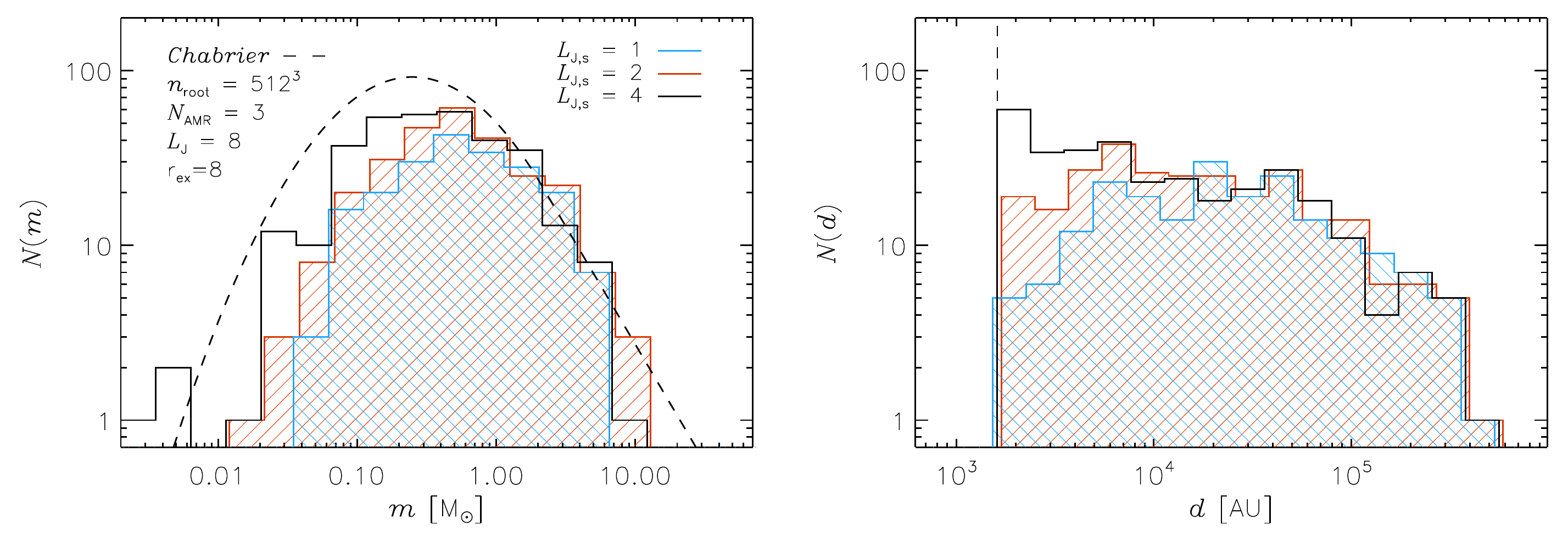}
\caption{IMF and neighbor statistics as a function of the density threshold for creation, quantified by the corresponding Jeans number for otherwise
well-resolved \emph{RUN28}, \emph{RUN30}, and \emph{RUN32}. The test runs have a $512^3$ root grid, $L_{\rm J}=$ 8, $r_{\rm ex}=8\Delta x$.
The number of AMR levels above the root grid is $N_{\rm AMR}=3$. Each panel shows three different values of the Jeans number at sink creation,
$L_{\rm J,s}=$ 4, 2, and 1, as black, red, and blue histograms, respectively. To the left is shown the IMF, and to the right the neighbor statistics.}
\label{fig:LJs_512}
\end{figure*}

To make sure that these results hold true at higher resolution, we consider again the simulations with a $512^3$ root grid, specifically {\it RUN28}, {\it RUN30}, 
and {\it RUN32}, where $L_{\rm J,s}=1$, 2, and 4, respectively. Fig.~\ref{fig:LJs_512} shows that the neighbor histogram yields a peak at $\rexcl$ when $L_{\rm J,s}=4$.
The largest value where the peak is suppressed is $L_{\rm J,s}=2$, as in the lower-resolution runs. Thus, in order to resolve the Jeans length while also avoiding 
spurious sink particles, the best choice is $L_{\rm J,s}=2$, which is the value we have adopted in the main simulations of this work.  
A similar conclusion was reached by \citet{Ostriker+13}.

\subsection{Impact of numerical resolution}
In this subsection we discuss how the numerical resolution impacts the IMF and the nearest-neighbor statistics.

\begin{figure*}[th]
\center
\includegraphics[width=6cm]{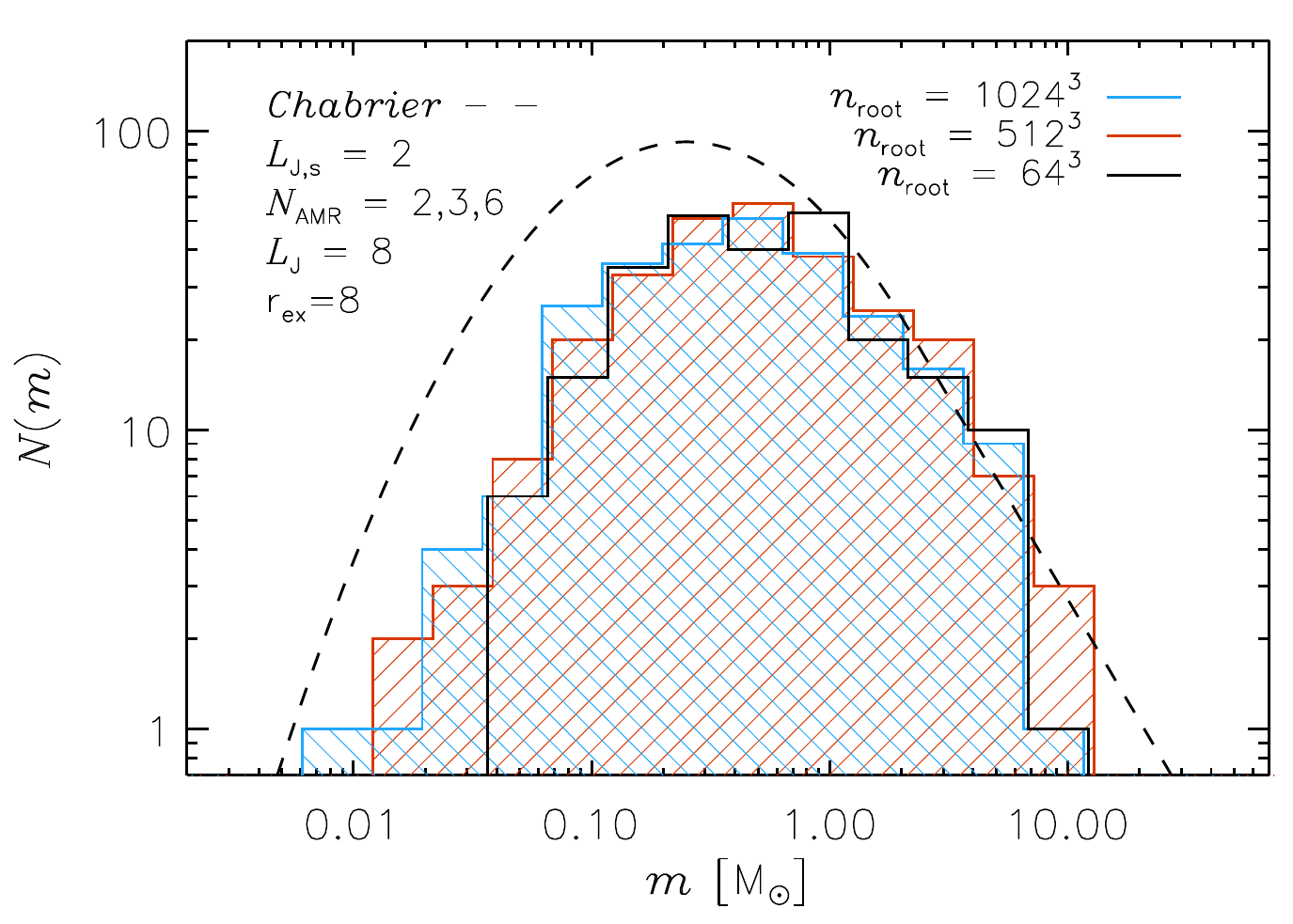}
\includegraphics[width=6cm]{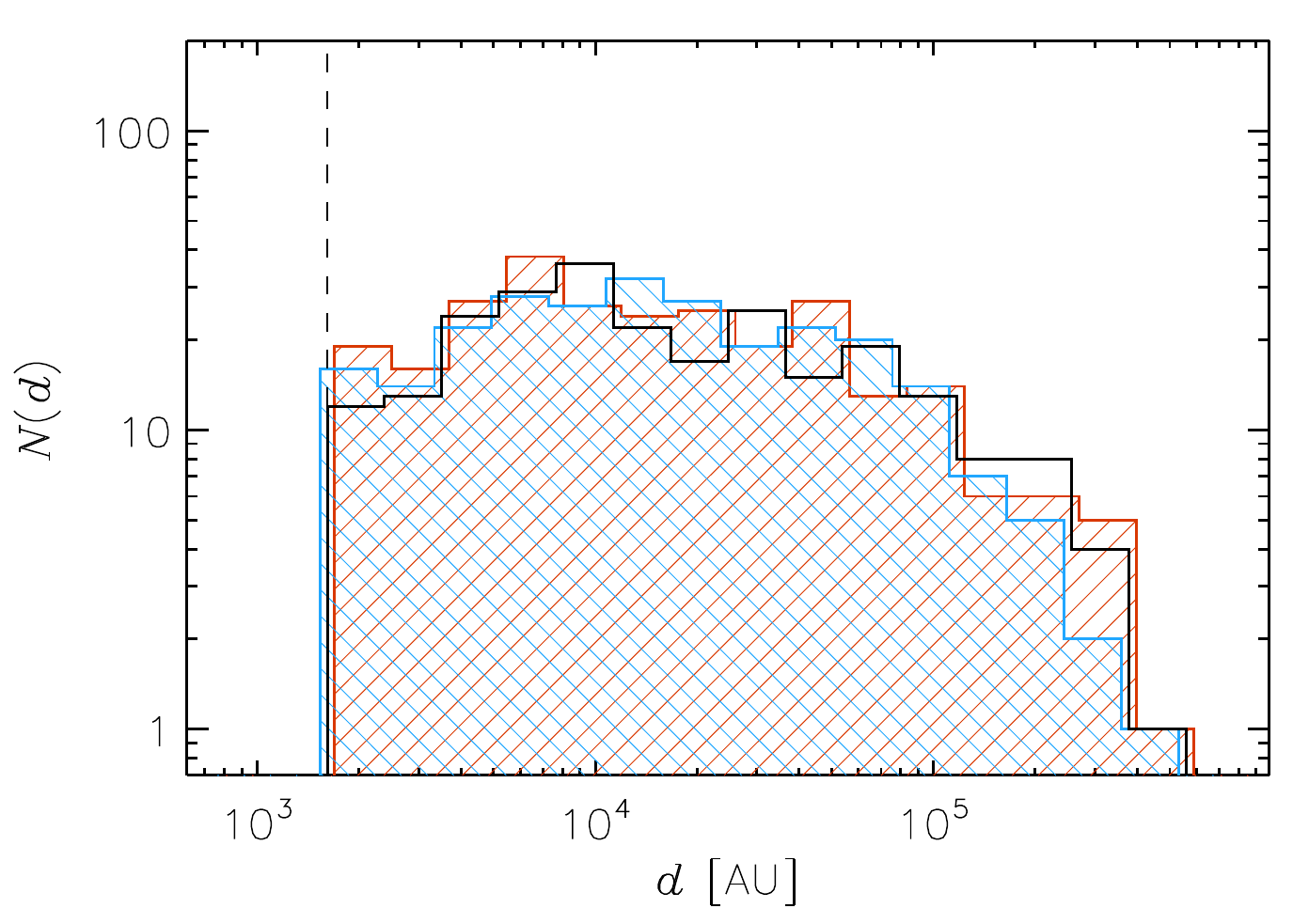}
\caption{IMF and neighbor statistics as a function of the root grid resolution for \emph{RUN33}, \emph{RUN31}, and \emph{RUN34}.
The test runs have $L_{\rm J}=8$, $r_{\rm ex}=8\Delta x$, and $L_{\rm J,s}=2$. Each panel shows three different values of the root grid resolution,
$N_{\rm root}=64^3, 512^3, 1024^3$, and number of AMR levels $N_{\rm AMR}$=6, 3, and 2, as black, red, and blue histograms, respectively.
To the left is shown the IMF, and to the right the neighbor statistics. The dashed vertical line marks the distance equal to $\rexcl$.}
\label{fig:root-grid}
\end{figure*}

\emph{Root grid size.} The impact of changing the root grid size while keeping all other parameters fixed can be seen in Fig.~\ref{fig:root-grid}. 
The root grid resolution impacts how well the turbulence is resolved throughout the volume, but if the number of cells on the refined levels is determined by 
the same Jeans number, the resulting neighbor statistics and IMF are almost identical. We only see a minor difference with a slight increase in the number of 
low-mass stars when using a higher-resolution root grid. This makes sense, since a better resolution of the turbulence facilitates the creation of rare peaks with
both high density and low mass.

\emph{Number of refinement levels.} The number of refinement levels controls the minimum cell size, and correspondingly the maximum density that can be 
reached, if everything else is kept fixed. The most realistic cases for our test runs are shown in the right column of Fig.~\ref{fig:imf_64} corresponding to \emph{RUN7}--\emph{RUN9} 
($N_{\rm AMR}=6$), \emph{RUN16}--\emph{RUN18} ($N_{\rm AMR}=7$), and \emph{RUN25}--\emph{RUN27} ($N_{\rm AMR}=8$), where in all cases relatively 
high densities are reached with a Jeans number at sink formation of $L_{\rm J,s}=2$. The increase in the number of levels of refinement results in a better resolution of 
low-mass stars. This is because higher densities are reached before sink formation, as required for small cores to start collapsing. However, the turbulence is 
not well sampled in these runs because of a too-small root grid resolution.

As found with the convergence runs in table \ref{tc}, by simultaneously increasing the root grid size and the maximal resolution, we can properly populate the 
low-mass end of the IMF. A very high number of AMR levels is in itself not enough to do it.

To summarize this section, we list a guiding set of rules that we have established for setting the numerical parameters of the sink particle model and the 
numerical resolution:
\begin{itemize}
\item{The numerical Jeans length at sink formation should be as high as possible, but in practice it is limited to a rather low ideal value, $L_{\rm J,s}=2$, to avoid 
the creation of spurious sink particles.}
\item{The exclusion radius should be kept as small as possible. As long as $L_{\rm J,s}=2$, the exclusion radius can be set equal to twice the accretion radius,
$\rexcl=2\racc$ ($r_{\rm ex}=8\Delta x$ in our main simulations where $\racc=4\Delta x$).}
\item{A sufficiently high resolution root grid, and a large enough minimum Jeans number $L_{\rm J}$ are needed to resolve both the turbulence and the gravitational 
collapse. A high number of levels of refinement makes it possible to resolve small, high-density cores that are the natal sites of low-mass stars. Therefore, to populate
the IMF toward the low-mass end, it is necessary to increase all three quantities, $N_{\rm root}$, $N_{\rm AMR}$, and $L_{\rm J}$, simultaneously.}
\end{itemize}
These principles have been followed to design the convergence runs listed in table \ref{tc}, and they have made it possible to demonstrate the trend toward a convergence 
of the numerical model.
\bibliographystyle{apj}
\bibliography{apj-jour,MC,padoan,nsf08}

\end{document}